\documentclass[a4paper,11pt]{article}
\usepackage{jheppub} 
\usepackage[T1]{fontenc} 
\usepackage[utf8]{inputenc}

\usepackage{mathrsfs}  

\usepackage{subcaption} 
\usepackage{cleveref} 
\usepackage{slashed} 
\usepackage{physics}
\usepackage{placeins} 
\usepackage[normalem]{ulem}
\usepackage{eucal}

\def\nn{\nonumber}

\newcommand{\mdm}[1]{m_{\chi_{#1}}}
\newcommand{\mom}{\mathbb{M}}
\newcommand{\ana}{\mathcal{A}}

\newcommand{\Lag}{\mathcal{L}} 
\newcommand{\hc}{\text{h.c.}} 
\newcommand{\g}{\gamma}

\newcommand{\eV}{\,\text{eV}}
\newcommand{\gev}{\,\text{GeV}} 

\let\matheul\mathcal

\DeclareMathAlphabet{\mathcal}{OMS}{cmsy}{m}{n}

\graphicspath{{./figures/}}
\title{\boldmath General one-loop expressions for the diagonal and transition electromagnetic moments of spin-1/2 fermions }
\author[a]{Alejandro Ibarra}
\author[a,b]{Merlin Reichard}
\affiliation[a]{Technical University of Munich, TUM School of Natural Sciences, Physics Department, 85748 Garching, Germany}
\affiliation[b]{School of Physics, Korea Institute for Advanced Study,
85 Hoegi-ro, Seoul 02455, Republic of Korea}
\emailAdd{ibarra@tum.de, mreichard@kias.re.kr
}
\abstract{Electrically neutral spin-1/2 fermions can interact via Yukawa or gauge couplings with electrically charged particles, leading to quantum-induced couplings between the neutral fermion and the photon. In this paper, we present general expressions for the diagonal and transition electromagnetic moments of Dirac and Majorana spin-1/2 fermions generated at the one-loop level. 
We also briefly discuss the implications of these results for several neutral spin-1/2 fermions of phenomenological interest, such as neutral components of $SU(2)_L$ multiplets (which are potential candidates for dark matter) or Standard Model neutrinos.
}
\begin{document} 
\maketitle
\flushbottom

\section{Introduction}

Electrically neutral fermions appear in many extensions of the Standard Model. Further, if stable in cosmologically time scales, they constitute promising candidates for the dark matter of our Universe.  Although these particles do not couple directly to the photon through their electric charge, electromagnetic interactions are generically generated at the quantum level whenever the neutral fermion interacts with electrically charged particles. The resulting electromagnetic moments therefore provide a sensitive probe of the interactions and symmetries of the underlying theory, and may play an important role in laboratory searches, astrophysical observations and dark matter search experiments.

The most general electromagnetic vertex between two on-shell spin-$1/2$ fermions can be parametrized in terms of the charge, magnetic dipole, electric dipole and anapole form factors, or equivalently, at small momentum transfer, in terms of the electric charge and charge radius, magnetic and electric dipole moments, and anapole moment~\cite{Bernstein:1963qh}. These interactions can be diagonal, when the initial and final fermions coincide, or transitional, when they correspond to different mass eigenstates. Their properties depend strongly on whether the fermions are Dirac or Majorana particles and on the discrete symmetries of the underlying interaction. In particular, while a neutral Dirac fermion can in general possess all four electromagnetic moments, the only allowed diagonal electromagnetic moment of a Majorana fermion is the anapole moment. Majorana fermions can nevertheless possess non-vanishing transition moments.

Electromagnetic moments have been calculated in a variety of specific models and in several phenomenologically relevant limits (see {\it e.g}~\cite{Kopp:2014tsa,Garny:2015wea,Ibarra:2015fqa,Sandick:2016zut,Hisano:2018bpz,Baker:2018uox,Herrero-Garcia:2018koq,Ibarra:2022nzm,Ibarra:2024mpq}). However, many applications require expressions valid for arbitrary masses of the external and internal particles, non-degenerate external states, general left- and right-handed couplings, and both scalar and vector particles in the loop. Such general expressions are particularly useful for theories containing several neutral fermions, where transition moments may be relevant, as well as for models in which the neutral state belongs to a non-trivial electroweak multiplet. Vector-boson contributions also require some care, since the transverse and longitudinal degrees of freedom have to be consistently included in order to obtain a physical, gauge-independent result.

In this work we derive general one-loop expressions for the diagonal and transition electromagnetic moments of electrically neutral spin-$1/2$ fermions (Dirac or Majorana), induced by their interaction with charged fermions and charged vectors or scalars. For the calculation of the moments induced by gauge vector bosons we will employ the background field method~\cite{Denner:1994nn,Papavassiliou:1994yi,Hashimoto:1994ct,BINOSI20091}, to ensure that the result is finite and gauge-independent.

As an illustration, we apply our results to neutral fermions belonging to non-trivial representations of $SU(2)_L$, and we calculate their electromagnetic multipoles induced by the interactions with the $W^\pm$ bosons (and the Goldstone bosons). We consider both a neutral state belonging to a single electroweak multiplet and the more general situation in which the neutral components of two different multiplets mix. In the latter case, electroweak symmetry breaking also induces mixing in the charged sectors and couplings to the charged Goldstone bosons, which can be incorporated directly into our general expressions. We also recover the known electromagnetic moments of Dirac and Majorana neutrinos in the Standard Model ~\cite{Salam:1957st,Fujikawa:1980yx,Kayser:1982br,Shrock:1982sc,Bernabeu:2000hf,Papavassiliou:2002te,Papavassiliou:2003rx}  and show how the same formulas can be used to calculate additional contributions induced by new particles or interactions.

The paper is organized as follows. In Section 
\ref{sec:Dirac} we derive the one-loop electromagnetic moments for Dirac fermions coupled to charged scalars and vector bosons. In Section \ref{sec:Majorana} we extend the analysis to Majorana fermions. In Section \ref{sec:SU2-multiplets} we apply the general results to electroweak multiplets and to neutrinos. We summarize our conclusions in Section \ref{sec:Conclusions}. The general loop functions and several useful limiting expressions are collected in the appendices, together with the effective description of the mixing of different $SU(2)_L$ multiplets.

\section{Multipole moments at one-loop for spin-1/2 Dirac fermions}
\label{sec:Dirac}

We consider two Dirac fermions, $\chi_i$ and $\chi_f$, with masses $\mdm{1}$ and $\mdm{2}$ respectively, and identical gauge quantum numbers. The most general interaction vertex of these two on-shell fermions with the photon can be parametrized as~\cite{Nieves:1981zt,Kayser:1983wm,Kayser:1984ge,Nowakowski:2004cv,Mohapatra:2004,Giunti:2014ixa})
\begin{equation}\label{eq:vertex_parameterization}
\mathcal{M}^\mu_{fi}=(\g^\mu -q^\mu \slashed{q}/q^2)\left[(f_{Q,fi}(q^2) +f_{A,fi}(q^2)q^2\g^5\right]  +i \sigma^{\mu\nu}q_\nu\left[f_{M,fi}(q^2) +i f_{E,fi}(q^2)\g^5\right] \;,
\end{equation}
where $i$ ($f$) labels the initial (final) external mass eigenstate and  $f_Q$, $f_A$, $f_M$, and $f_E$ are respectively the charge, anapole, magnetic dipole and electric dipole form factor. For small momentum transfer, the form factors reduce to their corresponding static analogues:
\begin{align}
&f_{Q,fi}= e Q_{fi} + q^2 b_{fi}+{\cal O}(q^4),
&f_{A,fi}= \ana_{fi}+{\cal O}(q^2),\\
&f_{M,fi} = \mu_{fi} +{\cal O}(q^2),
&f_{E,fi}= d_{fi}+{\cal O}(q^2),
\end{align}
where $Q_{fi}$ is the charge, $b_{fi}$ the charge radius, $\mu_{fi}$ the magnetic dipole moment, $d_{fi}$ the electric dipole moment, and $\ana_{fi}$ the anapole moment. If $f=i$, the moment is referred to as \textit{diagonal} moment, whereas for $f\neq i$, it is a \textit{transition} or \textit{off-diagonal} moment.

The above form factors and moments are generated at the quantum level if the spin 1/2 fermions $\chi_i$ interact with charged particles. We consider in this paper the Yukawa interaction with a fermion $f$ and complex scalar $S$, with electric charges $Q_f$ and $Q_S$,
\begin{equation}
    \Lag = \overline\chi_i \left[c^i_L P_L + c^i_R P_R\right]  f  S^* + \hc
\label{eq:Lag-scalar}
\end{equation}
as well as the gauge interaction with the fermion $f$ and a gauge boson $V^\mu$, wich electric charges $Q_f$ and $Q_V$
\begin{equation}
    \Lag = \overline\chi_i \g^\mu \left[v^i_L P_L + v^i_R P_R\right] f V^\dagger_\mu +\hc.
\label{eq:Lag-vector}
\end{equation}
Here, $P_L$ and $P_R$ project the Dirac fermion into its left- and right components, respectively, while the complex coupling constants $c^i_{L/R}$ and $v^i_{L/R}$ parameterize the coupling strengths of the scalar or vector to the left- and right components of the Dirac fermions  $\chi_i$.

\begin{figure}[t]
    \centering
        \includegraphics[width=0.24\linewidth]{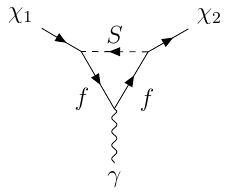}
        \includegraphics[width=0.24\linewidth]{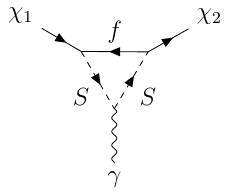}
        \includegraphics[width=0.24\linewidth]{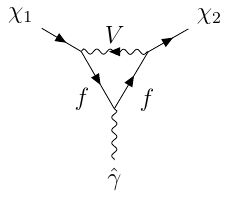}
        \includegraphics[width=0.24\linewidth]{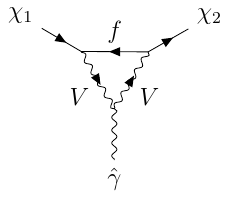}
    \caption{One loop diagrams giving rise to the electromagnetic interactions between the fermions $\chi_1$ and $\chi_2$ via scalar loops and vector loops. For the latter, the diagrams with the longitudinal modes of $V$ should also be included in the evaluation. For the case in which $\chi_{1,2}$ obey the Majorana condition, the corresponding conjugated diagrams also contribute.}
    \label{fig:scalar_vector_transition_general}
\end{figure}

These interactions generate at the quantum level a coupling of the fermion $\chi_i$ to the fermion $\chi_j$ and a photon, via the one-loop diagrams shown in \cref{fig:scalar_vector_transition_general}. In order to ensure the gauge independence in our calculation we will use the background field method for the diagrams with gauge bosons in the loop. This is achieved by  replacing the photon field with its background field analog, $\gamma\rightarrow \hat\gamma$, and adjusting the vertices accordingly. In the background Feynman gauge, the vertices which are adjusted at the one-loop level read \cite{Denner:1994xt}
\begin{equation}
    i\hat{\Gamma}^{\mu\nu\rho}_{\gamma VV}(k_1,k_2,k_3) = -ie\left[g_{\nu\rho}(k_3-k_2)_\mu + g_{\mu\nu}(k_2-k_1+k_3)_\rho +g_{\rho\mu} (k_1-k_3-k_2)_\nu \right]
\end{equation}
and
\begin{equation}
    i\hat{\Gamma}^{\mu\nu\rho}_{\gamma GV}(k_1,k_2,k_3) = 0,
\end{equation}
where $G$ denotes the Goldstone mode associated with the spontaneous symmetry breaking giving rise to the mass of the gauge boson $V$.

We then extract the electromagnetic moments from \cref{fig:scalar_vector_transition_general} by projecting out the relevant Lorentz structure and taking the limit $q^2\rightarrow 0$. The contribution from scalar (S) and vector (V) loops to the electromagnetic moments of two Dirac particles read, respectively,~\footnote{
The Feynman diagrams were evaluated with the help of \texttt{Package-X} \cite{Patel:2015tea,Patel:2016fam} and \texttt{FeynCalc} \cite{MERTIG1991345,Shtabovenko:2016sxi,Shtabovenko:2020gxv}.  The Feynman diagrams in the Figures were drawn with \texttt{TikZ-Feynman} \cite{ELLIS2017103}.}
\begin{align}\label{eq:general_scalar_result}
(\mom_{fi}^S)^D = \frac{e Q_f}{32 \pi^2 }\biggr\{&\left[c_L^f (c_L^i)^* \pm c_R^f (c_R^i)^* \right] \matheul{F}_\mom^S\left(\frac{m_f}{\mdm1}, \frac{m_S}{\mdm1}, \frac{\mdm2}{\mdm1}\right)\nn \\
+ &\left[c_L^f (c_R^i)^* \pm c_R^f(c_L^i)^*\right]  \matheul{G}_\mom^S\left(\frac{m_f}{\mdm1}, \frac{m_S}{\mdm1}, \frac{\mdm2}{\mdm1}\right)\biggr\},
\end{align}
and
\begin{align}\label{eq:general_vector_result}
(\mom_{fi}^V)^D = \frac{eQ_f}{32 \pi^2 }\biggr\{&\left[v_L^f (v_L^i)^* \pm v_R^f (v_R^i)^* \right] \matheul{F}_\mom^V\left(\frac{m_f}{\mdm1}, \frac{m_V}{\mdm1}, \frac{\mdm2}{\mdm1}\right)\nn \\
+ &\left[v_L^f (v_R^i)^* \pm v_R^f (v_L^i)^*\right]  \matheul{G}_\mom^V\left(\frac{m_f}{\mdm1}, \frac{m_V}{\mdm1}, \frac{\mdm2}{\mdm1}\right)\biggr\}.
\end{align}
Here, $\mom = \mu,d,\ana,b$ are the EM moments, and the positive sign corresponds to $\mom = \mu, b$, and the negative sign for $\mom = d,\ana$. 
Note that  \cref{eq:general_vector_result} only accounts for the contribution from the transverse components of the vector boson; the contribution from the longitudinal component (the Goldstone mode) is  model-dependent and must be included using \cref{eq:general_scalar_result}.
For the inverse process, $\chi_2 \rightarrow \chi_1 \gamma$, the hermicity of the electromagnetic current implies that \cite{Giunti:2014ixa}  
\begin{equation}\label{eq:hermicity}
    \mom^{S,V} = (\mom^{S,V})^\dagger,
\end{equation}
such that the loop functions for the ``forward'' and ``backward'' directions are related as
\begin{align}\label{eq:hermicity_loop_functions}
\matheul{F}^{S,V}_\mom \left(\frac{m_f}{\mdm1},\frac{m_{S/V}}{\mdm1},\frac{\mdm{2}}{\mdm1}\right) &= \matheul{F}^{S,V*}_\mom \left(\frac{m_f}{\mdm{2}},\frac{m_{S/V}}{\mdm{2}},\frac{\mdm1}{\mdm{2}}\right),\\
\matheul{G}^{S,V}_\mom  \left(\frac{m_f}{\mdm1},\frac{m_{S/V}}{\mdm1},\frac{\mdm{2}}{\mdm1}\right) &= \pm \matheul{G}^{S,V*}_\mom\left(\frac{m_f}{\mdm{2}},\frac{m_{S/V}}{\mdm{2}},\frac{\mdm1}{\mdm{2}}\right).
\end{align}
Therefore, without loss of generality, we focus on the ``forward'' direction for the remaining part of the discussion.
 For anti-fermions, the EM moments $\overline{\mom}$ are related to $\mom$ via CPT-invariance~\cite{Giunti:2014ixa}, such that
\begin{align}
\begin{gathered}  
\overline{Q}^{S,V} = - (Q^{S,V})^*,\\
\overline{\mu}^{S,V} = -(\mu^{S,V})^*,\quad  \overline{d}^{S,V} = - (d^{S,V})^*,\\
\overline{\ana}^{S,V} =  (\ana^{S,V})^*, \quad \overline{b}^{S,V} =  -(b^{S,V})^*.
\end{gathered}  
\end{align}

The explicit expressions for the loop functions $\matheul{F}_\mom^{S,V}$ and $\matheul{G}_\mom^{S,V}$ are presented  in \cref{appendix:Loopfunctions_q_zero}, and particularized for the diagonal case in \cref{app:loop_functions_diagonal}, 
for large mass splittings in \cref{app:loop_functions_large_mass_splitting}, and finally for the limit of small external masses in \cref{app:loop_functions_light_external_states}. 

Without loss of generality, we assume that only the Yukawa couplings to the right-chiral fermions are carrying a CP-violating phase, denoted by $\phi_i$ for the scalar mediator and $\Phi_i$ for the vector mediator, \textit{i.e.} we substitute in \cref{eq:general_scalar_result,eq:general_vector_result} $c_R^i \rightarrow c_R^i e^{i\phi_i}$ and $v_R^i\rightarrow v_R^i e^{i\Phi_i}$ and consider all Yukawa couplings to be real. Further, we can express the real parameters $c^i_{L/R}$ ($v^i_{L/R}$) in polar coordinates by introducing the P-violating angle $\theta_i$ ($\Theta_i$), such that
\begin{equation}
    c_L^i = c_i \sin\theta_i,\quad c^i_R = c_i \cos\theta_i,\quad
    v_L^i = v_i \sin\Theta_i, \quad v_R^i = v_i \cos\Theta_i.
\end{equation}
In this parameterization, \cref{eq:general_scalar_result} becomes
\begin{align}
(\mom_{fi}^S)^D = \frac{e Q_f c_i c_f}{32 \pi^2 }\biggr\{&\left[\sin\theta_i \sin\theta_f \pm e^{i(\phi_f -\phi_i)}  \cos\theta_i \cos\theta_f \right] \matheul{F}_\mom^S\left(\frac{m_f}{\mdm1}, \frac{m_S}{\mdm1}, \frac{\mdm2}{\mdm1}\right)\nn \\
+ &\left[e^{-i\phi_i} \cos\theta_i\sin\theta_f \pm e^{i \phi_f} \cos\theta_f \sin\theta_i \right]  \matheul{G}_\mom^S\left(\frac{m_f}{\mdm1}, \frac{m_S}{\mdm1}, \frac{\mdm2}{\mdm1}\right)\biggr\},
\end{align}
and correspondingly \cref{eq:general_vector_result} in this new parameterization can be obtained by replacing $S\rightarrow V$, $c_{i,f}\rightarrow v_{i,f}$, $\theta_{i,f}\rightarrow \Theta_{i,f}$ and $\phi_{i,f}\rightarrow \Phi_{i,f}$ in the above equation. 

\begin{figure}
    \centering
    \includegraphics[width=.49\linewidth]{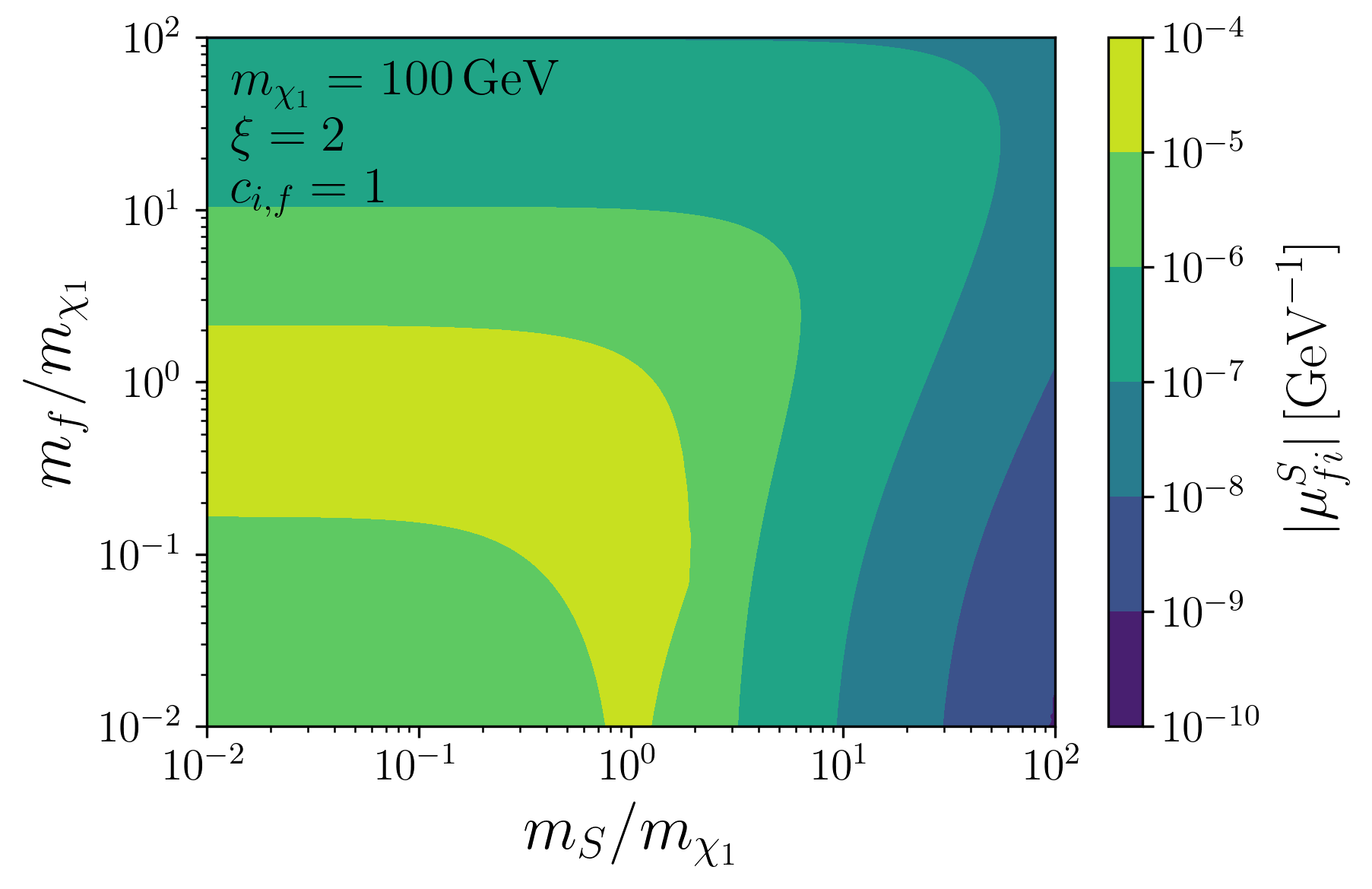}
    \includegraphics[width=.49\linewidth]{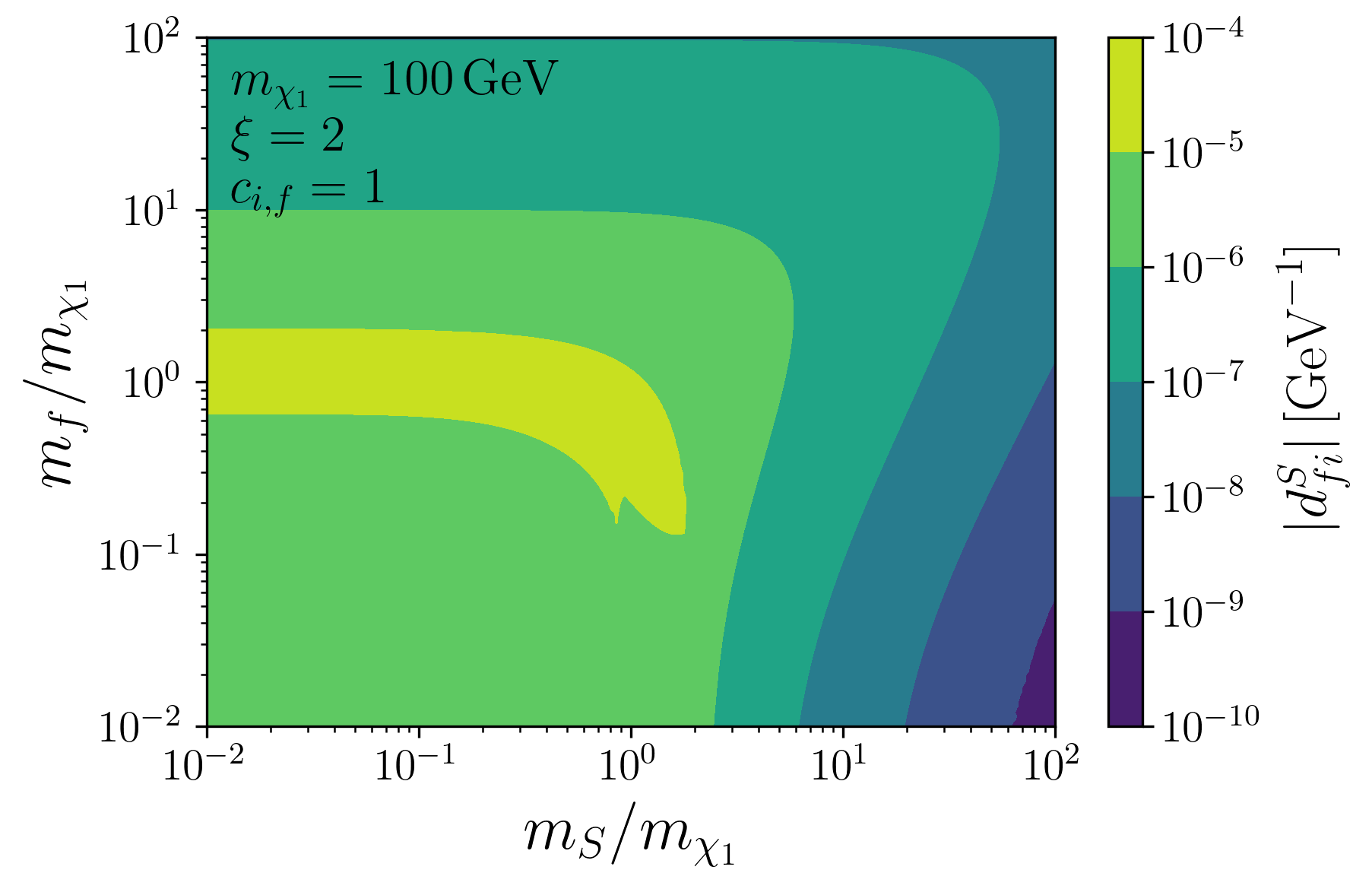}
    \includegraphics[width=.49\linewidth]{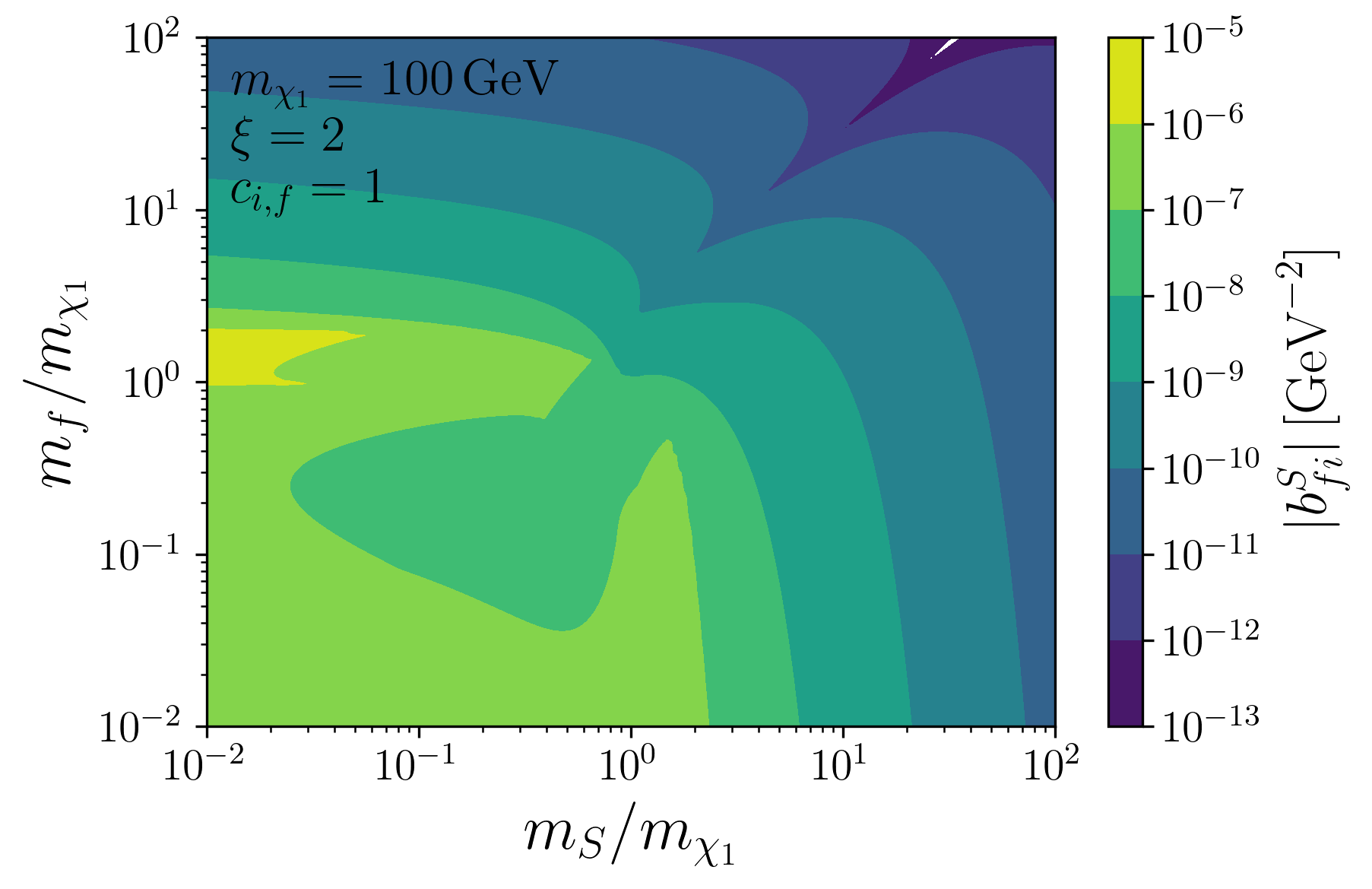}
    \includegraphics[width=.49\linewidth]{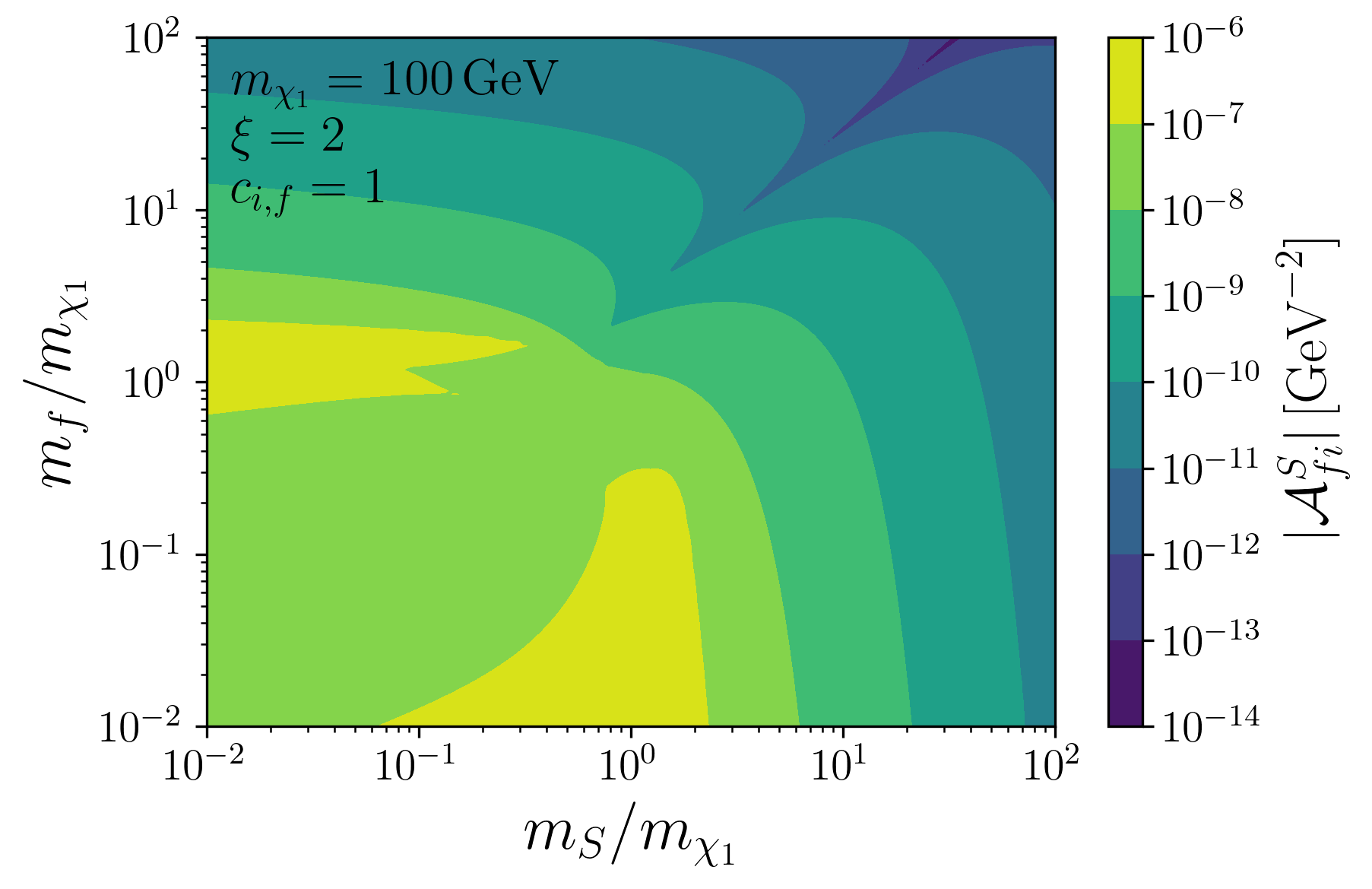}
    \caption{Scalar contribution to the transition EM moments for fixed $\mdm2/\mdm1 = 2$ maximized with respect to $\theta_{i,f}$ and $\phi_{i,f}$.}
    \label{fig:scalar_max_EMmoments}
\end{figure}

\begin{figure}
    \centering
    \includegraphics[width=.49\linewidth]{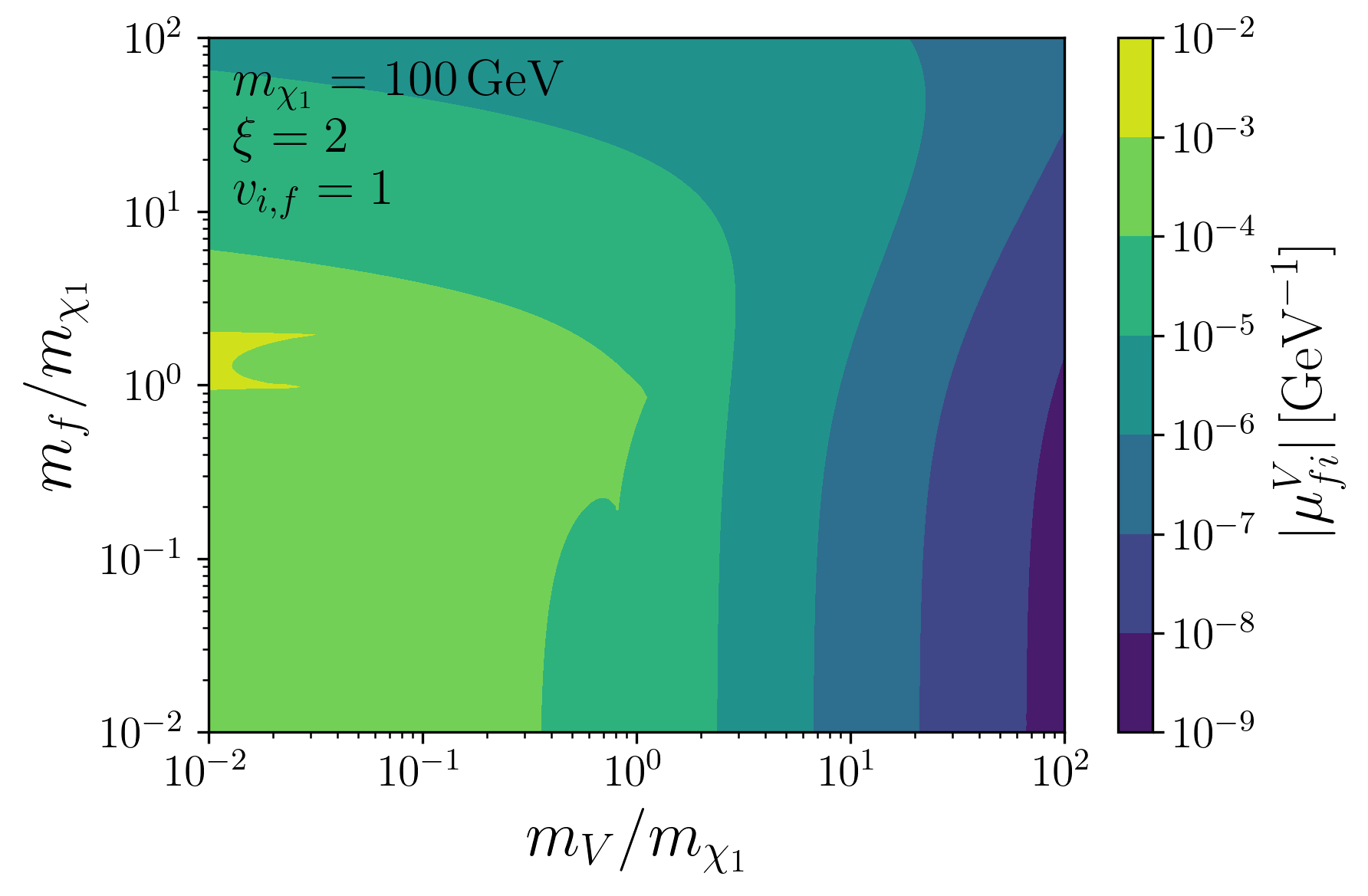}
    \includegraphics[width=.49\linewidth]{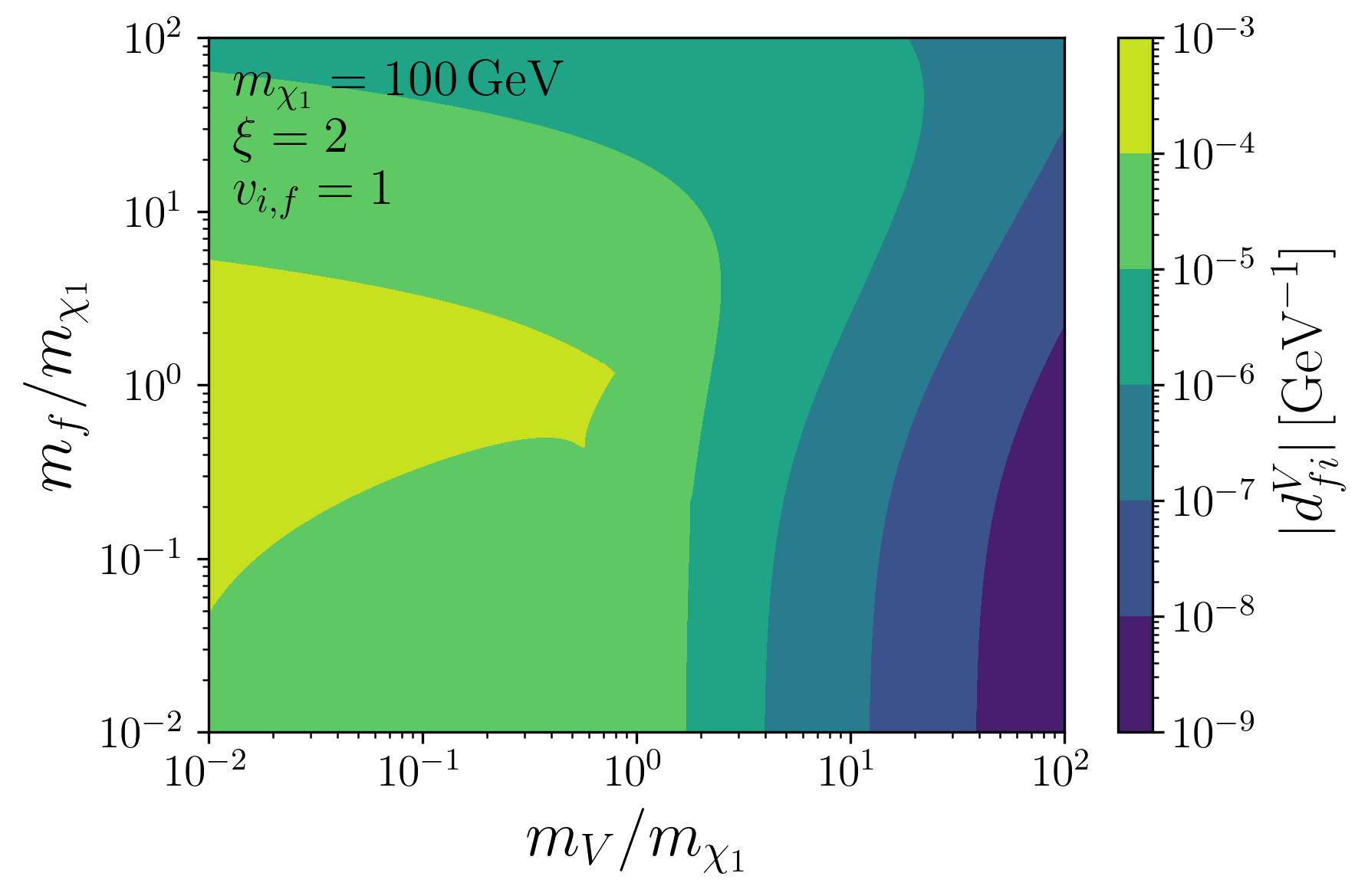}
    \includegraphics[width=.49\linewidth]{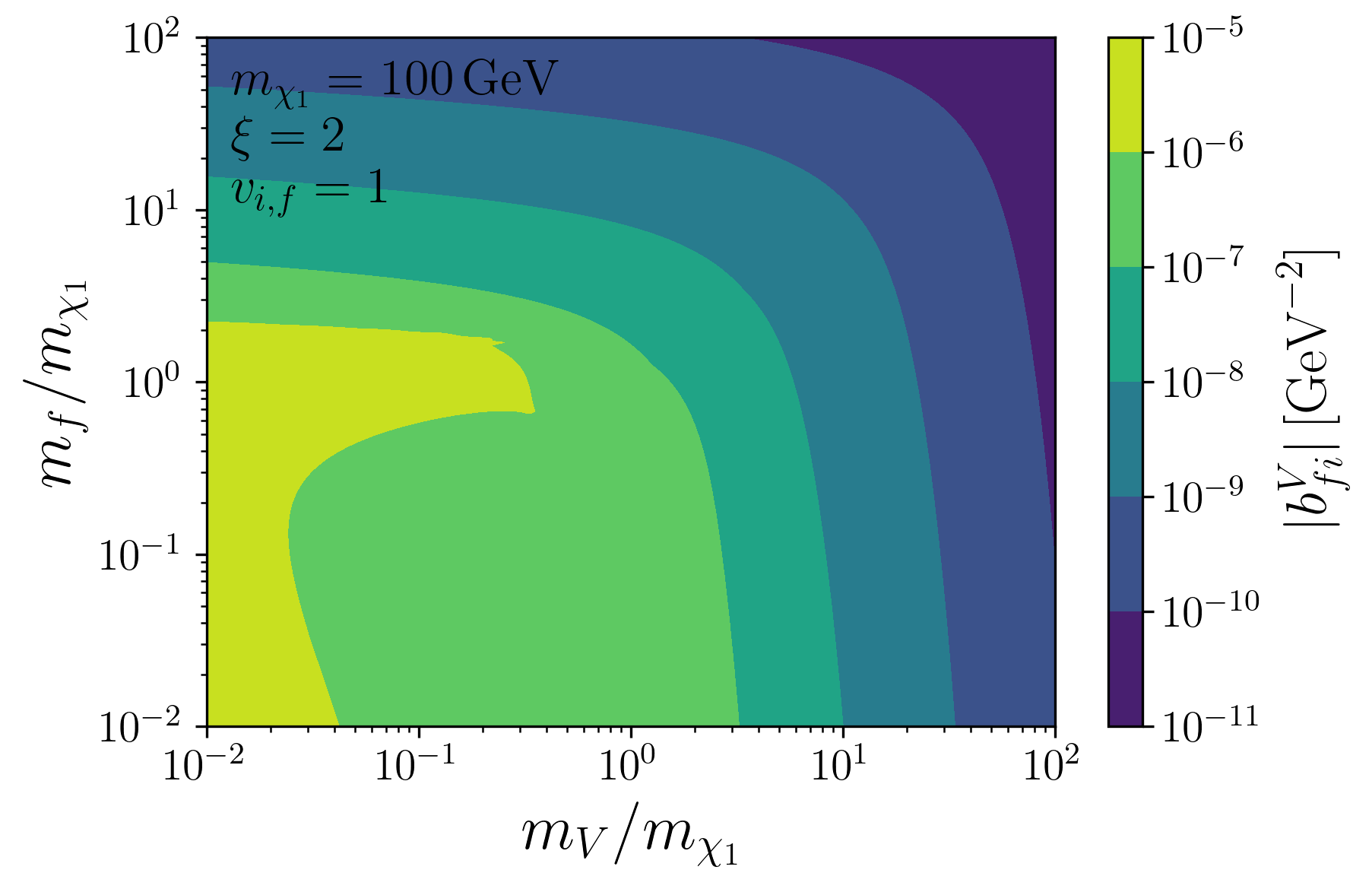}
    \includegraphics[width=.49\linewidth]{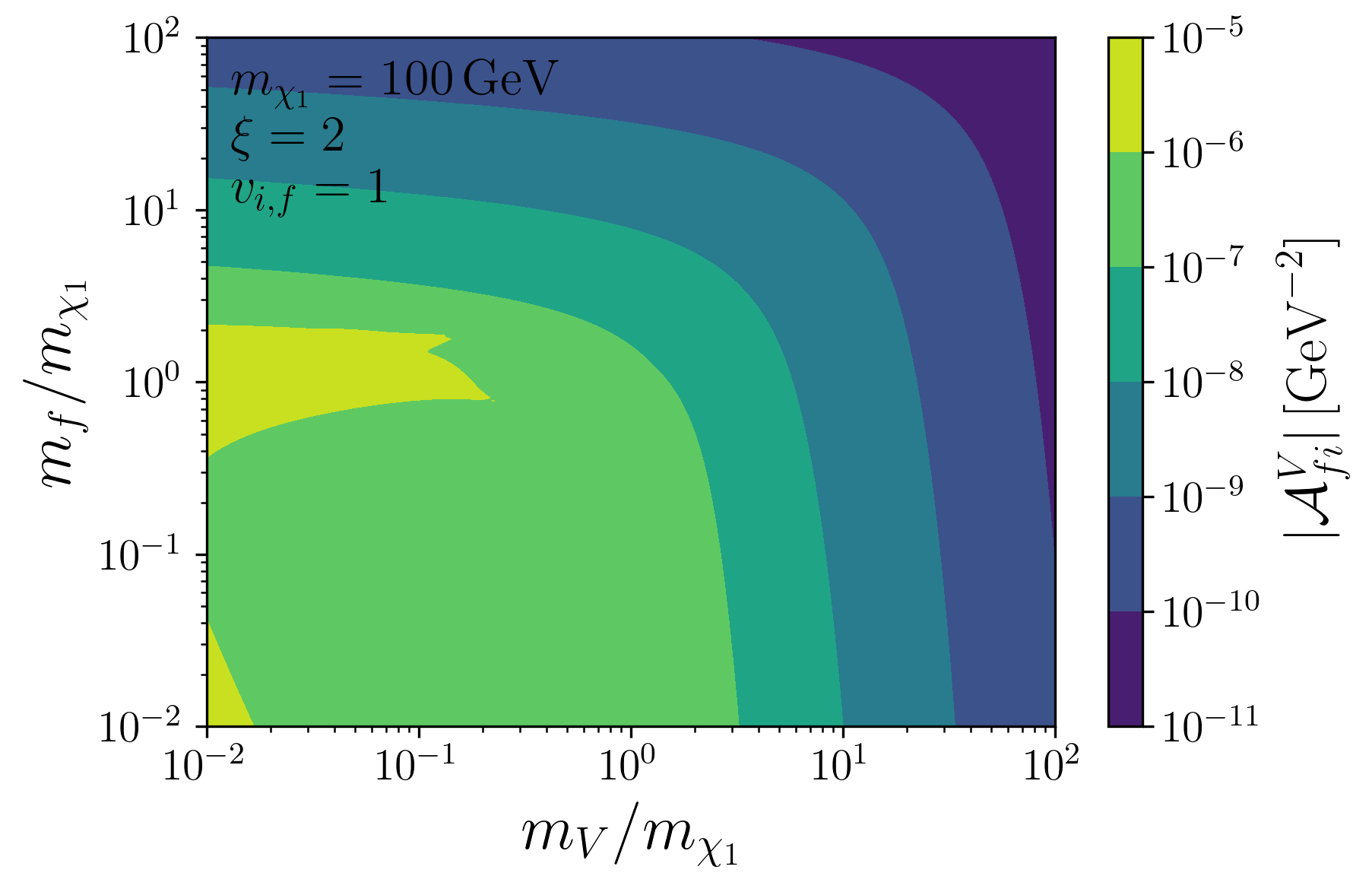}
    \caption{Same as \cref{fig:scalar_max_EMmoments} but for the vector contribution.}
    \label{fig:vector_max_EMmoments}
\end{figure}

We present in \cref{fig:scalar_max_EMmoments} (\cref{fig:vector_max_EMmoments}) the scalar (vector) transition EM moments for $\mdm1 = 100\gev$, $\mdm2 = \xi\,\mdm1=200\gev$ and $c_{i,f} =1$ ($v_{i,f} =1$) as a function of $m_S/\mdm1$ ($m_V/\mdm1$) and $m_f/\mdm2$; for each point, we maximize the EM moment with respect to $\theta_{i,f}$ and $\phi_{i,f}$ ($\Theta_{i,f}$ and $\Phi_{i,f}$). We find that the EM moments are typically enhanced if one of the  masses in the loop is close to $m_{\chi_{1,2}}$. Further, the larger the mass-splitting between an internal state and an external, the larger the suppression of the EM moment is, in line with the decoupling theorem \cite{Appelquist:1974tg}. 

\begin{figure}
    \centering
    \includegraphics[width=.49\linewidth]{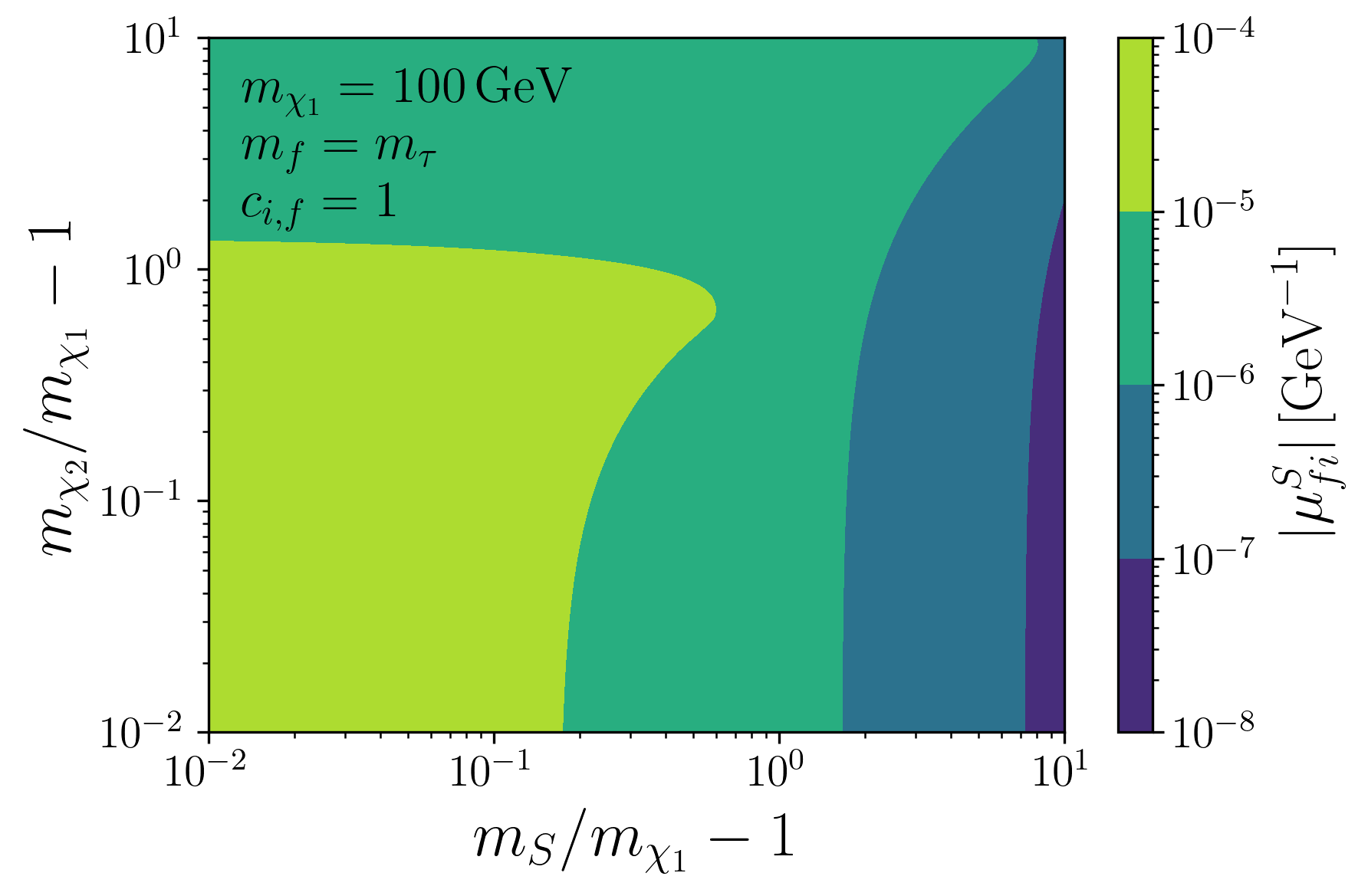}
    \includegraphics[width=.49\linewidth]{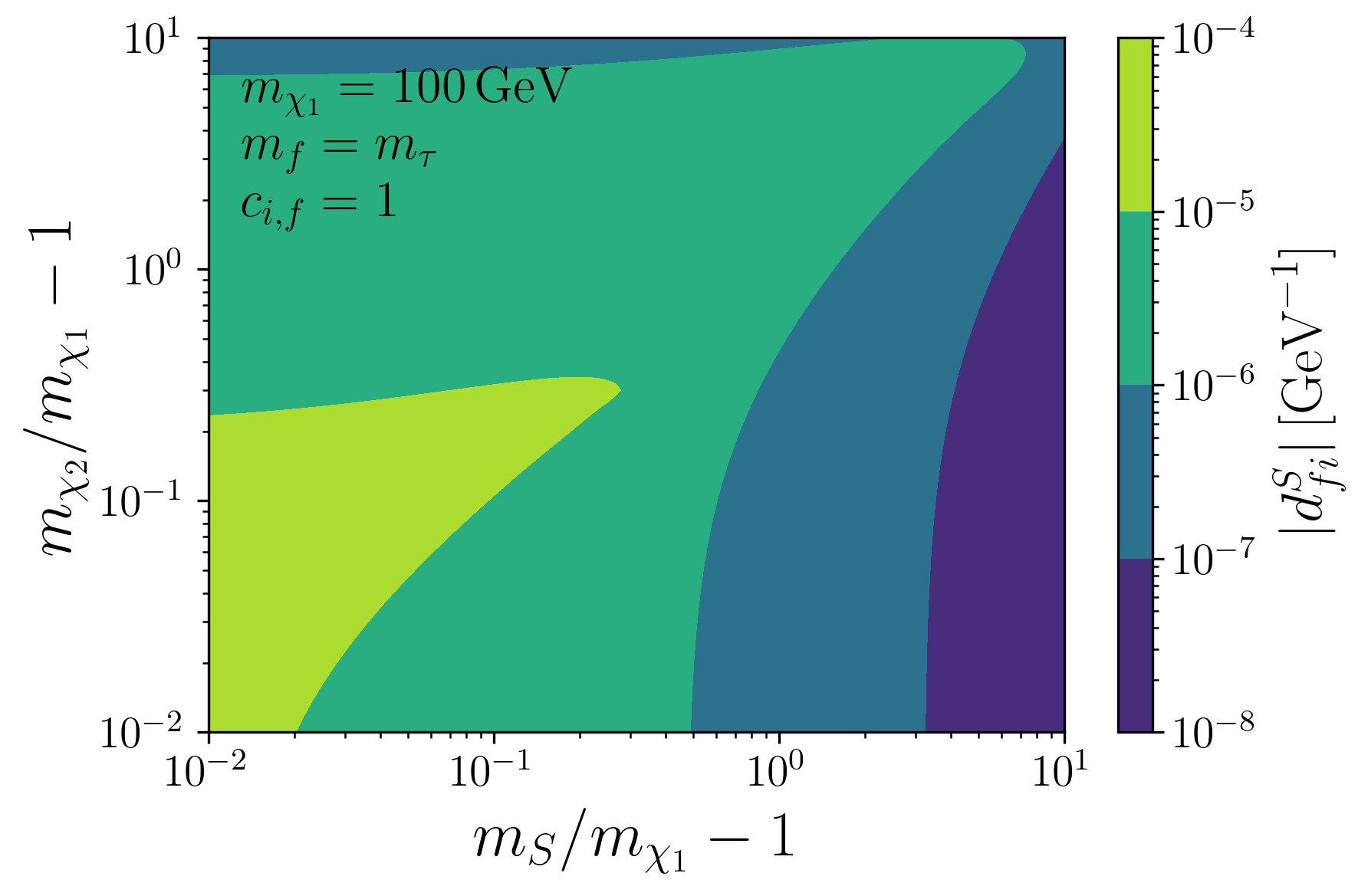}
    \includegraphics[width=.49\linewidth]{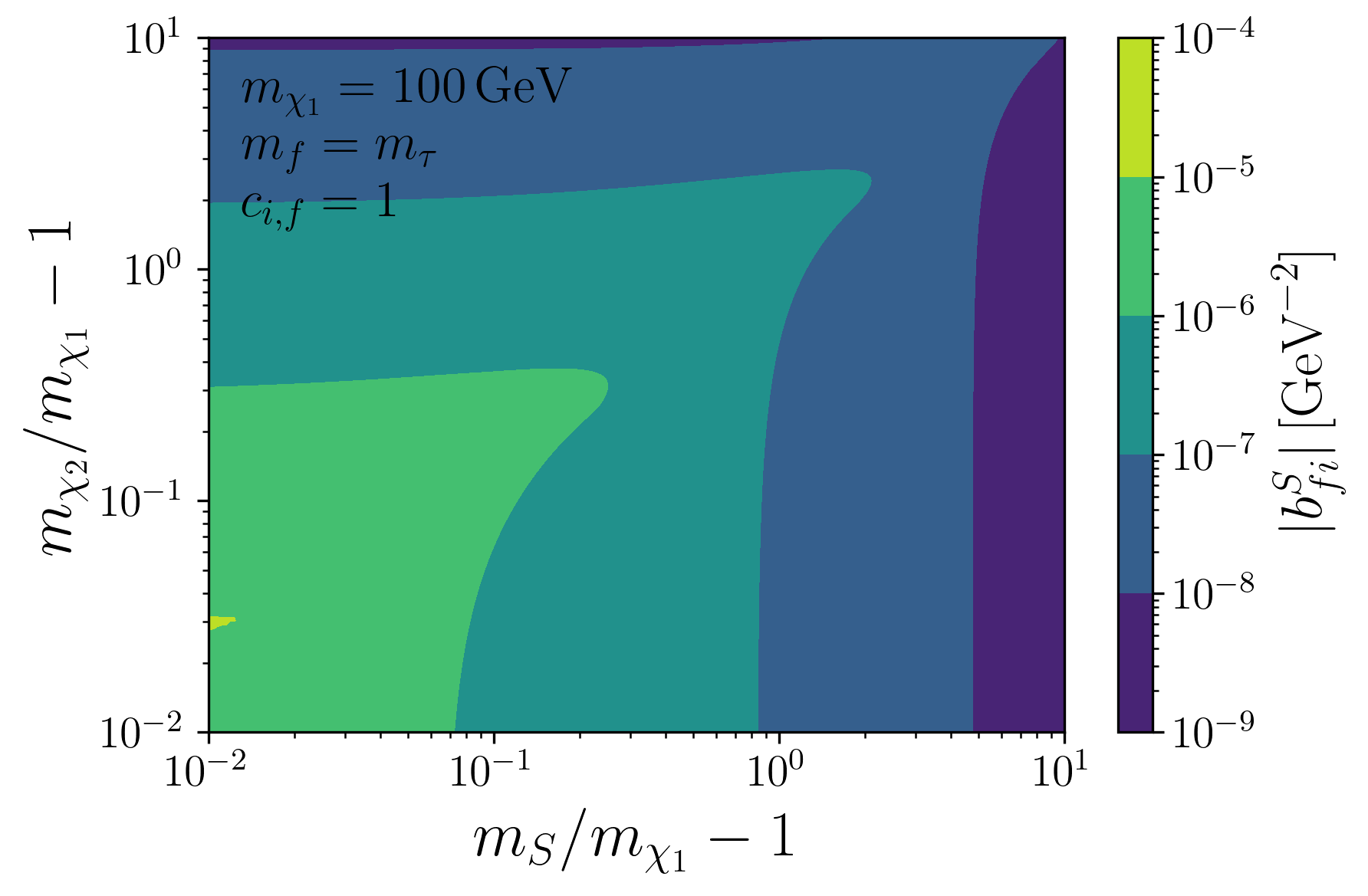}
    \includegraphics[width=.49\linewidth]{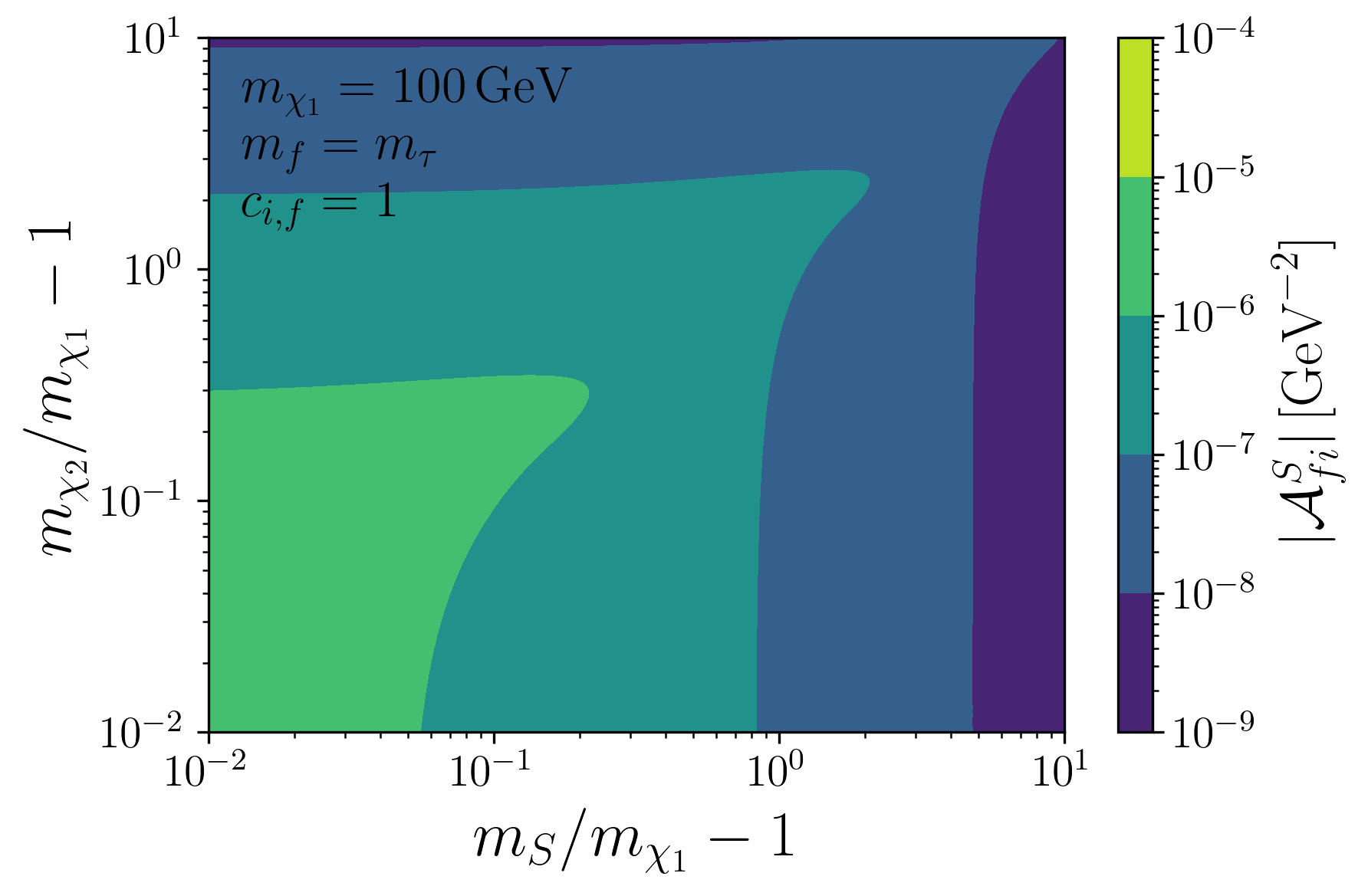}
    \caption{Scalar contribution to the transition EM moments for fixed $m_f = m_\tau$ maximized with respect to $\theta_{i,f}$ and $\phi_{i,f}$.}
    \label{fig:scalar_max_xi_EMmoments}
\end{figure}

\begin{figure}
    \centering
    \includegraphics[width=.49\linewidth]{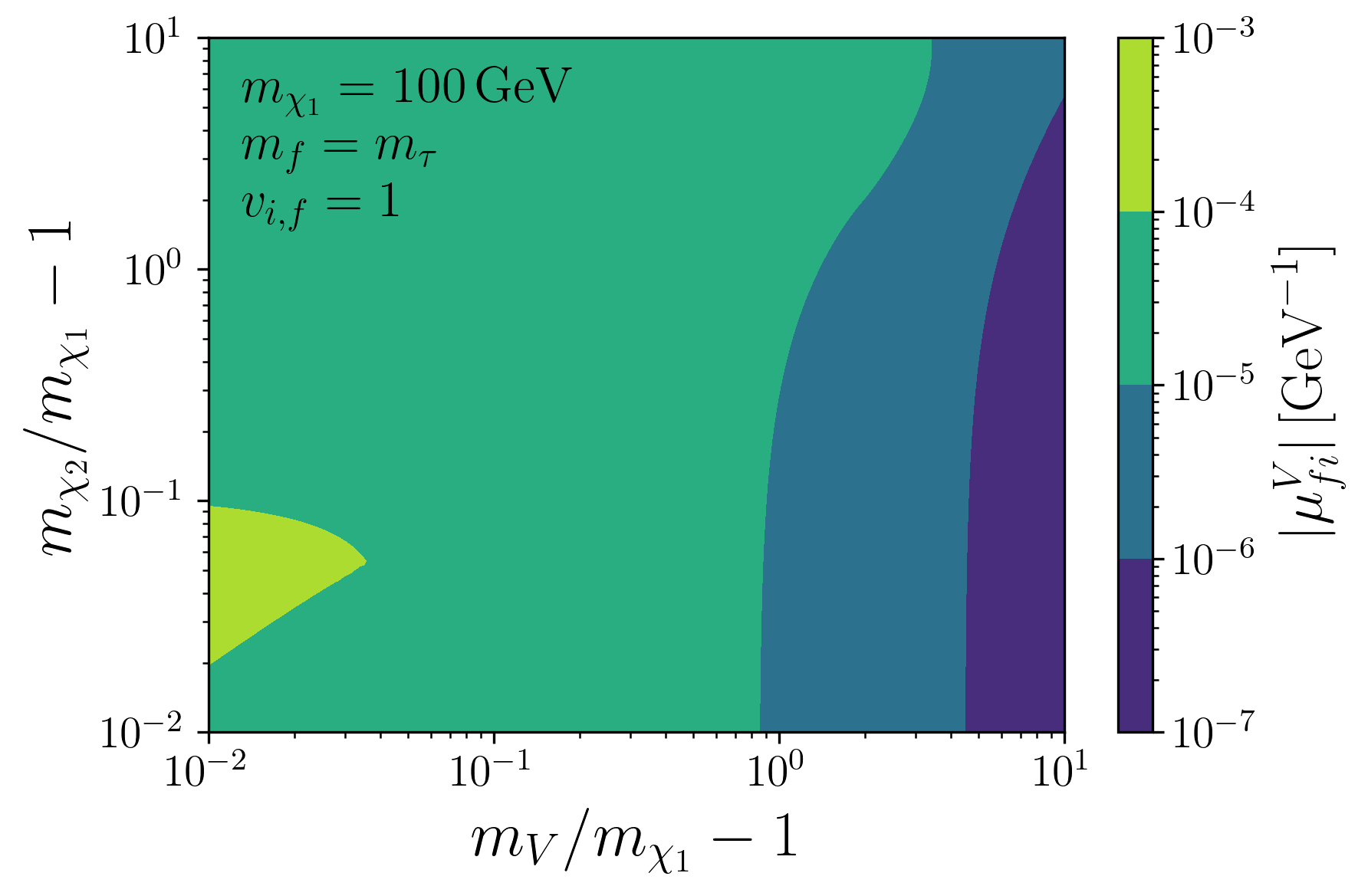}
    \includegraphics[width=.49\linewidth]{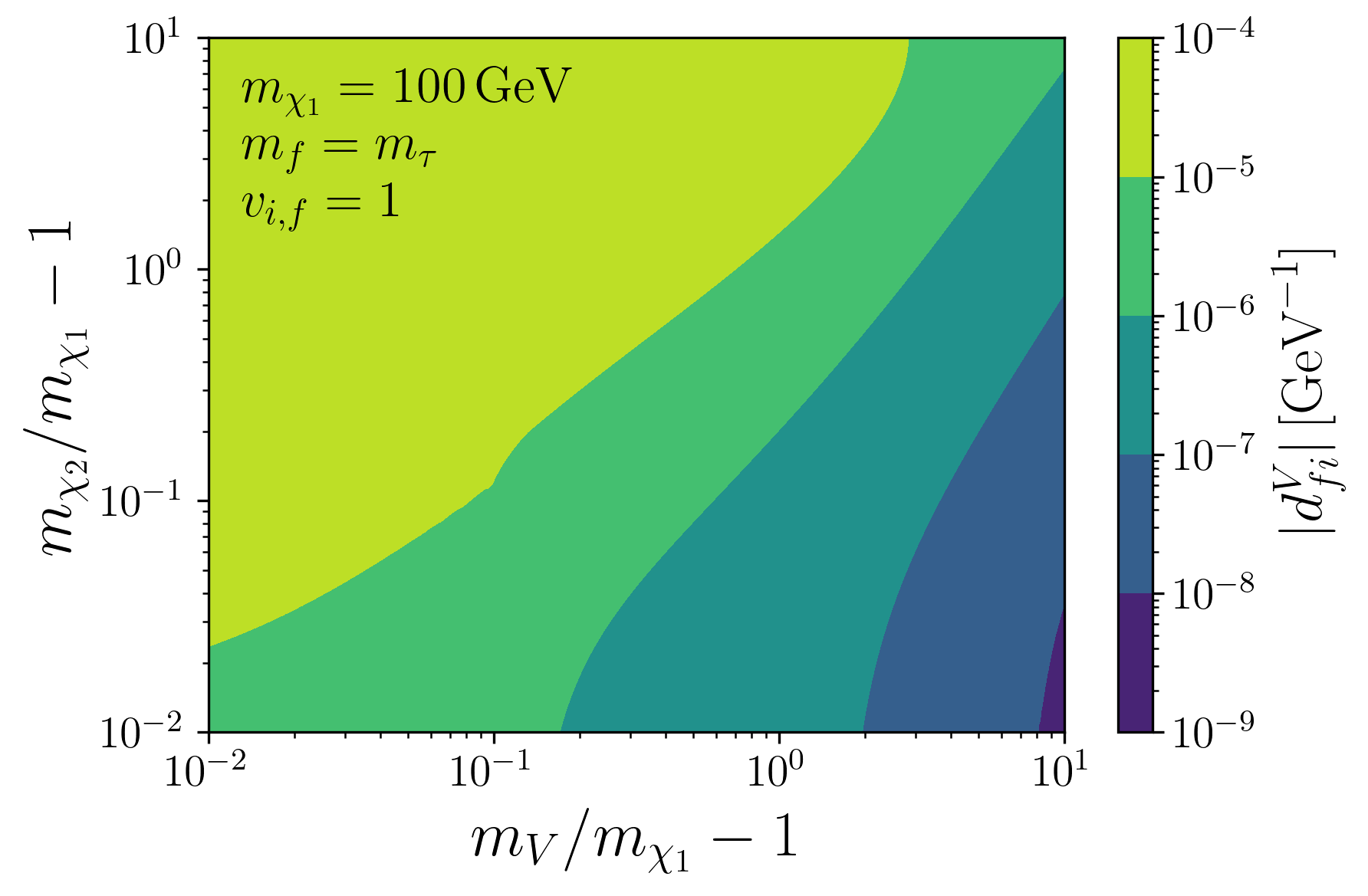}
    \includegraphics[width=.49\linewidth]{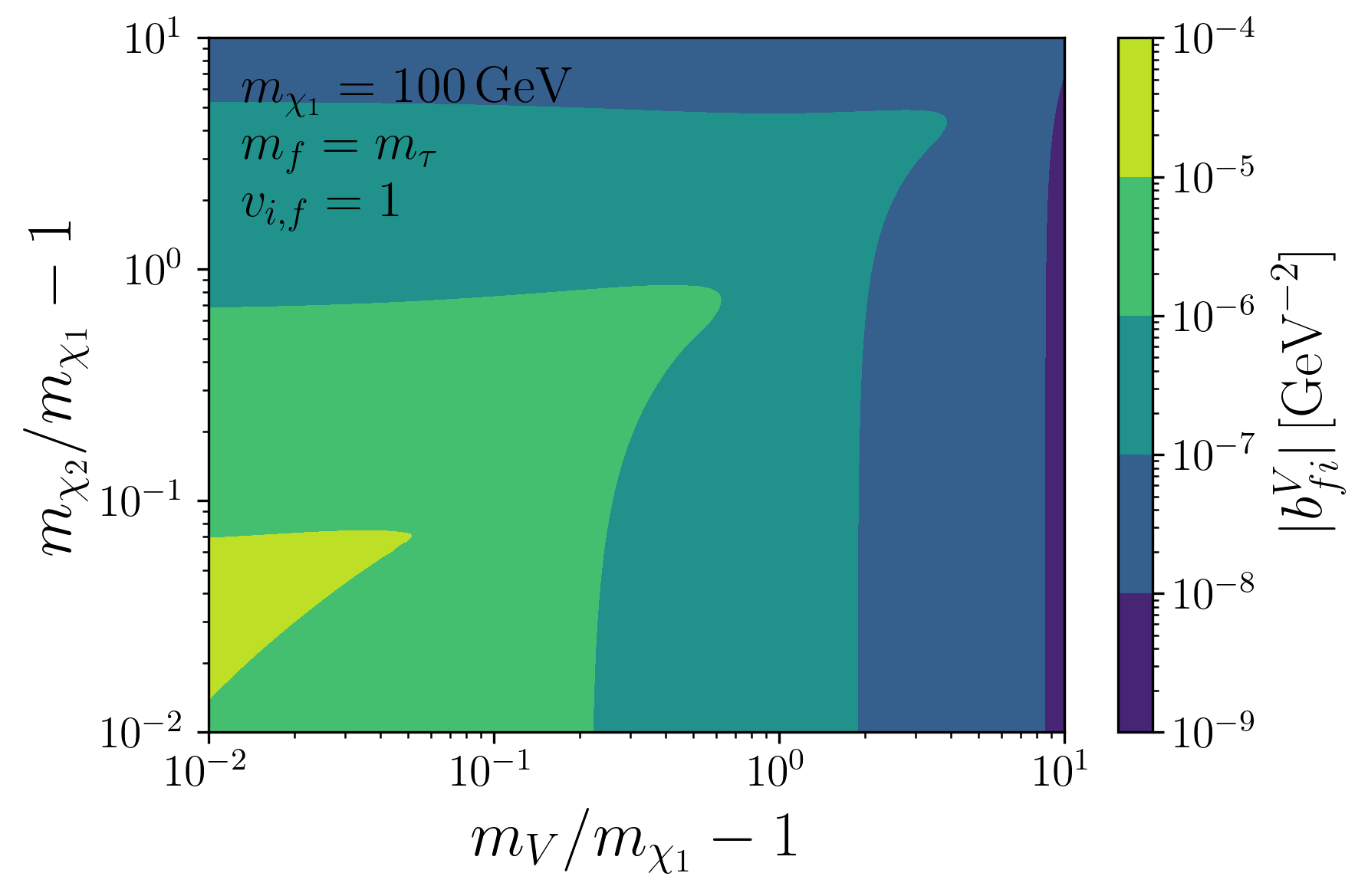}
    \includegraphics[width=.49\linewidth]{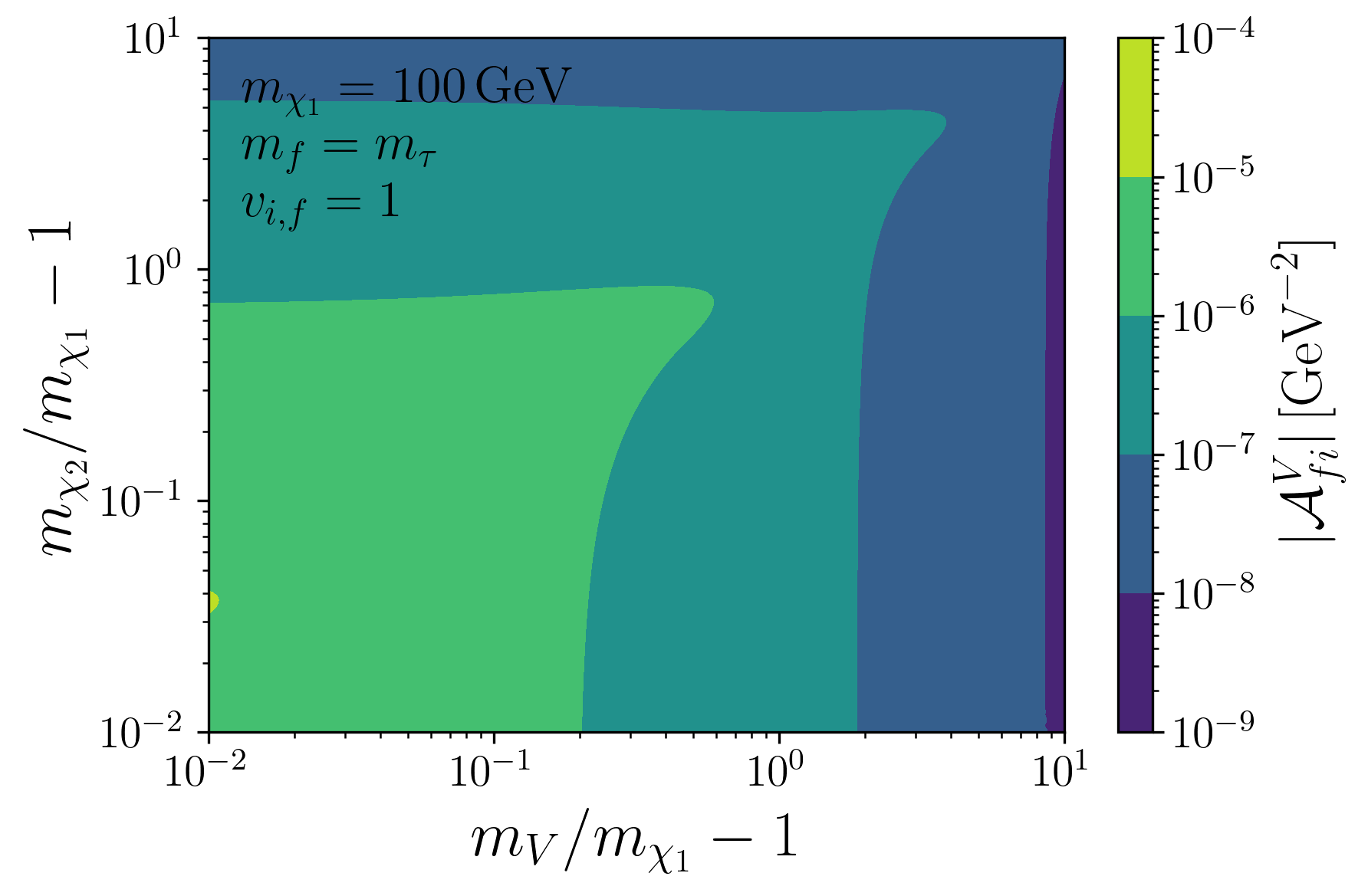}
    \caption{Same as \cref{fig:scalar_max_xi_EMmoments} but for the vector contribution.}
    \label{fig:vector_max_xi_EMmoments}
\end{figure}

Besides, we show in \cref{fig:scalar_max_xi_EMmoments,fig:vector_max_xi_EMmoments} the maximized EM moments in the plane spanned by $m_{S,V}/\mdm1-1$ and $\mdm2/\mdm1 -1$ for a tau-lepton in the loop, \textit{i.e.} $m_f=m_\tau$. We find a generic enhancement of the moments when the mass-splitting between the external states is comparable to $m_\tau/\mdm1$. The notable exception being the vector contribution to the electric dipole moment, $|d_{ji}^V|$, shown in the top right corner of \cref{fig:vector_max_xi_EMmoments}, where the enhancement is found to hold also for larger values of $\xi = \mdm2/\mdm1$. This behavior is understood by comparing with the analytical approximations of the loop functions in the limit of a large hierarchy (see \cref{app:loop_functions_large_mass_splitting}). For $\xi \gg 1$, the dipole loop functions become identical in magnitude, such that also the maximized dipole moments coincide in this limit.

In the diagonal limit ($j=i$) we find
\begin{align}
(\mom_{ii}^S)^D = \frac{e Q_f c_i^2}{32 \pi^2 }\biggr\{&\left[\sin^2\theta_i \pm \cos^2\theta_i \right] \matheul{F}_\mom^S\left(\frac{m_f}{\mdm1}, \frac{m_S}{\mdm1},1\right)\nn \\
+ & \cos\theta_i \sin\theta_i \left[e^{-i\phi_i}  \pm e^{i \phi_i} \right]  \matheul{G}_\mom^S\left(\frac{m_f}{\mdm1}, \frac{m_S}{\mdm1}, 1\right)\biggr\},
\end{align}
with $\matheul{F}^{S,V}_d(m_f/\mdm1, m_S/\mdm1, 1) = \matheul{G}^{S,V}_\ana(m_f/\mdm1, m_S/\mdm1, 1) =0$.
Note that for maximal P violation ($\sin\theta_i =0$ or $\cos\theta_i = 0$) or for maximal CP violation ($\sin\phi_i = 1$) the term $\sim\matheul{G}_{\mu,b}$ does not contribute. Furthermore, when CP (P) is conserved,  the electric dipole moment (anapole moment) vanishes identically, as expected.

\section{Multipole moments at one-loop for spin-1/2 Majorana fermions}
\label{sec:Majorana}

The calculation for the case that $\chi_i$ and $\chi_f$ are Majorana fermions ($\chi_i = \chi_i^c = i\gamma_2\chi^*$) follows along similar lines as for the Dirac case, \textit{i.e.} we project out the form factor from the relevant diagrams at the one-loop level. The key difference is that in addition to the diagrams shown in \cref{fig:scalar_vector_transition_general}, the reversed processes must be considered, which we evaluated following the conventions of \cite{Denner:1992me,Denner:1992vza}.

For the case of two Majorana fermions one finds \cite{Shrock:1982sc}
\begin{align}\label{eq:General_Majorana_Dirac_correspondence}
    (\mom^{S,V}_{fi})^M = (\mom^{S,V}_{fi})^D \mp (\mom^{S,V}_{fi})^D(c^{i,f}_{L/R}\rightarrow (c^{i,f}_{L/R})^*, v^{i,f}_{L/R}\rightarrow (v^{i,f}_{L/R})^*),
\end{align}
where the upper sign is for $\mom = \mu, b$ and the lower sign for $\mom=d,\ana$.
In the following, we focus on the scalar contribution, as the vector contribution can be obtained trivially by making the appropriate substitutions. Explicitly, \cref{eq:General_Majorana_Dirac_correspondence} reduces to
\begin{align}\label{eq:General_Majorana_Scalar_a}
     \left(\mom^{S}_{fi}\right)^M = i\frac{eQ_f}{16\pi^2}\biggr\{&\Im{c_L^f (c_L^i)^* + c_R^f (c_R^i)^* } \matheul{F}_\mom^{S}\left(\frac{m_f}{\mdm1}, \frac{m_S}{\mdm1}, \frac{\mdm2}{\mdm1}\right)\nn \\
+ &\Im{c_L^f (c_R^i)^* + c_R^f (c_L^i)^*}  \matheul{G}_\mom^{S}\left(\frac{m_f}{\mdm1}, \frac{m_S}{\mdm1}, \frac{\mdm2}{\mdm1}\right)\biggr\}
\end{align}
for $\mom = \mu, b$, and for the electric dipole moment and anapole
\begin{align}\label{eq:General_Majorana_Scalar_b}
    \left(\mom^{S}_{fi}\right)^M = \frac{eQ_f}{16\pi^2}\biggr\{&\Re{c_L^f (c_L^i)^* - c_R^f (c_R^i)^* } \matheul{F}_\mom^{S}\left(\frac{m_f}{\mdm1}, \frac{m_S}{\mdm1}, \frac{\mdm2}{\mdm1}\right)\nn \\
+ &\Re{c_L^f (c_R^i)^* - c_R^f (c_L^i)^*}  \matheul{G}_\mom^{S}\left(\frac{m_f}{\mdm1}, \frac{m_S}{\mdm1}, \frac{\mdm2}{\mdm1}\right)\biggr\}
\end{align}
with similar expressions for the vector contribution $(\mom^V_{fi})^M$. If we employ the same parameterization as introduced in the discussion of the Dirac EM moments, we find for $\mom = \mu,b$
\begin{align}\label{eq:General_Majorana_Scalar_a_alt}
    \left(\mom^{S}_{fi}\right)^M = i\frac{eQ_f c_i c_f}{16\pi^2}\biggr\{& -\cos\theta_i\cos\theta_f \sin(\phi_i-\phi_f) \matheul{F}_\mom^{S}\left(\frac{m_f}{\mdm1}, \frac{m_S}{\mdm1}, \frac{\mdm2}{\mdm1}\right)\nn \\
 & (-\cos\theta_i \sin\theta_f \sin\phi_i + \cos\theta_f \sin\theta_i \sin\phi_f)\matheul{G}_\mom^{S}\left(\frac{m_f}{\mdm1}, \frac{m_S}{\mdm1}, \frac{\mdm2}{\mdm1}\right)\biggr\},
\end{align}
and for $\mom = d,\ana$
\begin{align}\label{eq:General_Majorana_Scalar_b_alt}
    \left(\mom^{S}_{fi}\right)^M = \frac{eQ_f c_i c_f}{16\pi^2}\biggr\{& (-\cos\theta_i\cos\theta_f \cos(\phi_i-\phi_f)+\sin\theta_i \sin\theta_f) \matheul{F}_\mom^{S}\left(\frac{m_f}{\mdm1}, \frac{m_S}{\mdm1}, \frac{\mdm2}{\mdm1}\right)\nn \\
+ & (-\cos\theta_f \sin\theta_i \cos\phi_f - \cos\theta_i \sin\theta_f \cos\phi_i)\matheul{G}_\mom^{S}\left(\frac{m_f}{\mdm1}, \frac{m_S}{\mdm1}, \frac{\mdm2}{\mdm1}\right)\biggr\}.
\end{align}
In the diagonal limit ($f=i$), the only non-zero EM moment is the anapole moment~\cite{Nieves:1981zt,Kayser:1983wm,Kayser:1982br,Schechter:1981hw,Giunti:2014ixa}, for which the scalar contribution reads
\begin{equation}
    \left(\ana^S_{ii}\right)^M = -\frac{eQ_f c_i^2}{16\pi^2} \cos(2\theta_i) \matheul{F}_\ana^{S}\left(\frac{m_f}{\mdm1}, \frac{m_S}{\mdm1}, 1\right).
\end{equation}
Correspondingly the vector contribution can be obtained by replacing $c_i\rightarrow v_i$, $\theta_i \rightarrow \Theta_i$ and $S\rightarrow V$.

\section{Multipole moments of $SU(2)_L$ multiplets}
\label{sec:SU2-multiplets}

We now apply the general results of Sections \ref{sec:Dirac} and \ref{sec:Majorana} to neutral fermions belonging to non-trivial representations of $SU(2)_L$. We first consider the case where the neutral state belongs to a single $n$-plet and subsequently generalize the discussion to the case where the neutral components of two different multiplets mix. 

\subsection{One $n$-plet}
\label{sec:one_nplet}

We consider a $SU(2)_L$ multiplet of dimension $n$ and hypercharge $Y_n$, that we denote $\chi_n$. The components of $\chi_n$ have electric charge $Q=T_3+Y_n$, therefore a neutral component $\chi_n^0$ exists provided  the representation contains the state with $ T_3=-Y_n$. The gauge invariant mass term reads
\begin{align}
\mathcal{L}_{\rm mass} &= -\frac{1}{2} M_n
\left(\overline{\chi_n^{\,c}}\,\chi_n\right)_{\mathbf{1}}
+\text{h.c.}=
-\frac{1}{2}M_n\,
\overline{(\chi^0)^c}\chi^0
-
M_n\sum_{Q=1}^{(n-1)/2}
\overline{(\chi^{Q})^c}\chi^{-Q}
+\text{h.c.}
\end{align}
where the subindex 1 indicates the singlet combination, and $\chi_n^c=C\overline{\chi_n}^{\,T}$. 

The interaction between $\chi_n^0$, the charged component $\chi_n^-$ and the transverse components of the $W$ boson can be written as
\begin{equation} 
{\cal L}_{W}=\overline{\chi_n^0}\gamma^\mu\left(v_LP_L+v_RP_R\right)\chi_n^-W_\mu^+ + {\rm h.c.}\,,
\label{eq:LW_nplet} 
\end{equation}
Here, $v_L$ and $v_R$ can be chosen real and read:
\begin{equation}
v_L=\frac{g}{\sqrt{2}}C_{n,-}\,,\qquad v_R=\frac{g}{\sqrt{2}}C_{n,+}\,,
\label{eq:vLR_nplet} 
\end{equation}
where $C_{n,\pm}$ are $SU(2)_L$ Clebsch-Gordan coefficients given by
\begin{equation} 
C_{n,-}=\frac{1}{2}\sqrt{n^2-(1+2Y_n)^2}\,,\qquad  C_{n,+}=\frac{1}{2}\sqrt{n^2-(1-2Y_n)^2}\,.
\label{eq:CG}
\end{equation}

Further, the interaction with the charged Goldstone bosons $G^\pm$ is
\begin{equation}
{\cal L}_{G}=\overline{\chi_n^0}\left(c_LP_L+c_RP_R\right)\chi_n^-G^+ +{\rm h.c.}\,, 
\label{eq:LG_nplet}
\end{equation}
where the gauge invariance relates the Goldstone couplings to the charged-current couplings as
\begin{equation} 
c_L=\frac{m_{\chi_n^0}}{m_W}v_L-\frac{m_{\chi_n^-}}{m_W}v_R\,,\qquad c_R=\frac{m_{\chi_n^0}}{m_W}v_R-\frac{m_{\chi_n^-}}{m_W}v_L\,. 
\label{eq:cLR_nplet} 
\end{equation}

Since $\chi^0$ is a Majorana fermion, only the anapole moment is a priori non-vanishing. The contribution induced by the charged fermion, the $W$ boson and the charged Goldstone boson is
\begin{equation} {\cal A}_n=-\frac{e}{16\pi^2}\Big[\big(|v_L|^2-|v_R|^2\big)\matheul{F}_{\cal A}^V(\mu_n,\eta_n,1)+\big(|c_L|^2-|c_R|^2\big)\matheul{F}_{\cal A}^S(\mu_n,\eta_n,1)\Big]\,. \label{eq:An_WG} \end{equation}
where we have defined
\begin{equation} 
\mu_n=\frac{m_{\chi_n^-}}{m_{\chi_n^0}}\,,\qquad \eta_n=\frac{m_W}{m_{\chi_n^0}}\,.
\end{equation}

Using Eqs.~\eqref{eq:vLR_nplet}, \eqref{eq:cLR_nplet} and \eqref{eq:CG}, this expression can be cast in a more compact form
\begin{equation} 
{\cal A}_n=-\frac{eg^2Y_n}{16\pi^2}\matheul{F}_{\cal A}(\mu_n,\eta_n)\,.
\label{eq:An_complete}
\end{equation}
with 
\begin{equation} \matheul{F}_{\cal A}(\mu,\eta)\equiv \matheul{F}_{\cal A}^V(\mu,\eta,1)+\frac{1-\mu^2}{\eta^2}\matheul{F}_{\cal A}^S(\mu,\eta,1)\,.\label{eq:FAcomplete}
\end{equation}

When $Y_n=0$,  clearly $ {\cal A}_n=0$. This is to be expected, since the interactions of the neutral component to the components with charge $+1$ and $-1$ are identical. This allows to accommodate $\chi_n^+$ and $\chi_n^-$ into a Dirac fermion, leading to an interaction that preserves parity. This is in fact the most generic situation, since a single chiral $SU(2)_L$ multiplet admits a gauge-invariant mass term involving no additional fermionic fields only if $Y=0$ and the representation is real, corresponding to odd $n$. Multiplets with $Y\neq0$ can nevertheless become massive through electroweak-symmetry-breaking interactions involving additional fields. In this case, a non-zero anapole moment given by Eq.~\eqref{eq:An_complete} would be generated. One notable example is the Standard Model neutrino, which is the neutral component of a $SU(2)_L$ doublet with hypercharge $Y=-1/2$, and that will be discussed in detailed in Subsection \ref{sec:neutrino}. Other scenarios may exist if the particle content of the model contains additional multiplets, as we discuss in what follows.

\subsection{Mixing of an $n$-plet and an $m$-plet}
\label{sec:nm_multiplets}

We consider an $n$-plet and an $m$-plet of $SU(2)_L$, $\chi_n$ and $\chi_m$, with hypercharges $Y_n$ and $Y_m$, respectively, and isospins $j_n=(n-1)/2$ and $j_m=(m-1)/2$. We assume that both multiplets contain an electrically neutral component, denoted by $\chi_n^0$ and $\chi_m^0$, respectively, and that the underlying theory provides the corresponding gauge-invariant mass terms. The components of the two multiplets can mix after electroweak symmetry breaking. We distinguish below between Dirac- and Majorana-type mixing.

\subsubsection{Dirac-type mixing}
\label{sec:nm_dirac}

At the renormalizable level, Dirac-type mixing through a single Higgs doublet is possible only when $\chi_n\otimes\chi_m$ contains a representation with isospin $J=1/2$. This requires $|j_n-j_m|=1/2$, or equivalently $|n-m|=1$. Taking without loss of generality $m=n+1$, the interaction can be written as
\begin{equation}
{\cal L}_{\rm mix}^{D}
=
-\lambda_D\left(\overline{\chi_n}P_L\chi_m\right)_{1/2}\cdot\Phi
+{\rm h.c.}\,,
\qquad
\Phi=H\ {\rm or}\ \widetilde H\,,
\label{eq:LmixrenD}
\end{equation}
where the subscript $1/2$ denotes the $SU(2)_L$ isospin of the fermion bilinear. Hypercharge conservation requires
\begin{equation}
-Y_n+Y_m+Y_\Phi=0\,,
\label{eq:YmixrenD}
\end{equation}
with $Y_H=1/2$ or $Y_{\widetilde H}=-1/2$. If $\chi_n\otimes\chi_m$ does not contain an isospin-$1/2$ representation, mixing can instead arise from higher-dimensional operators,
\begin{equation}
{\cal L}_{\rm mix}^{D,{\rm EFT}}
=
-\frac{\lambda_D}{\Lambda^{p-1}}
\left(\overline{\chi_n}P_L\chi_m\right)_J
\cdot
\left(H^k\widetilde H^{\,p-k}\right)_J
+{\rm h.c.}\,,
\label{eq:LmixEFTD_main}
\end{equation}
with
\begin{equation}
-Y_n+Y_m+\frac{2k-p}{2}=0\,.
\end{equation}
The group-theory structure of these operators is discussed in Appendix~\ref{sec:EFT_multiplets}.

If the interactions preserve a conserved fermion number, the neutral components mix into Dirac mass eigenstates. Since a general Dirac mass matrix is diagonalized by independent rotations of its left- and right-handed components, we write
\begin{equation}
P_L\chi_i^0=\sum_{r=n,m}N^L_{ir}P_L\chi_r^0\,,
\qquad
P_R\chi_i^0=\sum_{r=n,m}N^R_{ir}P_R\chi_r^0\,.
\label{eq:NmatrixD}
\end{equation}
Similarly, for the negatively charged states,
\begin{equation}
P_L\chi_\alpha^-=\sum_{r=n,m}U_{\alpha r}P_L\chi_r^-\,,
\qquad
P_R\chi_\alpha^-=\sum_{r=n,m}V_{\alpha r}P_R\chi_r^-\,.
\label{eq:chargedmixingD}
\end{equation}
The precise field content required to form the Dirac states depends on the ultraviolet realization of the multiplet masses.

The charged-current interaction in the fermion mass basis can then be written as
\begin{equation}
{\cal L}_W
=
\sum_{i,\alpha}
\overline{\chi_i^0}\gamma^\mu
\left(v^L_{i\alpha}P_L+v^R_{i\alpha}P_R\right)
\chi_\alpha^-W_\mu^+
+{\rm h.c.}\,,
\label{eq:LW_nmD}
\end{equation}
where
\begin{equation}
v^L_{i\alpha}
=
\frac{g}{\sqrt2}
\sum_{r=n,m}N^L_{ir}C_{r,-}U_{\alpha r}^*\,,
\qquad
v^R_{i\alpha}
=
\frac{g}{\sqrt2}
\sum_{r=n,m}N^R_{ir}C_{r,+}V_{\alpha r}^*\,,
\label{eq:vLR_nmD}
\end{equation}
with
\begin{equation}
C_{r,-} = \frac{1}{2}\sqrt{r^2-(1+2Y_r)^2}\,,
\qquad
C_{r,+} = \frac{1}{2}\sqrt{r^2-(1-2Y_r)^2}\,.
\label{eq:Cnpm}
\end{equation}

The interactions with the charged Goldstone boson are analogously written as
\begin{equation}
{\cal L}_G = \sum_{i,\alpha} \overline{\chi_i^0} \left(c^L_{i\alpha}P_L+c^R_{i\alpha}P_R\right) \chi_\alpha^-G^+ +{\rm h.c.}\,,
\label{eq:LG_nmD}
\end{equation}
where, in a convention in which the fermion masses are real and positive, the Ward identity gives
\begin{equation}
c^L_{i\alpha}=\frac{m_{\chi_i^0}}{m_W}v^L_{i\alpha} -\frac{m_{\chi_\alpha^-}}{m_W}v^R_{i\alpha}\,, \qquad c^R_{i\alpha} = \frac{m_{\chi_i^0}}{m_W}v^R_{i\alpha} -\frac{m_{\chi_\alpha^-}}{m_W}v^L_{i\alpha}\,.
\label{eq:cLR_nmD}
\end{equation}

The electromagnetic moments of the neutral mass eigenstates are obtained by inserting the couplings in Eqs.~(\ref{eq:vLR_nmD}) and (\ref{eq:cLR_nmD}) into the general expressions derived in Section ~\ref{sec:Dirac}

\subsubsection{Majorana-type mixing}
\label{sec:nm_majorana}

Majorana-type mixing can instead be generated by
\begin{equation}
{\cal L}_{\rm mix}^{M}
=
-\lambda_M
\left(\overline{\chi_n^{\,c}}P_L\chi_m\right)_{1/2}\cdot\Phi
+{\rm h.c.}\,,
\qquad
\Phi=H\ {\rm or}\ \widetilde H\,,
\label{eq:LmixrenM}
\end{equation}
which is again possible at the renormalizable level for $|j_n-j_m|=1/2$, or equivalently $|n-m|=1$. In this case hypercharge conservation requires
\begin{equation}
Y_n+Y_m+Y_\Phi=0\,.
\label{eq:YmixrenM}
\end{equation}
More generally, Majorana-type mixing can arise from
\begin{equation}
{\cal L}_{\rm mix}^{M,{\rm EFT}}
=
-\frac{\lambda_M}{\Lambda^{p-1}}
\left(\overline{\chi_n^{\,c}}P_L\chi_m\right)_J
\cdot
\left(H^k\widetilde H^{\,p-k}\right)_J
+{\rm h.c.}\,,
\label{eq:LmixEFTM_main}
\end{equation}
with
\begin{equation}
Y_n+Y_m+\frac{2k-p}{2}=0\,.
\end{equation}
These operators are discussed in detail in Appendix~\ref{sec:EFT_multiplets}.

After electroweak symmetry breaking, the neutral components mix into Majorana mass eigenstates,
\begin{equation}
\chi_i^0
=
N_{in}\chi_n^0+N_{im}\chi_m^0\,.
\label{eq:NmatrixM}
\end{equation}
The charged states are diagonalized as
\begin{equation}
P_L\chi_\alpha^-=\sum_{r=n,m}U_{\alpha r}P_L\chi_r^-\,,
\qquad
P_R\chi_\alpha^-=\sum_{r=n,m}V_{\alpha r}P_R\chi_r^-\,.
\label{eq:chargedmixingM}
\end{equation}
The charged-current interaction takes the form
\begin{equation}
{\cal L}_W
=
\sum_{i,\alpha}
\overline{\chi_i^0}\gamma^\mu
\left(v^L_{i\alpha}P_L+v^R_{i\alpha}P_R\right)
\chi_\alpha^-W_\mu^+
+{\rm h.c.}\,,
\label{eq:LW_nmM}
\end{equation}
where
\begin{equation}
v^L_{i\alpha}
=
\frac{g}{\sqrt2}
\sum_{r=n,m}N_{ir}C_{r,-}U_{\alpha r}^*\,,
\qquad
v^R_{i\alpha}
=
\frac{g}{\sqrt2}
\sum_{r=n,m}N_{ir}^*C_{r,+}V_{\alpha r}^*\,.
\label{eq:vLR_nmM}
\end{equation}
The Clebsch--Gordan coefficients $C_{r,\pm}$ are given in Eq.~(\ref{eq:Cnpm}).

The charged-Goldstone interaction is
\begin{equation}
{\cal L}_G
=
\sum_{i,\alpha}
\overline{\chi_i^0}
\left(c^L_{i\alpha}P_L+c^R_{i\alpha}P_R\right)
\chi_\alpha^-G^+
+{\rm h.c.}\,,
\label{eq:LG_nmM}
\end{equation}
with
\begin{equation}
c^L_{i\alpha}
=
\frac{m_{\chi_i^0}}{m_W}v^L_{i\alpha}
-\frac{m_{\chi_\alpha^-}}{m_W}v^R_{i\alpha}\,,
\qquad
c^R_{i\alpha}
=
\frac{m_{\chi_i^0}}{m_W}v^R_{i\alpha}
-\frac{m_{\chi_\alpha^-}}{m_W}v^L_{i\alpha}\,.
\label{eq:cLR_nmM}
\end{equation}

We can now apply the general expressions of Sec.~\ref{sec:Majorana}. For diagonal matrix elements the Majorana condition implies $\mu_{ii}=d_{ii}=b_{ii}=0$ and the only allowed diagonal electromagnetic moment is the anapole moment. For $j=i$, the coupling combinations multiplying $\matheul{G}_{\cal A}^{S,V}$ vanish identically, and one obtains
\begin{equation}
{\cal A}_{ii}=
-\frac{e}{16\pi^2}\sum_\alpha\Big[\big(|v^L_{i\alpha}|^2-|v^R_{i\alpha}|^2\big)\matheul{F}_{\cal A}^V(\mu_{i\alpha},\eta_i,1)+\big(|c^L_{i\alpha}|^2-|c^R_{i\alpha}|^2\big)\matheul{F}_{\cal A}^S(\mu_{i\alpha},\eta_i,1)\Big]\,,
\label{eq:A_nmM}
\end{equation}
where
\begin{equation}
\mu_{i\alpha}=\frac{m_{\chi_\alpha^-}}{m_{\chi_i^0}}\,,\qquad\eta_i= \frac{m_W}{m_{\chi_i^0}}\,.
\label{eq:ratios_nmM}
\end{equation}
Using Eq.~(\ref{eq:cLR_nmM}), one finds
\begin{equation}
|c^L_{i\alpha}|^2-|c^R_{i\alpha}|^2 =
\frac{m_{\chi_i^0}^2-m_{\chi_\alpha^-}^2}{m_W^2}\left(|v^L_{i\alpha}|^2-|v^R_{i\alpha}|^2\right)\,,
\label{eq:DeltaC_nmM}
\end{equation}
and therefore
\begin{equation}
{\cal A}_{ii}=-\frac{e}{16\pi^2}\sum_\alpha\left(|v^L_{i\alpha}|^2-|v^R_{i\alpha}|^2\right)\matheul{F}_{\cal A}(\mu_{i\alpha},\eta_i)\,,
\label{eq:A_nmM_compact}
\end{equation}
where
\begin{equation}
\matheul{F}_{\cal A}(\mu,\eta)=\matheul{F}_{\cal A}^V(\mu,\eta,1)+\frac{1-\mu^2}{\eta^2}\matheul{F}_{\cal A}^S(\mu,\eta,1)\,.
\label{eq:FA_nmM}
\end{equation}
Explicitly,
\begin{equation}
\big|v^L_{i\alpha}\big|^2-\big|v^R_{i\alpha}\big|^2=\frac{g^2}{2}\Bigg[\bigg|\sum_{r=n,m}N_{ir}C_{r,-}U_{\alpha r}^*\bigg|^2-\bigg|\sum_{r=n,m}N_{ir}^*C_{r,+}V_{\alpha r}^*\bigg|^2\Bigg]\,.
\label{eq:DeltaV_nmM}
\end{equation}

The transition moments with $i\neq j$ are in general non-vanishing. In particular, the Majorana condition allows transition magnetic and electric dipole moments, transition charge radii and transition anapole moments. Their values are controlled by the relative phases and chiral structure of the couplings in Eqs.~(\ref{eq:vLR_nmM}) and (\ref{eq:cLR_nmM}). In a CP-conserving theory, the allowed transition moments are further restricted by the relative CP parities of the two Majorana states.

\subsection{Neutrino electromagnetic moments}
\label{sec:neutrino}

Neutrinos are the only electrically neutral spin-$1/2$ fermions in the Standard Model. Although they do not couple to the photon at tree level, electromagnetic interactions are generated at the quantum level through their couplings to the $W$ boson and the charged leptons. Additional contributions may arise in extensions of the Standard Model, making neutrino electromagnetic moments sensitive probes of new particles and interactions \cite{Pal:1981rm,Enqvist:1989vt,Sher:2002ij,Mohapatra:2004,Beg:1977xz,Maalampi:1985fa,Chattopadhyay:1986cj,Duncan:1987ki,Mohapatra:1987nx,Duncan:1987ki,Zhang:2024ijy,Ruan:2024fgf,Bolanos-Carrera:2023ppu,Barbieri:1988fh,Babu:1990wv,Babu:1990hu,Barr:1990um,Pal:1991qr,Lindner:2017uvt}. In this section we show how the general expressions derived in this paper reproduce the known electromagnetic moments of Dirac and Majorana neutrinos in the minimally extended Standard Model. Since the neutrino masses are much smaller than the masses of the particles running in the loop, we use the expressions derived in \cref{app:loop_functions_light_external_states} for $\mdm{1,2}\ll m_f,m_S,m_V$.

The interactions relevant for the one-loop electromagnetic vertex are the charged-current interaction with the $W$ boson and the corresponding interaction with the charged Goldstone boson. In the neutrino interaction and mass bases they read~\cite{Pal:1981rm}
\begin{align} 
\Lag &= \frac{g}{\sqrt{2}}\sum_\ell\left[\overline{\nu_\ell}\gamma^\mu P_L\ell W_\mu^+-\overline{\nu_\ell}\left(\frac{m_\ell}{m_W}P_R-\frac{m_\alpha}{m_W}P_L\right)\ell G^+\right]+\hc \nn\\ 
&=\frac{g}{\sqrt{2}}\sum_{\ell,\alpha}U_{\ell\alpha}^*\overline{\nu_\alpha}\left[\gamma^\mu P_L\ell W_\mu^+-\left(\frac{m_\ell}{m_W}P_R-\frac{m_\alpha}{m_W}P_L\right)\ell G^+\right]+\hc\,.
\end{align}
Here $\nu_\ell$ ($\ell=e,\mu,\tau$) denote the neutrino interaction eigenstates and $\nu_\alpha$ ($\alpha=1,2,3$) the neutrino mass eigenstates, with masses $m_\alpha$. Further, $g=e/\sin\theta_W$ is the weak coupling constant and $U$ is the leptonic mixing matrix. Comparing with Eqs.~(\ref{eq:Lag-scalar}) and (\ref{eq:Lag-vector}), one finds
\begin{align}
v_L^\alpha &=\frac{g}{\sqrt{2}}U_{\ell\alpha}^*\,,\qquad &&v_R^\alpha=0\,,\nn\\ c_L^\alpha &=\frac{g}{\sqrt{2}}U_{\ell\alpha}^*\frac{m_\alpha}{m_W}\,,\qquad &&c_R^\alpha=-\frac{g}{\sqrt{2}}U_{\ell\alpha}^*\frac{m_\ell}{m_W}\,. 
\label{eq:couplings_constants_SM} 
\end{align}

We first consider Dirac neutrinos. Using the analytical expressions for the dipole moments in the limit of light external masses presented in \cref{app:loop_functions_light_external_states}, together with the couplings in \cref{eq:couplings_constants_SM}, one obtains
\begin{align}
\mu_{\beta\alpha} &= \frac{3eG_F}{16\sqrt{2}\pi^2}(m_\alpha+m_\beta)\left(\delta_{\alpha\beta}-\frac{1}{2}\sum_{\ell=e,\mu,\tau}U_{\ell\beta}^*U_{\ell\alpha}\frac{m_\ell^2}{m_W^2}\right),\nn\\ 
d_{\beta\alpha} &= -i\frac{3eG_F}{16\sqrt{2}\pi^2}(m_\alpha-m_\beta)\left(\delta_{\alpha\beta}-\frac{1}{2}\sum_{\ell=e,\mu,\tau}U_{\ell\beta}^*U_{\ell\alpha}\frac{m_\ell^2}{m_W^2}\right). 
\end{align}
Here we have retained the leading terms in the expansion in $m_\ell^2/m_W^2$ and summed over all charged leptons in the loop, using $\sum_\ell U_{\ell\beta}^*U_{\ell\alpha}=\delta_{\alpha\beta}$.\footnote{This relation may change in the presence of additional sterile neutrinos; see Ref.~\cite{Giunti:2014ixa}.} These expressions reproduce the results of Ref.~\cite{Shrock:1982sc}. The transition dipole moments are suppressed by $m_\ell^2/m_W^2$ relative to the diagonal magnetic dipole moments as a consequence of the GIM mechanism~\cite{Glashow:1970gm}.

The corresponding anapole moments and charge radii are~\cite{Rosado:1999yn,Bernabeu:2000hf,Cabral-Rosetti:2002zyl,Ibarra:2022nzm}
\begin{align}
\ana_{\beta\alpha} &=-\frac{eG_F}{24\sqrt{2}\pi^2}\left[3\delta_{\alpha\beta}-2\sum_{\ell=e,\mu,\tau}U_{\ell\beta}^*U_{\ell\alpha}\log\left(\frac{m_\ell^2}{m_W^2}\right)\right],\nn\\ 
b_{\beta\alpha} &=\frac{eG_F}{24\sqrt{2}\pi^2}\left[3\delta_{\alpha\beta}-2\sum_{\ell=e,\mu,\tau}U_{\ell\beta}^*U_{\ell\alpha}\log\left(\frac{m_\ell^2}{m_W^2}\right)\right].
\end{align}
In contrast to the dipole moments, the transition dimension-$6$ moments are only logarithmically suppressed by the charged-lepton masses. The charge radius and anapole moment are, up to a sign and corrections of order $m_\alpha m_\beta/m_W^2$, equivalent. Since active neutrinos are relativistic and predominantly left-chiral, one may set $\gamma^5\simeq-1$ in \cref{eq:vertex_parameterization}. The dimension-$6$ part of the electromagnetic vertex can therefore be expressed either in terms of a charge radius or an anapole interaction, with $(b^\nu_{\alpha\beta})_{\rm eff}=b^\nu_{\alpha\beta}-\ana^\nu_{\alpha\beta}\simeq2b^\nu_{\alpha\beta}$. As discussed below, the Majorana anapole moment is twice the corresponding Dirac result, so that the dimension-$6$ electromagnetic interactions of relativistic Dirac and Majorana neutrinos are indistinguishable up to mass-suppressed corrections \cite{Zralek:1997sa}, in agreement with the well-known ``neutrino confusion theorem''~\cite{Kayser:1981nw,Kayser:1982br}.

For the diagonal moments, the above expressions reduce to
\begin{align}
\mu_{\alpha\alpha} &\simeq 9.5\times10^{-17}\left(\frac{m_\alpha}{1\,\eV}\right)\gev^{-1}\simeq3.2\times10^{-19}\left(\frac{m_\alpha}{1\,\eV}\right)\mu_B\,,\nn\\ 
d_{\alpha\alpha} &=0\,. 
\end{align}
The vanishing of the diagonal electric dipole moment follows directly from the factor $m_\alpha-m_\beta$ in the expression above and does not require assuming CP conservation. The diagonal anapole moments and charge radii satisfy
\begin{equation} 
\ana_{\alpha\alpha}=\sum_{\ell=e,\mu,\tau}|U_{\ell\alpha}|^2\ana_{\ell\ell}=-b_{\alpha\alpha}\,,
\end{equation}
where~\cite{Bernabeu:2000hf,Cabral-Rosetti:2002zyl}
\begin{align}
\ana_{\ell\ell} &\simeq\text{diag}\left(-5.37,-3.11,-1.93\right)\times10^{-7}\,\gev^{-2}\nn\\ 
&\simeq\text{diag}\left(-2.07,-1.20,-0.74\right)\times10^{-34}\,\text{cm}^2\,. 
\end{align}

For Majorana neutrinos, the electromagnetic moments follow directly from the general results of Section \ref{sec:Majorana}. Using the same approximations as above, the transition dipole moments are~\cite{Schechter:1981hw,Shrock:1982sc}
\begin{align}
\mu^{\nu^M}_{\beta\alpha} &\simeq-i\frac{3eG_F}{16\sqrt{2}\pi^2}(m_\alpha+m_\beta)\sum_{\ell=e,\mu,\tau}\Im\left(U_{\ell\beta}^*U_{\ell\alpha}\right)\frac{m_\ell^2}{m_W^2}\,,\nn\\ 
d^{\nu^M}_{\beta\alpha} &\simeq i\frac{3eG_F}{16\sqrt{2}\pi^2}(m_\alpha-m_\beta)\sum_{\ell=e,\mu,\tau}\Re\left(U_{\ell\beta}^*U_{\ell\alpha}\right)\frac{m_\ell^2}{m_W^2}\,, 
\end{align}
while the dimension-$6$ moments are
\begin{align}
\ana^{\nu^M}_{\beta\alpha} &=-\frac{eG_F}{12\sqrt{2}\pi^2}\left[3\delta_{\alpha\beta}-2\sum_{\ell=e,\mu,\tau}\Re\left(U_{\ell\beta}^*U_{\ell\alpha}\right)\log\left(\frac{m_\ell^2}{m_W^2}\right)\right],\nn\\ b^{\nu^M}_{\beta\alpha} 
&=-i\frac{eG_F}{6\sqrt{2}\pi^2}\sum_{\ell=e,\mu,\tau}\Im\left(U_{\ell\beta}^*U_{\ell\alpha}\right)\log\left(\frac{m_\ell^2}{m_W^2}\right).
\end{align}

As expected from the general properties of Majorana fermions discussed in \cref{sec:Majorana}, the diagonal magnetic dipole moment, electric dipole moment and charge radius vanish,
\begin{equation} 
\mu^{\nu^M}_{\alpha\alpha}=d^{\nu^M}_{\alpha\alpha}=b^{\nu^M}_{\alpha\alpha}=0\,, 
\end{equation}
while the diagonal anapole moment is non-zero and is twice the corresponding Dirac result,
\begin{equation} 
\ana^{\nu^M}_{\alpha\alpha}=2\ana_{\alpha\alpha}\,.
\end{equation}
For $\alpha\neq\beta$, all four types of transition moments are in principle possible. If CP is conserved, however, their allowed structure is determined by the relative CP parities of $\nu_\alpha$ and $\nu_\beta$: depending on these parities, only a transition electric or magnetic dipole moment and, correspondingly, a transition anapole moment or charge radius can be non-zero \cite{Nieves:1981zt,Schechter:1981hw,Pal:1981rm,Kayser:1984ge,Giunti:2014ixa}.

Additional contributions to the neutrino electromagnetic moments can arise from new charged particles or interactions associated, for instance, with the mechanism responsible for neutrino mass generation \cite{Pal:1981rm,Barr:1990um,Pal:1991qr}. Such contributions can be calculated directly using the general Dirac and Majorana results derived in Sections \ref{sec:Dirac} and \ref{sec:Majorana}.

\section{Conclusions}
\label{sec:Conclusions}

In this work, we have calculated the one-loop contributions to all diagonal and transition electromagnetic moments of charge-neutral Dirac-or Majorana fermions stemming from scalar and vector mediators in the loop. For the diagrams with vector degrees of freedom, we opted for utilizing the background field method to ensure a gauge-independent and physical amplitude. We provide analytical formulas for phenomenologically interesting cases, including a large hierarchy between the external fermions, small mass splitting between the external fermions, and a large hierarchy between the UV physics generating the EM moments and the fermion masses. Thereby, the general results we provide are applicable to a variety of phenomenologically interesting models. 

We have applied our results to calculate the EM moments of neutral components of $SU(2)_L$ multiplets (which constitute candidates for dark matter) and active neutrinos. While we find the known SM-like results in the latter case, the general formulas can be used to calculate additional contributions to their EM moments due to BSM physics, helping to interpret new experimental results on the neutrino multipoles in light of new physics. Meanwhile, for the former models our findings indicate very strong constraints on a subclass of the minimal dark matter framework from dark matter direct detection experiments.

\acknowledgments
The work of AI and MR was supported by the Collaborative Research Center SFB1258 and by the Deutsche Forschungsgemeinschaft (DFG, German Research Foundation) under Germany’s Excellence Strategy - EXC-2094 - 390783311. The work of MR is supported by the KIAS Individual Grant PG108401. MR would like to thank Martin Napetschnig for helpful discussions.

 .

\appendix

\section{General expressions for the loop functions}\label{appendix:Loopfunctions_q_zero}

In this appendix, we present analytic expressions for the loop functions appearing in the EM interactions for neutral fermions $\chi_1$ and $\chi_2$ (with masses $\mdm1$ and $\mdm{2}$ respectively) at the one-loop level in full generality.

We present the results in terms of the scalar three point Passarino-Veltman function~\cite{Passarino:1978jh}
\begin{align}
C_0(p_1^2,q^2, p_2^2 , m_0, m_1, m_2) &=  \frac{1}{i \pi^2} \int d^n k \frac{1}{(k^2-m_0^2)}\frac{1}{(k+p_1)^2 - m_1^2} \frac{1}{(k+p_2)^2-m_2^2},
\end{align}
and
\begin{equation}
\widetilde{F}(\xi,\mu,\eta) = \frac{\sqrt{\lambda(\xi^2,\eta^2,\mu^2)}}{\xi^2}\, \log\left(\frac{\mu^2+\eta^2-\xi^2 + \sqrt{\lambda(\xi^2,\eta^2,\mu^2)}}{2\eta\mu}+i\epsilon \right),
\end{equation}
where
\begin{equation}
\lambda(a,b,c) = a^2+b^2+c^2-2ab -2ac -2 bc
\end{equation}
is the Källén function.

\begin{figure}
    \centering
   \includegraphics[width=.49\linewidth]{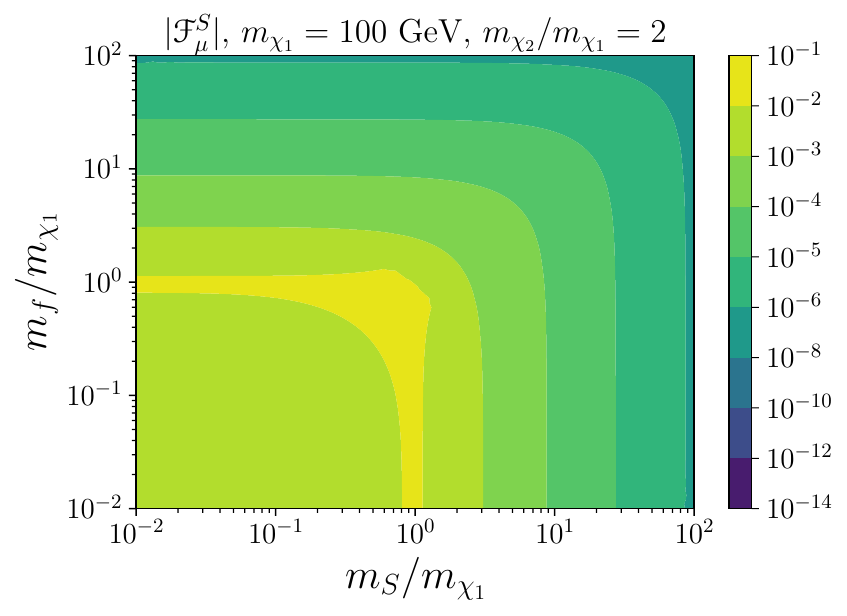}
  \includegraphics[width=.49\linewidth]{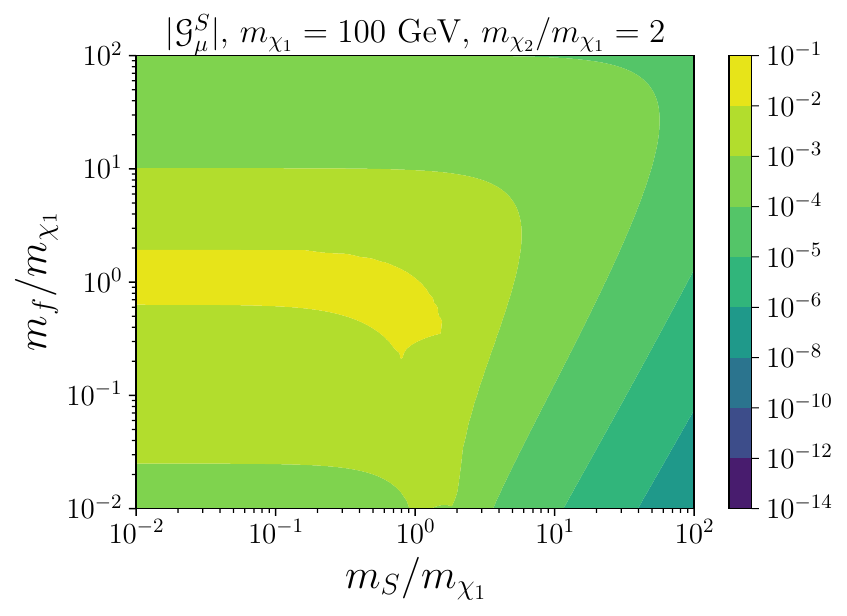}
 \includegraphics[width=.49\linewidth]{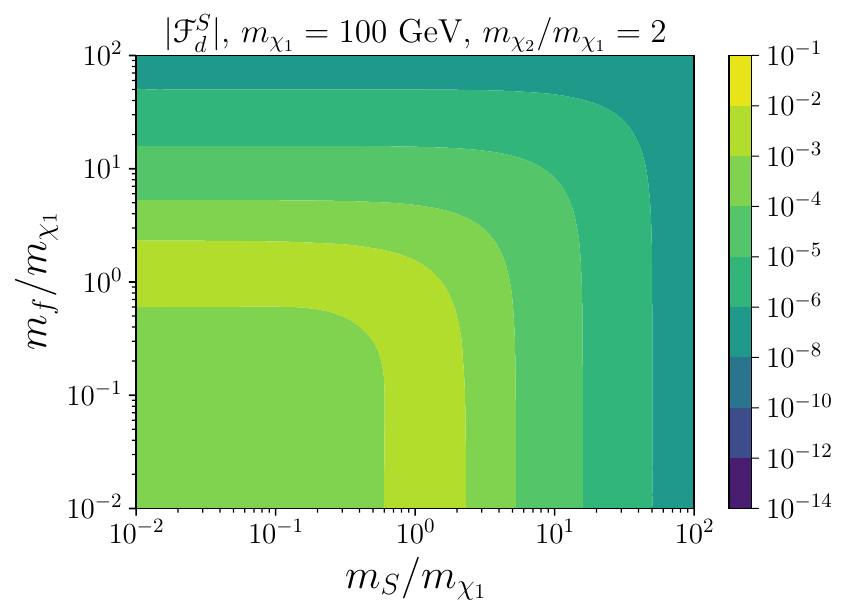}
  \includegraphics[width=.49\linewidth]{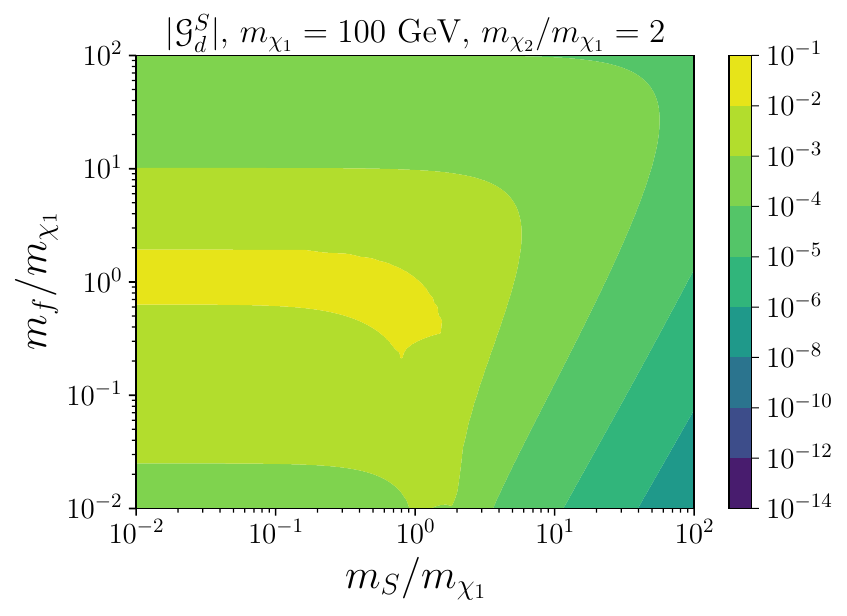}
   \includegraphics[width=.49\linewidth]{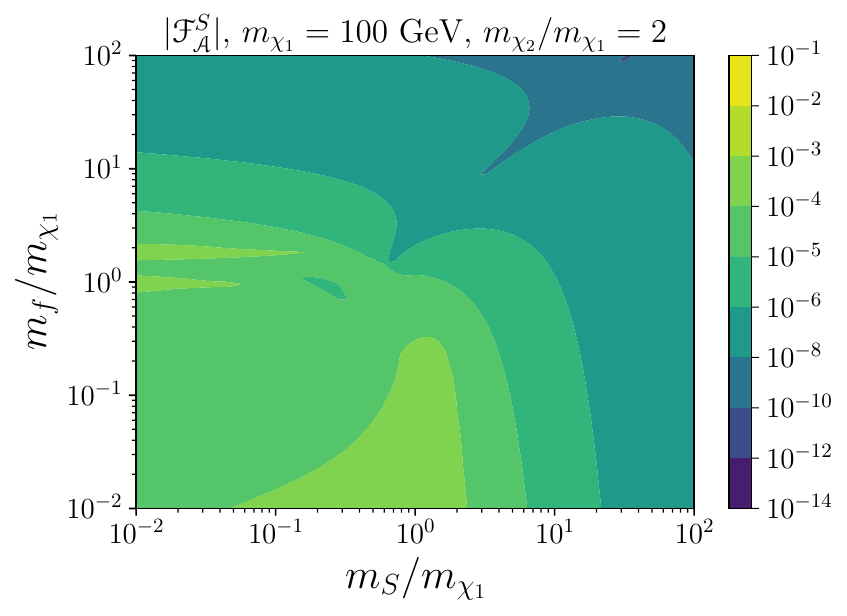}
  \includegraphics[width=.49\linewidth]{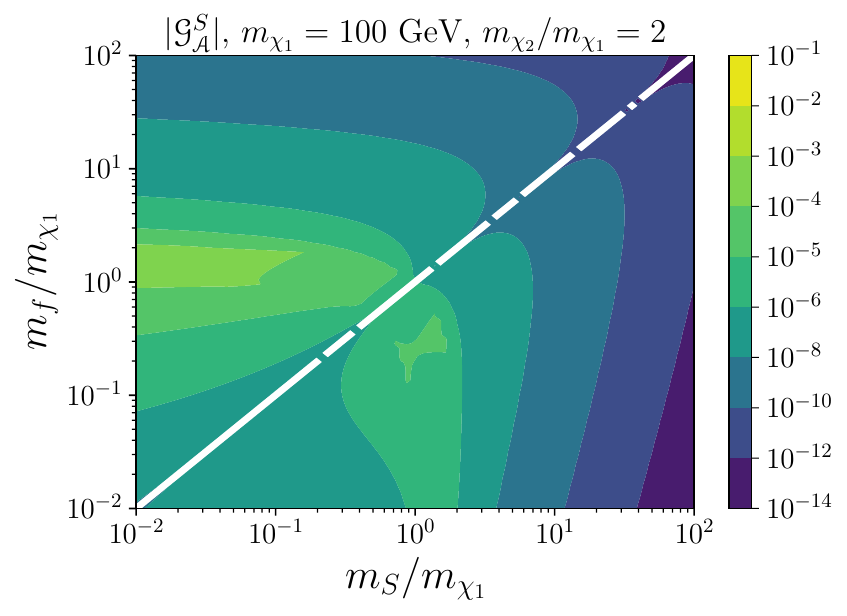}
   \includegraphics[width=.49\linewidth]{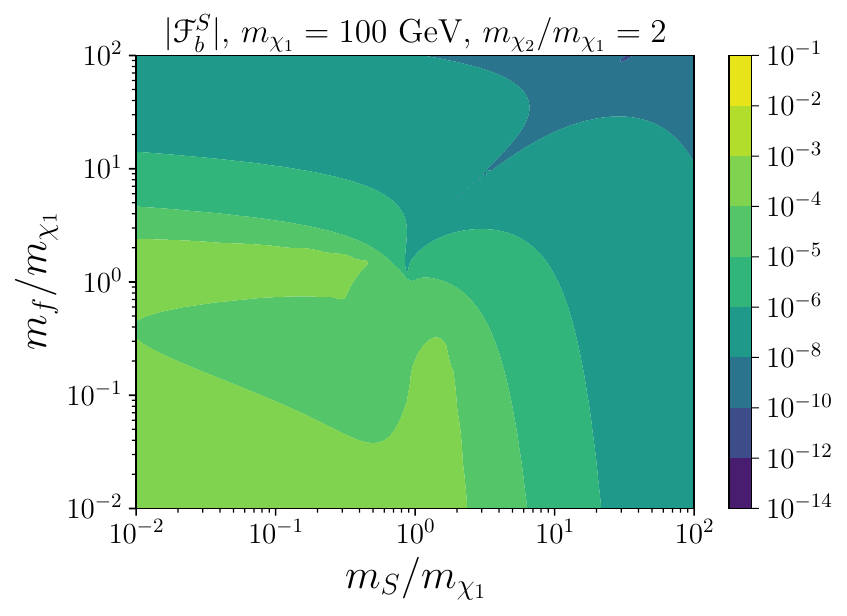}
  \includegraphics[width=.49\linewidth]{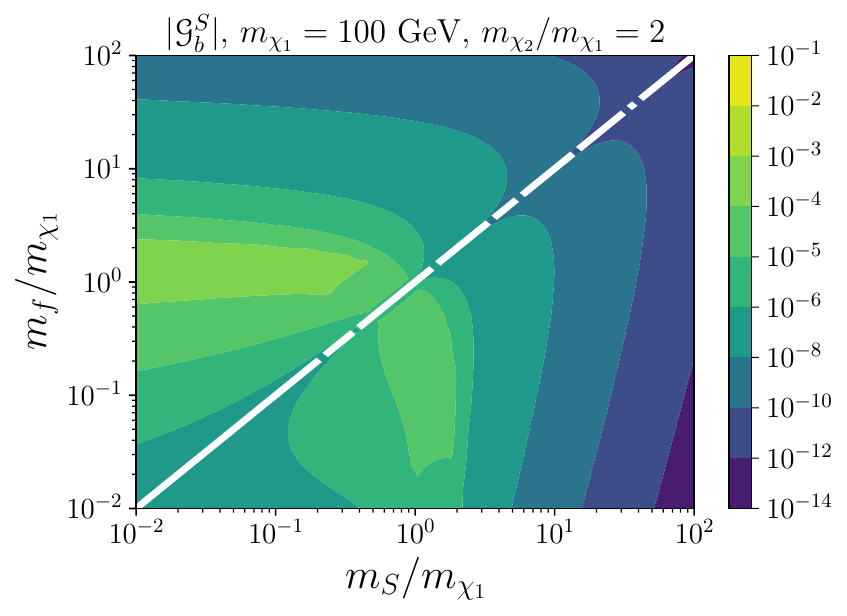}
    \caption{Scalar loop functions for fixed $\mdm1 = 100\gev$ and $\mdm{2} = 2 \mdm1$; the dipole loop functions have units of $\gev^{-1}$, the anapole and charge radius loop functions have units of $\gev^{-2}$.}
    \label{app:fig:Scalar_loop_functions}
\end{figure}

\begin{figure}
    \centering
   \includegraphics[width=.49\linewidth]{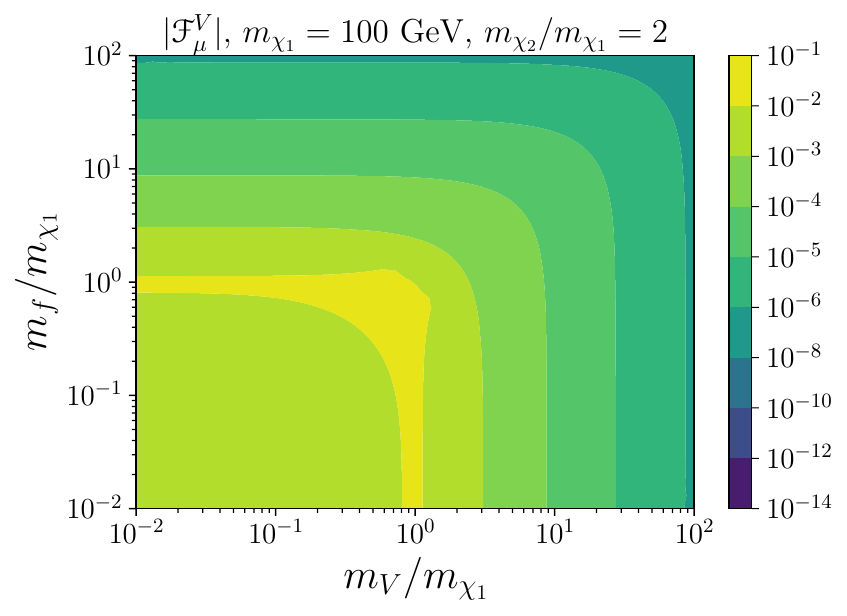}
  \includegraphics[width=.49\linewidth]{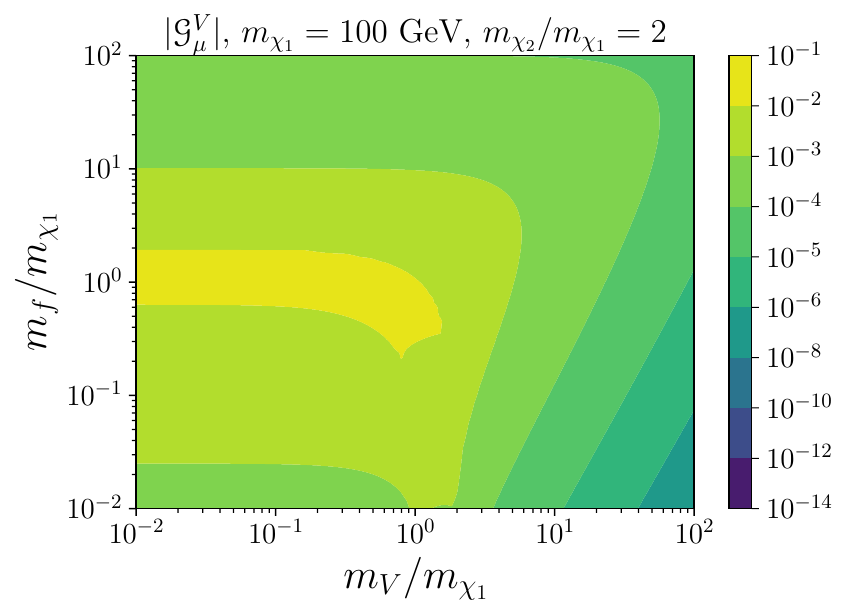}
 \includegraphics[width=.49\linewidth]{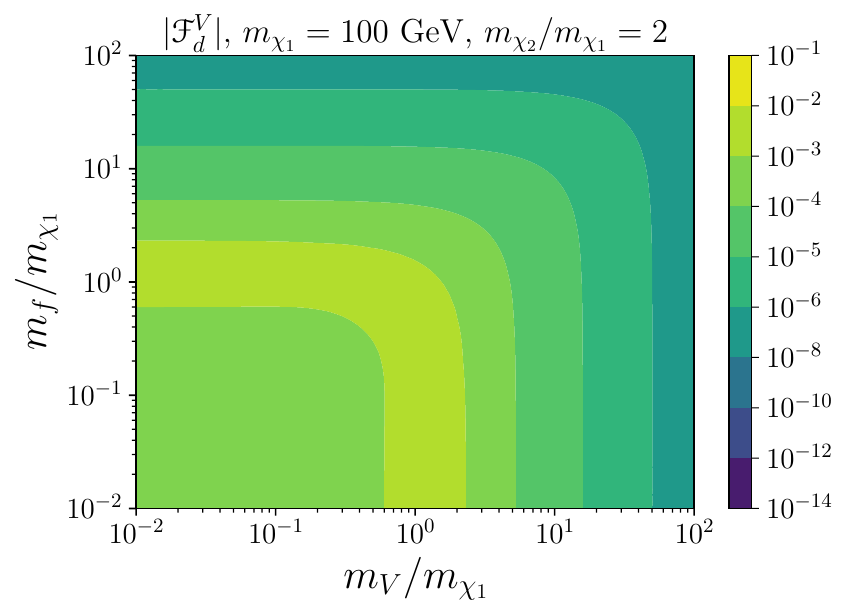}
  \includegraphics[width=.49\linewidth]{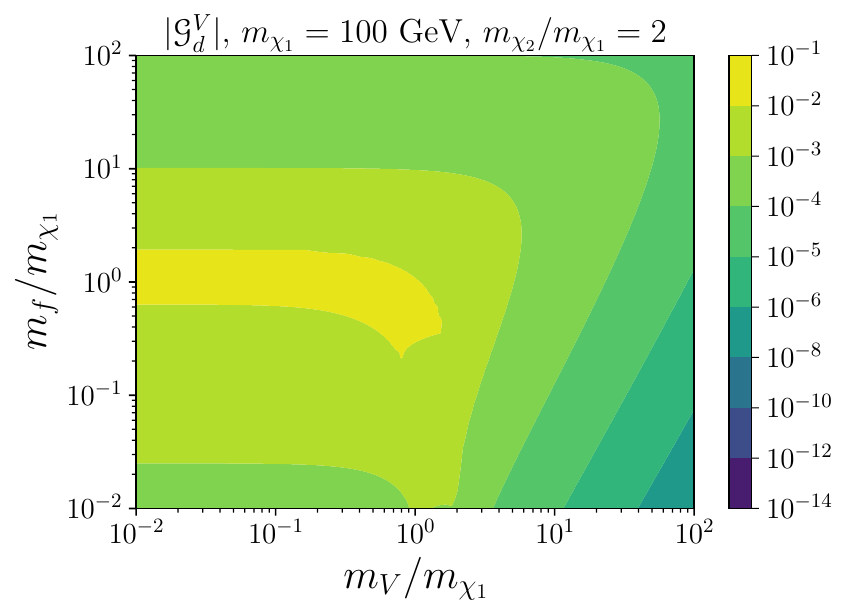}
   \includegraphics[width=.49\linewidth]{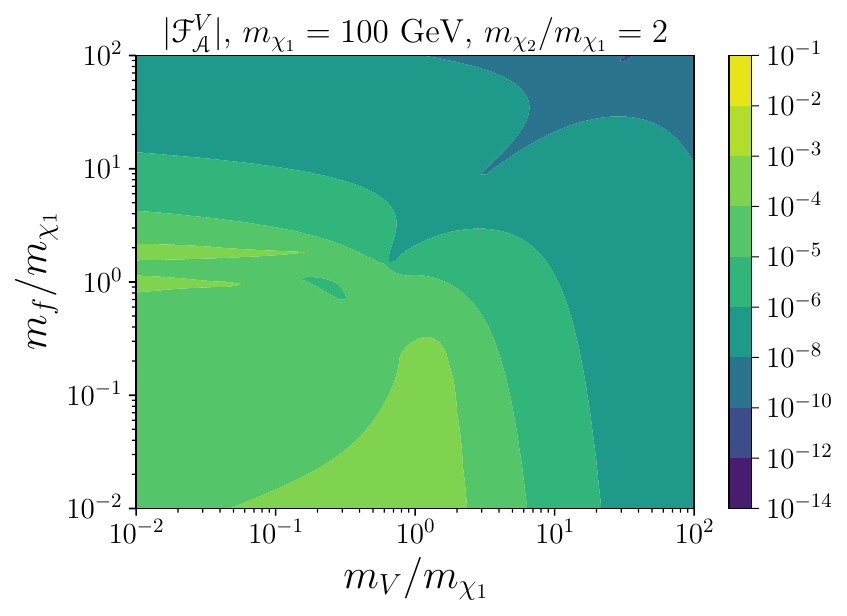}
  \includegraphics[width=.49\linewidth]{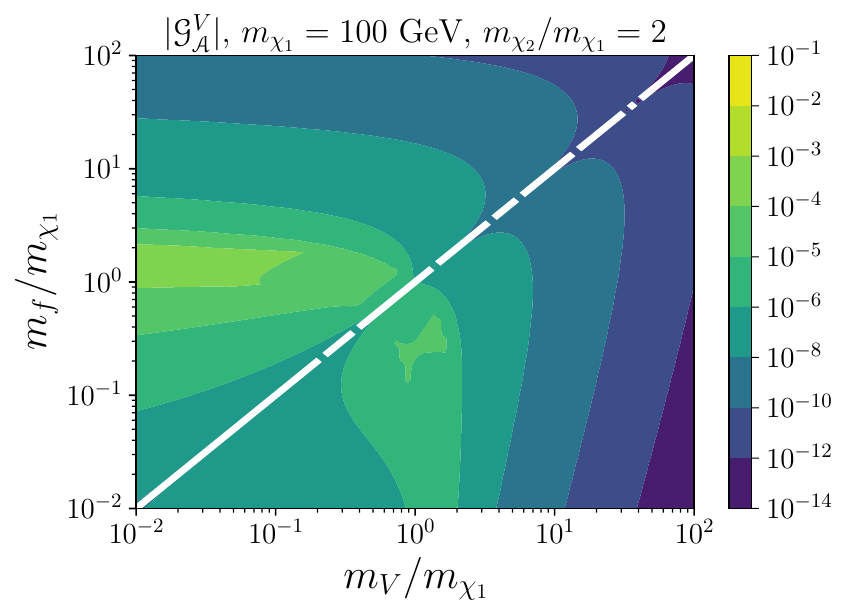}
   \includegraphics[width=.49\linewidth]{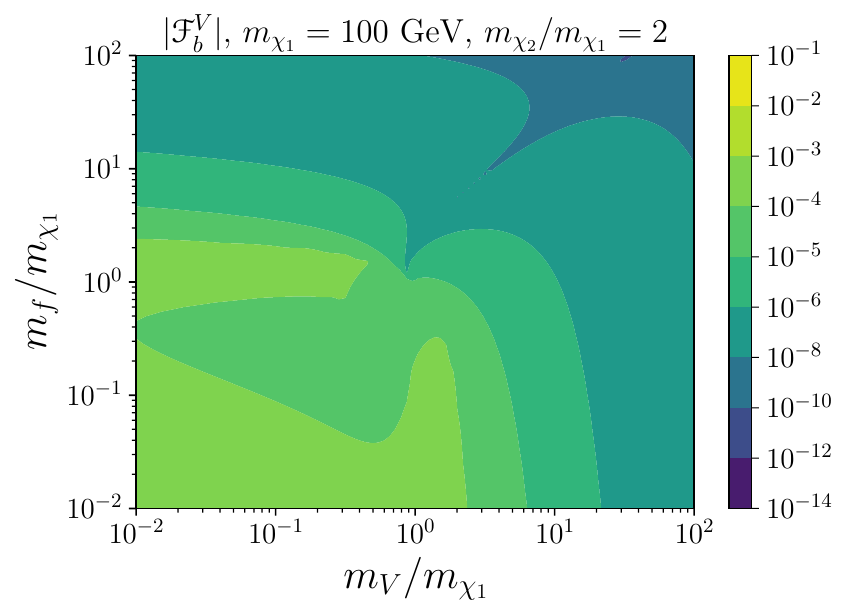}
  \includegraphics[width=.49\linewidth]{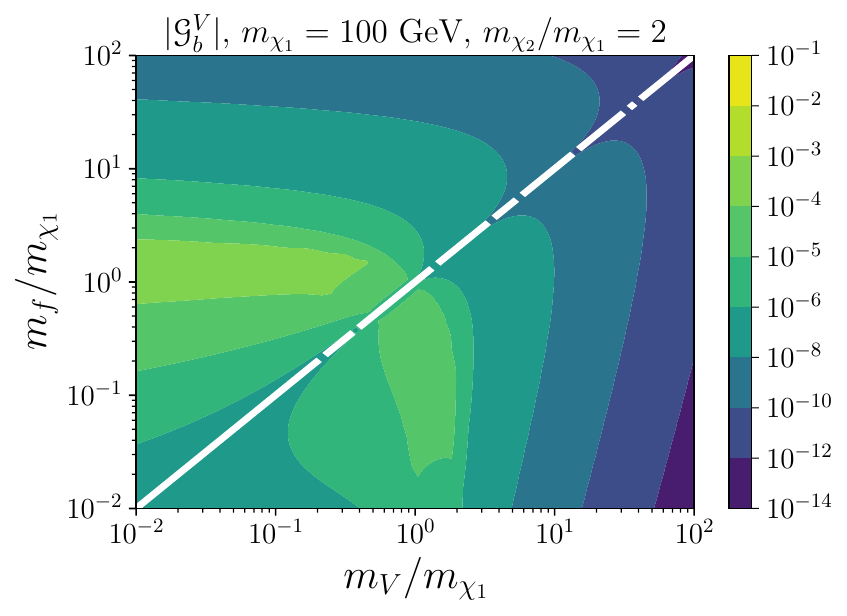}
    \caption{Same as \cref{app:fig:Scalar_loop_functions}, but for the vector loop functions.}
    \label{app:fig:Vector_loop_functions}
\end{figure}

Further, for the presentation of the general results we parameterize the loop functions $\matheul{F}^{S/V}_i$ and $\matheul{G}^{S/V}_i$ as
\begin{equation}
\matheul{F}_i^{S/V}(\mu,\eta,\xi) = \frac{\overline{\matheul{F}}_i^{S/V}(\mu,\eta,\xi) }{\mdm1^{(i)} (1+\xi)^2}\quad\text{and}\quad \matheul{G}_i^{S/V}(\mu,\eta,\xi) = \frac{\overline{\matheul{G}}_i^{S/V}(\mu,\eta,\xi) }{\mdm1^{(i)} (1+\xi)^2},
\end{equation}
where we defined $\mdm1^{(\mu)} = \mdm1^{(d)} = \mdm1$, $\mdm1^{(\ana)} = \mdm1^{(b)} = \mdm1^{2}$. In the following we report the analytical results for these reduced loop functions $\overline{\matheul{F}}_i^{S/V}$ and $\overline{\matheul{G}}_i^{S/V}$ and a numerical evaluation in \cref{app:fig:Scalar_loop_functions,app:fig:Vector_loop_functions} using \texttt{LoopTools}~\cite{Hahn:1998yk}. 

\paragraph{Scalar loop functions}
\begin{align}
\overline{\matheul{F}}_\mu^S(\mu,\eta,\xi) = \frac{1}{\xi(\xi-1)}\biggr\{&   \xi^2 \left[ \widetilde{F}(\xi,\mu,\eta) -  \widetilde{F}(1,\mu,\eta)     \right] \nn\\
&+(1-\xi^2)\biggr[\xi + (\eta^2-\mu^2)\log(\frac{\mu}{\eta}) + \mdm1^2\xi \times \nn\\
&\times\biggr(\eta^2 C_0(0,\mdm1^2,\mdm1^2\xi^2, \mdm1 \eta, \mdm1 \eta, \mdm1 \mu) + (\mu \leftrightarrow \eta)\biggr)
\biggr]\biggr\}
\end{align}
\begin{align}
\overline{\matheul{G}}_\mu^S(\mu,\eta,\xi) = - \mdm1^2 \mu (1+\xi)^2 C_0(0,\mdm1^2, \mdm1^2\xi^2, \mdm1 \mu, \mdm1 \mu, \mdm1 \eta)
\end{align}
\begin{align}
\overline{\matheul{F}}_d^S(\mu,\eta,\xi) = i\frac{ (1+\xi)}{\xi(1-\xi)^2} \biggr\{&   \xi^2 \left[ \widetilde{F}(\xi,\mu,\eta) -  \widetilde{F}(1,\mu,\eta)     \right] \nn\\
&-(1-\xi^2)\biggr[\xi + (\mu^2-\eta^2)\log(\frac{\mu}{\eta}) + \mdm1^2\xi \times \nn\\
&\times\biggr(\eta^2 C_0(0,\mdm1^2,\mdm1^2\xi^2, \mdm1 \eta, \mdm1 \eta, \mdm1 \mu) + (\mu \leftrightarrow \eta)\biggr)
\biggr]\biggr\}
\end{align}
\begin{equation}
\overline{\matheul{G}}_d^S(\mu,\eta,\xi)= i\overline{\matheul{G}}_\mu^S(\mu,\eta,\xi).
\end{equation}
\begin{align} 
\overline{\matheul{F}}_\ana^S(\mu,\eta,\xi) =\frac{1}{(\xi - 1)^3} \biggr\{&(1+\xi)\left[(1+2\xi) \widetilde{F}(1,\mu,\eta)- \xi(2+\xi) \widetilde{F}(\xi,\mu,\eta)\right]\nn \\
&+ \frac{1-\xi}{\xi}\biggr[ 3\xi (1+\xi)^2 + \kappa^\xi_1 \log(\frac{\mu}{\eta}) \nn\\
&+\kappa^\xi_2 C_0(0,\mdm1^2, \mdm1^2\xi^2,\mdm1 \eta, \mdm1 \eta, \mdm1 \mu)\nn\\
&+ \kappa^\xi_3 C_0(0,\mdm1^2, \mdm1^2\xi^2,\mdm1 \mu, \mdm1 \mu, \mdm1 \eta) \biggr]
\biggr\}
\end{align}
with
\begin{align}
\kappa_1^\xi &=  2 \left(\xi  (\xi +4)+1\right) (\mu^2-\eta^2)+(\xi -1)^2 \xi\\
\kappa_2^\xi &= \mdm1^2 \xi \left[2 \eta ^4+ 2\eta ^2 \left((\xi +1)^2-2 \mu ^2\right)+2 \mu ^4-\mu ^2 (\xi +1)^2+\xi ^3+\xi\right]\\
\kappa_3^\xi &= \mdm1^2\xi \left[2 \eta ^4-2 \eta ^2 \left(2 \mu ^2+\xi ^2+1\right)+2 \mu ^4+\mu ^2 (\xi  (3 \xi +2)+3)+2 \xi ^2\right]
\end{align}
\begin{align}
\overline{\matheul{G}}_\ana^S(\mu,\eta,\xi) = \frac{\mu }{\xi-1} \biggr\{&2\log\left(\frac{\mu^2}{\eta^2}\right) +  \mdm1^2\biggr[\left(1 + 2 \mu^2 - 2 \eta^2 + \xi^2\right) \times\nn \\ &  C_0\left(0, \mdm1^2, \mdm1^2 \xi^2,  \mdm1 \mu, \mdm1 \mu, \mdm1 \eta\right) -(\mu \leftrightarrow \eta)\biggr] \biggr\}
\end{align}

\begin{align}
\overline{\matheul{F}}_b^S(\mu,\eta,\xi) =-\frac{1}{(1-\xi)^2(1+\xi)} \biggr\{&(\xi-1)\left[(2\xi-1) \widetilde{F}(1,\mu,\eta)+(\xi-2)\xi \widetilde{F}(\xi,\mu,\eta)\right]\nn \\
&+ \frac{(\xi+1)}{\xi}\biggr[ 3\xi (1-\xi)^2 + \kappa^\xi_4 \log(\frac{\mu}{\eta}) \nn\\
&+\kappa^\xi_5 C_0(0,\mdm1^2, \mdm1^2\xi^2,\mdm1 \eta, \mdm1 \eta, \mdm1 \mu)\nn\\
&+ \kappa^\xi_6 C_0(0,\mdm1^2, \mdm1^2\xi^2,\mdm1 \mu, \mdm1 \mu, \mdm1 \eta) \biggr]
\biggr\}
\end{align}
with
\begin{align}
\kappa_4^\xi &= 2 \left(\xi  (\xi -4)+1\right) (\eta^2-\mu^2)+(\xi +1)^2 \xi\\
\kappa_5^\xi &= \mdm1^2 \xi  \left[2 \eta ^4+2 \eta ^2 \left((\xi -1)^2-2 \mu ^2\right)+\left(\mu ^2+\xi \right) \left(2 \mu ^2-\xi ^2-1\right) \right] \\
\kappa_6^\xi &= \mdm1^2 \xi  \left[2 \eta ^4-2 \eta ^2 \left(2 \mu ^2+\xi ^2+1\right)+2 \mu ^4+\mu ^2 (\xi  (3 \xi -2)+3)+2 \xi ^2\right]
\end{align}

\begin{align}
\overline{\matheul{G}}_b^S(\mu,\eta,\xi) =-\mu \frac{ (1+\xi) }{(1 - \xi)^2} \biggr\{&2\log\left(\frac{\mu^2}{\eta^2}\right) + \mdm1^2 \biggr[\left(1 + 2 \mu^2 - 2 \eta^2 + \xi^2\right) \times\nn \\ &  C_0\left(0, \mdm1^2, \mdm1^2 \xi^2,  \mdm1 \mu, \mdm1 \mu, \mdm1 \eta\right) -(\mu \leftrightarrow \eta)\biggr] \biggr\}
\end{align}

\paragraph{Vector loop functions}
\begin{align}
\overline{\matheul{F}}_\mu^V\left(\mu,\eta,\xi\right) = \frac{2}{\xi(\xi-1)} \biggr\{ &\xi^2 \left( -\widetilde{F}(1,\mu,\eta) + \widetilde{F}(\xi,\mu,\eta) \right)\nn \\ 
& + (\xi^2-1)\biggr[ -\xi + (\eta^2 - \mu^2) \log(\frac{\eta}{\mu}) \nn \\
&+\mdm1^2 \xi \biggr(\left((1+\xi^2)-\eta^2\right) C_0(0,\mdm1^2,\mdm1^2\xi^2,\mdm1\eta,\mdm1\eta,\mdm1\mu)\nn\\
&-\mu^2C_0(0,\mdm1^2,\mdm1^2\xi^2, \mdm1\mu, \mdm1\mu,\mdm1\eta)\biggr)\biggr] \biggr\}
\end{align}
\begin{align}
\overline{\matheul{G}}_\mu^V\left(\mu,\eta,\xi\right) =- 4\mdm1^2\mu(1+\xi)^2C_0\left(0,\mdm1^2,\mdm1^2\xi^2,\mdm1\eta,\mdm1\eta,\mdm1\mu\right)
\end{align}
\begin{align}
\overline{\matheul{F}}_d^V\left(\mu,\eta,\xi\right) = i\frac{2 (1+\xi)}{\xi(1-\xi)^2} \biggr\{ &\xi^2 \left( \widetilde{F}(1,\mu,\eta) - \widetilde{F}(\xi,\mu,\eta) \right)\nn \\ 
& + (\xi^2-1)\biggr[ -\xi + (\eta^2 - \mu^2) \log(\frac{\mu}{\eta}) \nn \\
&-\mdm1^2 \xi \biggr(\left(\eta-\xi+1\right)\left(\eta+\xi-1\right)\times\nn\\
&\hphantom{-\mdm1^2\xi}\times C_0(0,\mdm1^2,\mdm1^2\xi^2,\mdm1\eta,\mdm1\eta,\mdm1\mu)\nn\\
&+\mu^2C_0(0,\mdm1^2,\mdm1^2\xi^2, \mdm1\mu, \mdm1\mu,\mdm1\eta)\biggr)\biggr] \biggr\}
\end{align}
\begin{align}
\overline{\matheul{G}}_d^V\left(\mu,\eta,\xi\right) &= 4i \mdm1^2\mu(1+\xi)^2C_0\left(0,\mdm1^2,\mdm1^2\xi^2,\mdm1\eta,\mdm1\eta,\mdm1\mu\right)\\
&= - i\overline{\matheul{G}}_\mu^V \left(\mu,\eta,\xi\right)
\end{align}
\begin{align}
\overline{\matheul{F}}_\ana^V\left(\mu,\eta,\xi\right) =- \frac{2 }{(1-\xi)^3} \biggr\{& (1+\xi)\left[\xi(2+\xi)\widetilde{F}(\xi,\mu,\eta)-(1+2\xi)\widetilde{F}(1,\mu,\eta)\right]  \nn\\
& + \frac{\xi-1}{\xi}\biggr[3\xi(1+\xi)^2+\kappa_7^\xi \log(\frac{\mu}{\eta}) \nn\\
& +\kappa_8^\xi C_0(0,\mdm1^2,\mdm1^2\xi^2,\mdm1\eta,\mdm1\eta,\mdm1\mu) \nn\\
&+ \kappa_9^\xi C_0(0,\mdm1^2,\mdm1^2\xi^2,\mdm1\mu,\mdm1\mu,\mdm1\eta) 
\biggr]\biggr\}
\end{align}
where we defined
\begin{align}
\kappa_7^\xi &= -2 \eta ^2 (\xi  (\xi +4)+1)+2 \mu ^2 (\xi  (\xi +4)+1)+(\xi -1)^2 \xi\\
\kappa_8^\xi &= \mdm1^2\xi \bigr[2 \eta ^4+2 \eta ^2 \left((\xi +1)^2-2 \mu ^2\right)+2 \mu ^4\nn\\
&\hphantom{=}-\mu ^2 (\xi +1)^2+\xi ^2 \left(-\xi ^2+\xi +2\right)+\xi -1\bigr]\\
\kappa_9^\xi &= \mdm1^2\xi \bigr[2 \eta ^4-2 \eta ^2 \left(2 \mu ^2+\xi ^2+1\right)+2 \mu ^4+\mu ^2 (\xi  (3
\xi +2)+3)+2 \xi ^2 \bigr]
\end{align}
\begin{align}
\overline{\matheul{G}}_\ana^V\left(\mu,\eta,\xi\right) = \frac{4 \mu}{\xi-1}\biggr\{&2\log(\frac{\mu^2}{\eta^2}) + \mdm1^2 
\biggr[(1-2\eta^2+2\mu^2+\xi^2)\times\nn\\
&C_0(0,\mdm1^2,\mdm1^2\xi^2,\mdm1\mu,\mdm1\mu,\mdm1\eta)
- (\mu\leftrightarrow\eta) \biggr]\biggr\}
\end{align}
\begin{align}
\overline{\matheul{F}}_b^V\left(\mu,\eta,\xi\right) = -\frac{2}{(1-\xi)^2(1+\xi)}\biggr\{&(\xi-1)\left[(2\xi-1)\widetilde{F}(1,\mu,\eta)+(\xi-2)\xi \widetilde{F}(\xi,\mu,\eta) \right] \nn\\
&+\frac{1+\xi}{\xi}\biggr[3(1-\xi)^2\xi+ \kappa_{10}^\xi \log(\frac{\mu}{\eta})  \nn\\
&+\kappa_{11}^\xi C_0(0,\mdm1^2,\mdm1^2\xi^2,\mdm1\eta,\mdm1\eta,\mdm1\mu)\nn\\
&+\kappa_{12}^\xi C_0(0,\mdm1^2,\mdm1^2\xi^2,\mdm1\mu,\mdm1\mu,\mdm1\eta)
\biggr]
\biggr\}
\end{align}
\begin{align}
\kappa_{10}^\xi &= 2 \eta^2 ((\xi-4) \xi+1)-2 \mu^2 ((\xi-4) \xi+1)+\xi (\xi+1)^2\\
\kappa_{11}^\xi &= \mdm1^2\xi \bigr[-2 \eta^4+2\eta^2 \left(2 \mu^2-(\xi-1)^2\right)-2\mu^4+\mu^2 (\xi-1)^2\\
&\qquad\qquad+\xi^2 \left(\xi^2+\xi-2\right)+\xi+1\bigr]\\
\kappa_{12}^\xi &= \mdm1^2\xi \bigr[2 \eta ^4-2 \eta ^2 \left(2 \mu ^2+\xi ^2+1\right)+2 \mu ^4+\mu ^2 (\xi  (3
\xi -2)+3)+2 \xi ^2\bigr]
\end{align}
\begin{align}
    \overline{\matheul{G}}^V_b = \frac{4 \mu  (\xi +1)}{(\xi -1)^2} \biggr\{&2  \log \left(\frac{\mu^2 }{\eta^2 }\right) 
        +\mdm1^2 \biggr[\left(1-2 \eta ^2+2 \mu ^2+\xi ^2\right) \times \\
        & C_0\left(0,\mdm1^2,\mdm1^2 \xi ^2, \mdm1 \mu,  \mdm1 \mu, 
        \mdm1 \eta\right)- (\mu\leftrightarrow\eta)\biggr]\biggr\}
\end{align}

\section{Limit of identical external states}\label{app:loop_functions_diagonal}
Here we summarize the loop functions for the diagonal case, \textit{i.e.} when $\xi = 1$. We introduce
\begin{equation}
\Delta = \lambda(1,\eta^2,\mu^2)=(\mu^2-\eta^2+1)^2-4\mu^2
\end{equation}
and
\begin{equation}
    f(\mu,\eta)  = \frac{F(1,\mu,\eta)}{\Delta} = \frac{1}{\sqrt{\Delta}}\, \log\left(\frac{\mu^2+\eta^2-1 + \sqrt{\Delta}}{2\eta\mu}+i\epsilon \right),
\end{equation}
and find find the following loop functions:
\paragraph{Scalar loop functions}
\begin{align}
\matheul{F}_\mu^S =  - \frac{1}{\mdm1}\biggr[&\left(\Delta +\eta ^2+\mu ^2-1\right) f(\mu,\eta)+ (\eta^2 -\mu^2 ) \log \left(\frac{\mu}{\eta}\right)+1\biggr]
\end{align}
\begin{align}
\matheul{G}_\mu^S =- \frac{\mu}{\mdm1}  \biggr[\left(-\eta ^2+\mu ^2-1\right) f(\mu,\eta)+ \log \left(\frac{\eta }{\mu }\right)\biggr]
\end{align}
\begin{align}
\matheul{F}_d^S = 0,\quad  \matheul{G}_d^S = i  \matheul{G}_\mu^S
\end{align}
\begin{align}
\matheul{F}_\ana^S = \frac{1}{6\mdm1^2}\biggr[\left(-3 \eta ^2+3 \mu ^2-1\right)f(\mu,\eta)+3   \log \left(\frac{\eta }{\mu}\right)\biggr],\quad \matheul{G}_\ana^S = 0
\end{align}
\begin{align}
\matheul{F}_b^S = \frac{1}{6\Delta \mdm1^2}\biggr[& \Delta  \left(8 \eta ^2-8 \mu ^2+1\right) \log \left(\frac{\eta }{\mu }\right)-2 \left(4 \Delta +\eta ^2+3 \mu ^2-1\right) \nn\\
&+f(\mu,\eta) \biggr(\mu ^2 \left(-7 \Delta +12 \eta ^2-4\right)+\Delta  \left(-8 \Delta -9 \eta ^2+5\right)+4 \mu ^4\biggr) \biggr]
\end{align}
\begin{align}
\matheul{G}_b^S = \frac{2 \mu}{3 \Delta \mdm1^2}  \biggr[& \Delta  \log \left(\frac{\mu }{\eta }\right)+\eta ^2-\mu ^2+f(\mu,\eta)\times\nn\\
&\times \left(-\mu ^2 \left(\Delta +2 \eta ^2+3\right)+(\Delta -1) \left(\eta ^2-1\right)+2 \mu ^4\right)\biggr]
\end{align}
\paragraph{Vector loop functions}
\begin{align}
\matheul{F}_\mu^V = -\frac{2}{\mdm1} \biggr[&\left(\Delta -\eta ^2+3 \mu^2+1\right)f(\mu,\eta)+ \left(\mu^2 -\eta^2 +2\right) \log\left(\frac{\eta}{\mu}\right)+1 \biggr]
\end{align}
\begin{align}
\matheul{G}_\mu^V = -\frac{4\mu}{\mdm1} \biggr[\left(\eta ^2-\mu ^2-1\right)f(\mu,\eta)+ \log \left(\frac{\mu }{\eta
}\right)\biggr]
\end{align}
\begin{align}
\matheul{F}_d^V = 0,\quad  \matheul{G}_d^V = -i \matheul{G}_\mu^V
\end{align}
\begin{align}
\matheul{F}_\ana^V = \frac{1}{3\mdm1^2}\biggr[\left(-3 \eta ^2+3 \mu ^2+7\right) f(\mu,\eta)+3 \log \left(\frac{\eta }{\mu }\right)\biggr],\quad \matheul{G}_\ana^V = 0
\end{align}
\begin{align}
\matheul{F}_b^V = \frac{1}{3 \Delta
    \mdm1^2}  \biggr[&\Delta  \left(-8 \eta ^2+8 \mu ^2+5\right) \log \left(\frac{\mu }{\eta }\right)-2 \left(4 \Delta +\eta ^2+3 \mu ^2-1\right)\nn\\
&- f(\mu,\eta)\biggr(\mu ^2 \left(13
\Delta -12 \eta ^2+4\right)+\Delta  \left(8 \Delta +3 \eta ^2+1\right)-4 \mu ^4\biggr)\biggr]
\end{align}
\begin{align}
\matheul{G}_b^V = \frac{8 \mu }{3 \Delta \mdm1^2} \biggr[& \Delta  \log \left(\frac{\eta }{\mu }\right)-\eta ^2+\mu ^2+ f(\mu,\eta)\times\nn\\
&\times \biggr(\mu ^2 \left(\Delta +2 \eta ^2+3\right)-(\Delta -1) \left(\eta ^2-1\right)-2 \mu ^4\biggr) 
\biggr]
\end{align}

\section{Large mass splitting between external states}\label{app:loop_functions_large_mass_splitting}
Here we consider the scenarios in which $\mdm1 \gg \mdm{2}$. Using the hermicity of the EM moments, \cref{eq:hermicity}, the loop functions for the opposite hierarchy can be obtained using \cref{eq:hermicity_loop_functions}. We introduce $\alpha = \sqrt{\Delta} -\eta^2+\mu^2+1$, $\beta = \sqrt{\Delta} +\eta^2-\mu^2+1$ and find:

\paragraph{Scalar loop functions}

\begin{align}
\matheul{F}^S_\mu = \frac{1}{\mdm1} \biggr[&\eta^2 \left(\text{Li}_2\left(\frac{2}{2-\alpha}\right)+\text{Li}_2\left(\frac{2}{\beta }\right)-\text{Li}_2\left(1-\frac{\mu ^2}{\eta ^2}\right)\right)\nn\\
&+\mu ^2
\left(\text{Li}_2\left(\frac{2}{2-\beta}\right)+\text{Li}_2\left(\frac{2}{\alpha }\right)-\text{Li}_2\left(1-\frac{\eta ^2}{\mu ^2}\right)\right)-1\biggr]
\end{align}

\begin{align}
\matheul{G}^S_\mu = \frac{\mu}{\mdm1}  \left(\text{Li}_2\left(\frac{2}{\alpha }\right)+\text{Li}_2\left(\frac{2}{2-\beta}\right)-\text{Li}_2\left(1-\frac{\eta ^2}{\mu ^2}\right)\right)
\end{align}

\begin{align}
\matheul{F}^S_d = i \matheul{F}^S_\mu,\quad \matheul{G}^S_d = i \matheul{G}^S_\mu
\end{align}

\begin{align}
\matheul{F}^S_\ana = \frac{1}{\mdm1^2}\biggr[&-\Delta  f(\mu,\eta)-\left(3 \mu ^2 -3\eta^2+1\right) \log
\left(\frac{\mu }{\eta }\right)-3\nn\\
&+ \left(2 \Delta +6 \eta ^2+3 \mu ^2-2\right)\biggr(\text{Li}_2\left(\frac{2}{2-\alpha}\right) +\text{Li}_2\left(\frac{2}{\beta }\right)- \text{Li}_2\left(1-\frac{\mu ^2}{\eta ^2}\right)\biggr)\nn\\
&+\left(2 \Delta +2\eta^2+7 \mu^2 - 2\right) \biggr(\text{Li}_2\left(\frac{2}{2-\beta}\right) + \text{Li}_2\left(\frac{2}{\alpha }\right)- \text{Li}_2\left(1-\frac{\eta ^2}{\mu ^2}\right) \biggr)
\biggr]
\end{align}

\begin{align}
\matheul{G}^S_\ana = \frac{\mu}{\mdm1^2}  \biggr[&4 \log \left(\frac{\eta }{\mu }\right) +(\alpha -\beta -1) \biggr( \text{Li}_2\left(\frac{2}{2-\alpha }\right) + \text{Li}_2\left(\frac{2}{\beta }\right)- \text{Li}_2\left(1-\frac{\mu ^2}{\eta ^2}\right)\biggr)\nn\\
&+(\alpha -\beta +1) \biggr(\text{Li}_2\left(\frac{2}{2-\beta}\right) + \text{Li}_2\left(\frac{2}{\alpha }\right)  -  \text{Li}_2\left(1-\frac{\eta ^2}{\mu ^2}\right)\biggr)\biggr]
\end{align}

\begin{align}
\matheul{F}^S_b =  \matheul{F}^S_\ana    ,\quad \matheul{G}^S_b =  \matheul{G}^S_\ana 
\end{align}

\paragraph{Vector loop functions}

\begin{align}
\matheul{F}^V_\mu    = \frac{2}{\mdm1}\biggr[& \mu ^2 \biggr(\text{Li}_2\left(\frac{2}{2-\beta}\right)+\text{Li}_2\left(\frac{2}{\alpha }\right)-\text{Li}_2\left(1-\frac{\eta ^2}{\mu ^2}\right)\biggr)\nn\\
&   \left(\eta ^2-1\right) \biggr[  \text{Li}_2\left(\frac{2}{2-\alpha}\right) + \text{Li}_2\left(\frac{2}{\beta }\right)- \text{Li}_2\left(1-\frac{\mu ^2}{\eta ^2}\right)\biggr]
-1\biggr]
\end{align}
\begin{align}
\matheul{G}^V_\mu  = \frac{4 \mu }{\mdm1} \left(\text{Li}_2\left(\frac{2}{2-\alpha }\right)+\text{Li}_2\left(\frac{2}{\beta }\right)-\text{Li}_2\left(1-\frac{\mu ^2}{\eta
    ^2}\right)\right)
\end{align}

\begin{align}
\matheul{F}^V_d = -i\matheul{F}^V_\mu   ,\quad \matheul{G}^V_d =  -i\matheul{G}^V_\mu 
\end{align}

\begin{align}
\matheul{F}^V_\ana    = \frac{2}{\mdm1^2} \biggr[&\Delta f(\mu,\eta)+\left(-3 \eta ^2+3 \mu ^2+1\right) \log \left(\frac{\mu }{\eta }\right) + 3\nn\\
&  - \lambda_1 \left(\text{Li}_2\left(\frac{2}{\alpha }\right)+\text{Li}_2\left(-\frac{2}{\beta
    -2}\right)-\text{Li}_2\left(1-\frac{\eta ^2}{\mu ^2}\right)\right)\nn\\
&+ \lambda_2
\left(\text{Li}_2\left(-\frac{2}{\alpha -2}\right)+\text{Li}_2\left(\frac{2}{\beta }\right)-\text{Li}_2\left(1-\frac{\mu ^2}{\eta ^2}\right)\right)\biggr]
\end{align}
with $\lambda_1 = 2 \left(\eta ^4-\eta ^2 \left(2 \mu ^2+1\right)+\mu ^4\right)+3 \mu ^2$ and $\lambda_2 =-2 \left(\eta ^4+\eta ^2 \left(1-2 \mu ^2\right)+\mu ^4\right)+\mu ^2+1$.

\begin{align}
\matheul{G}^V_\ana    =   \frac{4 \mu  }{\mdm1^2}\biggr[&4 \log \left(\frac{\eta }{\mu }\right)+
(\alpha -\beta -1) \biggr(\text{Li}_2\left(\frac{2}{2-\alpha}\right) + \text{Li}_2\left(\frac{2}{\beta }\right) - \text{Li}_2\left(1-\frac{\mu ^2}{\eta ^2}\right)\biggr)
\nn\\
&+(\alpha -\beta +1) \biggr(\text{Li}_2\left(\frac{2}{2-\beta}\right) + \text{Li}_2\left(\frac{2}{\alpha
}\right)-  \text{Li}_2\left(1-\frac{\eta ^2}{\mu ^2}\right)\biggr) \biggr]
\end{align}

\begin{align}
\matheul{F}^V_b = - \matheul{F}^V_\ana   ,\quad \matheul{G}^V_b =  -\matheul{G}^S_\ana
\end{align}

\section{Light external mass limit}\label{app:loop_functions_light_external_states}
In this appendix we expand the loop functions presented in \cref{appendix:Loopfunctions_q_zero} for the limit in which the masses of the external fermions are small compared to the relevant mass scale in the loop. For this, we keep the leading order terms in $\mdm1$ and $\mdm{2}$, and introduce the abbreviations $\rho = m_f/m_S$ and $r=m_f/m_V$. We find the following approximations:

\paragraph{Scalar loop functions}

\begin{align}
\matheul{F}_\mu^S= \frac{(\mdm1+\mdm{2}) \left(\rho ^4-2 \rho ^2 \log (\rho^2 )-1\right)}{4 m_S^2 \left(\rho ^2-1\right)^3},\quad 
\matheul{G}_\mu^S =  \frac{\rho  \left(\rho ^2- \log (\rho^2 )-1\right)}{m_S \left(\rho ^2-1\right)^2}
\end{align}

\begin{align}
\matheul{F}_d^S = i\frac{(\mdm1-\mdm{2}) \left(\rho ^4- 2 \rho ^2 \log(\rho^2)-1\right)}{4 m_S^2 \left(\rho ^2-1\right)^3},\quad
\matheul{G}_d^S=  i\frac{\rho  \left(\rho ^2- \log(\rho^2)-1\right)}{m_S\left(\rho ^2-1\right)^2}
\end{align}

\begin{align}
\matheul{F}_\ana^S &= \frac{-3 \rho ^2+ \left(\rho ^2+2\right) \log (\rho^2)+3}{6 m_S^2 \left(\rho ^2-1\right)^2} \nn \\
&-\frac{\mdm1 \mdm{2} }{36 m_S^4 \left(\rho ^2-1\right)^5} \left[-19 \rho ^6-9 \rho ^4+27 \rho ^2 +6 \left(\rho ^4+5 \rho ^2+2\right) \rho ^2 \log(\rho^2)+1\right]
\end{align}

\begin{equation}
\matheul{G}_\ana^S = \frac{\rho  (\mdm1-\mdm{2}) \left(-3 \rho ^4+ \left(\rho ^4+4 \rho ^2+1\right) \log (\rho^2 )+3\right)}{6 m_S^3 \left(\rho ^2-1\right)^4}
\end{equation}

\begin{align}
\matheul{F}_b^S &= \frac{-3 \rho ^2+ \left(\rho ^2+2\right) \log (\rho^2)+3}{6 m_S^2 \left(\rho ^2-1\right)^2} \nn \\
&+\frac{\mdm1 \mdm{2}}{36 m_S^4 \left(\rho ^2-1\right)^5} \left[-19 \rho ^6-9 \rho ^4+27 \rho ^2+ 6 \left(\rho ^4+5 \rho ^2+2\right) \rho ^2 \log(\rho^2)+1\right]
\end{align}

\begin{equation}
\matheul{G}_b^S = \frac{\rho  (\mdm1 +\mdm{2}) \left(-3 \rho ^4+ \left(\rho ^4+4 \rho ^2+1\right) \log(\rho^2)+3\right)}{6 m_S^3 \left(\rho ^2-1\right)^4}
\end{equation}

\paragraph{Vector loop functions}
\begin{align}
\matheul{F}_\mu^V  &= -\frac{(\mdm1 + \mdm{2}) \left(-5 r ^4+8 r ^2+\left(4 r ^4-2 r^2\right) \log (r^2)-3\right)}{2 m_V^2 \left(r ^2-1\right)^3}\\
\matheul{G}_\mu^V &= r\frac{4 \left(-r^2 +r^2 \log(r^2)+1\right)}{m_V \left(r^2-1\right)^2}
\end{align}

\begin{align}
\matheul{F}_d^V &= i\frac{ (\mdm1 -\mdm{2}) \left(-5 r ^4+8 r ^2+\left(4 r^4-2 r^2\right) \log (r^2)-3\right)}{2 m_V^2 \left(r ^2-1\right)^3}\\
\matheul{G}_d^V &= -r\frac{4 i \left(-r^2+ r^2 \log (r^2)+1\right)}{m_V \left(r ^2-1\right)^2}
\end{align}

\begin{align}
\matheul{F}_\ana^V &=  -\frac{3 \left(r ^2-1\right)+\left(2-5 r^2\right) \log(r^2)}{3 m_V^2 \left(r^2-1\right)^2} \nn\\
&+\frac{\mdm1 \mdm{2}}{18 m_V^4 \left(r^2-1\right)^5} \bigr[-19 r^6-9 r^4+27 r^2+6 \left(r^4+5 r^2+2\right) r^2 \log(r^2)+1\bigr]
\end{align}
\begin{equation}
\matheul{G}_\ana^V = \frac{2 r  (\mdm1 - \mdm{2}) \left(-3 r ^4+ \left(r^4+4 r^2+1\right) \log(r^2)+3\right)}{3 m_V^3 \left(r ^2-1\right)^4}
\end{equation}

\begin{align}
\matheul{F}_b^V &= \frac{3 \left(r^2-1\right)+\left(2-5 r ^2\right) \log(r^2)}{3 m_V^2 \left(r ^2-1\right)^2} \nn\\
& +\frac{\mdm1 \mdm{2}}{18 m_V^4 \left(r ^2-1\right)^5} \bigr[-19 r ^6-9 r ^4+27 r^2+6\left(r^4+5 r^2+2\right) r^2 \log (r^2)+1\bigr]
\end{align}
\begin{equation}
 \matheul{G}_b^V = -\frac{2 r  (\mdm1 + \mdm{2}) \left(-3 r ^4+\left(r^4+4 r^2+1\right) \log(r^2)+3\right)}{3 m_V^3 \left(r^2-1\right)^4}   
\end{equation}
\section{Effective theory of mass mixing of $SU(2)_L$ multiplets}
\label{sec:EFT_multiplets}

In full generality, mixing between two electroweak multiplets $\chi_n$ and $\chi_m$, with hypercharges $Y_n$ and $Y_m$ and isospins $j_n=(n-1)/2$ and $j_m=(m-1)/2$, respectively, can be generated by higher-dimensional operators involving Higgs doublets. Depending on the fermion bilinear, we distinguish Dirac-type and Majorana-type mixing. In both cases, the fermion bilinear and the product of Higgs fields are projected onto the same $SU(2)_L$ representation of isospin $J$, such that their product can be contracted to an $SU(2)_L$ singlet. We denote by $a$ the possible Higgs recoupling channels leading to the representation $J$.

For a definite ordered Higgs recoupling channel $a=(J_1,\ldots,J_{p-2})$, we define
\begin{align}
{\cal C}_a(m_1,\ldots,m_p) =& (-1)^{M_1}\sqrt{2J_1+1}
\begin{pmatrix} 
\frac{1}{2} & \frac{1}{2} & J_1\\m_1 & m_2 & -M_1 
\end{pmatrix}
\nonumber\\
&\times
\prod_{\ell=2}^{p-1}\left[(-1)^{J_{\ell-1}-\frac{1}{2}+M_\ell}\sqrt{2J_\ell+1}\begin{pmatrix} J_{\ell-1} & \frac{1}{2} & J_\ell\\ M_{\ell-1} & m_{\ell+1} & -M_\ell
\end{pmatrix} \right],
\label{eq:CaHiggs}
\end{align}
where
\begin{equation}
J_{p-1}\equiv J\,,
\qquad
M_\ell=\sum_{i=1}^{\ell+1}m_i\,.
\label{eq:Mell}
\end{equation}
Only Higgs recoupling channels compatible with the Bose symmetry of identical Higgs fields correspond to independent non-vanishing operators; channels that vanish upon symmetrization are to be omitted.

\subsection{Dirac-type mixing}
\label{sec:EFT_multiplets_D}

Dirac-type mixing between the two multiplets can be described by operators of the form
\begin{equation}
{\cal L}_{\rm mix}^{D,{\rm EFT}}=-\frac{\lambda_D}{\Lambda^{p-1}}\Big(\overline{\chi_n}P_L\chi_m\Big)_J\cdot\Big(H^k\widetilde H^{\,p-k}\Big)_J+{\rm h.c.}\,.
\label{eq:LmixEFTD}
\end{equation}
Since $\overline{\chi_n}$ carries hypercharge $-Y_n$, hypercharge conservation requires
\begin{equation}
-Y_n+Y_m+\frac{2k-p}{2}=0\,.
\label{eq:YconditionEFTD}
\end{equation}
The integer $p$ is the smallest number of Higgs doublets for which an allowed representation $J$ occurs in both products and Eq.~(\ref{eq:YconditionEFTD}) can be satisfied. The renormalizable interaction corresponds to $p=1$ and $J=1/2$, with $k=1$ for $\Phi=H$ and $k=0$ for $\Phi=\widetilde H$.

To express the conjugate $SU(2)_L$ representation carried by $\overline{\chi_n}$ in the same basis as the original representation, we use
\begin{equation}
|j,t\rangle^c=(-1)^{j-t}|j,-t\rangle\,.
\label{eq:SU2conjugate}
\end{equation}
After electroweak symmetry breaking, Eq.~(\ref{eq:LmixEFTD}) generates mixing between components with the same electric charge,
\begin{equation}
{\cal L}_{\rm mix}^{D,{\rm EFT}}\supset-\sum_Q\delta_Q^D\,\overline{\chi_n^Q}P_L\chi_m^Q+{\rm h.c.}\,,
\qquad 
\delta_Q^D=C_Q^{D,(J,a)}\,\lambda_D\,\frac{v^p}{2^{p/2}\Lambda^{p-1}}\,.
\label{eq:deltaQEFTD}
\end{equation}
In the convention of Eq.~(\ref{eq:SU2conjugate}), the group-theory coefficient can be written as
\begin{equation}
C_Q^{D,(J,a)}=(-1)^{j_n-(Q-Y_n)}(-1)^{\frac{n-m}{2}+J}
\begin{pmatrix}
\frac{n-1}{2} & \frac{m-1}{2} & J\\ -Q+Y_n & Q-Y_m & -Y_n+Y_m
\end{pmatrix}
C_{H,0}^{D,(J,a)}\,.
\label{eq:CQEFTD}
\end{equation}
The neutral mixing is obtained by setting $Q=0$.

The coefficient associated with the neutral Higgs configuration is
\begin{equation}
C_{H,0}^{D,(J,a)} = {\cal C}_a \left(\underbrace{-\frac{1}{2},\ldots,-\frac{1}{2}}_{k},\underbrace{+\frac{1}{2},\ldots,+\frac{1}{2}}_{p-k}\right),
\label{eq:CH0D}
\end{equation}
where the recoupling function ${\cal C}_a$ is defined in Eq.~(\ref{eq:CaHiggs}). The total third component carried by the Higgs product is
\begin{equation}
M_{p-1}=-Y_n+Y_m\,,
\label{eq:MfinalD}
\end{equation}
as follows from Eq.~(\ref{eq:YconditionEFTD}).

The same operator also generates interactions with the charged Goldstone bosons. The terms containing one $G^+$ and one neutral fermion can be written as
\begin{equation}
{\cal L}_{G^+}^{D,{\rm EFT}} \supset -\frac{\lambda_D}{\Lambda^{p-1}} \left(\frac{v}{\sqrt2}\right)^{p-1} \left[C_{G,1}^{D,+,(J,a)}\overline{\chi_n^0}P_L\chi_m^-+C_{G,2}^{D,+,(J,a)} \overline{\chi_n^+}P_L\chi_m^0\right]G^+ +{\rm h.c.}\,.
\label{eq:GoldstoneEFTD}
\end{equation}
The corresponding coefficients are
\begin{equation}
C_{G,1}^{D,+,(J,a)} = (-1)^{j_n+Y_n} (-1)^{\frac{n-m}{2}+J}
\begin{pmatrix}
\frac{n-1}{2} & \frac{m-1}{2} & J\\ Y_n & -1-Y_m & 1-Y_n+Y_m
\end{pmatrix}
C_{H,G^+}^{D,(J,a)}\,,
\label{eq:CG1EFTD}
\end{equation}
and
\begin{equation}
C_{G,2}^{D,+,(J,a)} = (-1)^{j_n-1+Y_n} (-1)^{\frac{n-m}{2}+J}
\begin{pmatrix}
\frac{n-1}{2} & \frac{m-1}{2} & J\\
-1+Y_n & -Y_m & 1-Y_n+Y_m
\end{pmatrix}
C_{H,G^+}^{D,(J,a)}\,.
\label{eq:CG2EFTD}
\end{equation}
The Higgs-sector coefficient $C_{H,G^+}^{D,(J,a)}$ is obtained by replacing one of the neutral Higgs components by the corresponding charged component and projecting the resulting Higgs configuration onto isospin $J$. The analogous $G^-$ interactions follow by the corresponding replacement in the Higgs product.

\subsection{Majorana-type mixing}
\label{sec:EFT_multiplets_M}

Majorana-type mixing is described by operators of the form
\begin{equation}
{\cal L}_{\rm mix}^{M,{\rm EFT}} = -\frac{\lambda_M}{\Lambda^{p-1}} \Big(\overline{\chi_n^{\,c}}P_L\chi_m\Big)_J \cdot \Big(H^k\widetilde H^{\,p-k}\Big)_J +{\rm h.c.}\,.
\label{eq:LmixEFTM}
\end{equation}
Hypercharge conservation now requires
\begin{equation}
Y_n+Y_m+\frac{2k-p}{2}=0\,.
\label{eq:YconditionEFTM}
\end{equation}
Again, $p$ denotes the smallest number of Higgs doublets for which an allowed representation $J$ occurs in both products and the hypercharge condition can be satisfied. For $p=1$, the interaction is renormalizable and $J=1/2$.

After electroweak symmetry breaking, Eq.~(\ref{eq:LmixEFTM}) generates mass terms between components with opposite electric charges,
\begin{equation}
{\cal L}_{\rm mix}^{M,{\rm EFT}} \supset -\sum_Q \delta_Q^M\,\chi_n^Q\chi_m^{-Q} +{\rm h.c.}\,,
\qquad 
\delta_Q^M=C_Q^{M,(J,a)}\,\lambda_M\,\frac{v^p}{2^{p/2}\Lambda^{p-1}}\,.
\label{eq:deltaQEFTM}
\end{equation}
The corresponding group-theory coefficient is
\begin{equation}
C_Q^{M,(J,a)}=(-1)^{\frac{n-m}{2}+J}
\begin{pmatrix}
\frac{n-1}{2} & \frac{m-1}{2} & J\\
Q-Y_n & -Q-Y_m & Y_n+Y_m
\end{pmatrix}
C_{H,0}^{M,(J,a)}\,.
\label{eq:CQEFTM}
\end{equation}
The neutral mixing is again obtained by setting $Q=0$.

The coefficient associated with the neutral Higgs configuration is
\begin{equation}
C_{H,0}^{M,(J,a)}= {\cal C}_a
\left( \underbrace{-\frac{1}{2},\ldots,-\frac{1}{2}}_{k}, \underbrace{+\frac{1}{2},\ldots,+\frac{1}{2}}_{p-k} \right),
\label{eq:CH0M}
\end{equation}
with
\begin{equation}
M_{p-1}=Y_n+Y_m\,.
\label{eq:MfinalM}
\end{equation}

The same EFT operator generates interactions with the charged Goldstone bosons. The terms involving one $G^+$ and one neutral fermion can be written as
\begin{equation}
{\cal L}_{G^+}^{M,{\rm EFT}} \supset -\frac{\lambda_M}{\Lambda^{p-1}} \left(\frac{v}{\sqrt2}\right)^{p-1}\left[ C_{G,1}^{M,+,(J,a)} \chi_n^0\chi_m^-+ C_{G,2}^{M,+,(J,a)} \chi_n^-\chi_m^0 \right]G^+ +{\rm h.c.}\,,
\label{eq:GoldstoneEFTM}
\end{equation}
where
\begin{equation}
C_{G,1}^{M,+,(J,a)} = (-1)^{\frac{n-m}{2}+J}
\begin{pmatrix}
\frac{n-1}{2} & \frac{m-1}{2} & J\\ -Y_n & -1-Y_m & 1+Y_n+Y_m
\end{pmatrix}
C_{H,G^+}^{M,(J,a)}\,,
\label{eq:CG1EFTM}
\end{equation}
and
\begin{equation}
C_{G,2}^{M,+,(J,a)} = (-1)^{\frac{n-m}{2}+J}
\begin{pmatrix}
\frac{n-1}{2} & \frac{m-1}{2} & J\\ -1-Y_n & -Y_m & 1+Y_n+Y_m
\end{pmatrix}
C_{H,G^+}^{M,(J,a)}\,.
\label{eq:CG2EFTM}
\end{equation}
The Higgs-sector coefficient is
\begin{equation}
C_{H,G^+}^{M,(J,a)} = \sum_{s=1}^{k} {\cal C}_a \left( m_1^{(s)},\ldots,m_p^{(s)} \right),
\label{eq:CHGpM}
\end{equation}
where
\begin{equation}
m_i^{(s)}
=
\begin{cases}
+\dfrac{1}{2}\,, & i=s\,,\\[2mm]
-\dfrac{1}{2}\,, & 1\leq i\leq k,\quad i\neq s\,,\\[2mm]
+\dfrac{1}{2}\,, & k<i\leq p\,.
\end{cases}
\label{eq:miGoldstoneM}
\end{equation}
The analogous $G^-$ interactions are obtained by replacing one $\widetilde H$ by its charged component and are described by the corresponding coefficients $C_{G,1}^{M,-,(J,a)}$ and $C_{G,2}^{M,-,(J,a)}$.

The distinction between the two classes of operators becomes particularly transparent after electroweak symmetry breaking. Dirac-type operators generate mass terms of the form
\begin{equation}
\overline{\chi_n^Q}\chi_m^Q\,,
\end{equation}
and therefore mix states carrying the same electric charge. If a conserved fermion number is preserved by the remaining interactions, the neutral mass eigenstates are Dirac fermions. Majorana-type operators instead generate terms of the form
\begin{equation}
\chi_n^Q\chi_m^{-Q}\,,
\end{equation}
including $\chi_n^0\chi_m^0$ in the neutral sector, and can therefore lead to Majorana neutral mass eigenstates.

\bibliographystyle{JHEP}
\bibliography{references}

\providecommand{\href}[2]{#2}\begingroup\raggedright\begin{thebibliography}{10}

\bibitem{Bernstein:1963qh}
J.~Bernstein, M.~Ruderman and G.~Feinberg, \emph{{Electromagnetic Properties of
  the neutrino}}, \href{https://doi.org/10.1103/PhysRev.132.1227}{\emph{Phys.
  Rev.} {\bfseries 132} (1963) 1227}.

\bibitem{Kopp:2014tsa}
J.~Kopp, L.~Michaels and J.~Smirnov, \emph{{Loopy Constraints on Leptophilic
  Dark Matter and Internal Bremsstrahlung}},
  \href{https://doi.org/10.1088/1475-7516/2014/04/022}{\emph{JCAP} {\bfseries
  04} (2014) 022} [\href{https://arxiv.org/abs/1401.6457}{{\ttfamily
  1401.6457}}].

\bibitem{Garny:2015wea}
M.~Garny, A.~Ibarra and S.~Vogl, \emph{{Signatures of Majorana dark matter with
  t-channel mediators}},
  \href{https://doi.org/10.1142/S0218271815300190}{\emph{Int. J. Mod. Phys. D}
  {\bfseries 24} (2015) 1530019}
  [\href{https://arxiv.org/abs/1503.01500}{{\ttfamily 1503.01500}}].

\bibitem{Ibarra:2015fqa}
A.~Ibarra and S.~Wild, \emph{{Dirac dark matter with a charged mediator: a
  comprehensive one-loop analysis of the direct detection phenomenology}},
  \href{https://doi.org/10.1088/1475-7516/2015/05/047}{\emph{JCAP} {\bfseries
  05} (2015) 047} [\href{https://arxiv.org/abs/1503.03382}{{\ttfamily
  1503.03382}}].

\bibitem{Sandick:2016zut}
P.~Sandick, K.~Sinha and F.~Teng, \emph{{Simplified Dark Matter Models with
  Charged Mediators: Prospects for Direct Detection}},
  \href{https://doi.org/10.1007/JHEP10(2016)018}{\emph{JHEP} {\bfseries 10}
  (2016) 018} [\href{https://arxiv.org/abs/1608.00642}{{\ttfamily
  1608.00642}}].

\bibitem{Hisano:2018bpz}
J.~Hisano, R.~Nagai and N.~Nagata, \emph{{Singlet Dirac Fermion Dark Matter
  with Mediators at Loop}},
  \href{https://doi.org/10.1007/JHEP12(2018)059}{\emph{JHEP} {\bfseries 12}
  (2018) 059} [\href{https://arxiv.org/abs/1808.06301}{{\ttfamily
  1808.06301}}].

\bibitem{Baker:2018uox}
M.~J. Baker and A.~Thamm, \emph{{Leptonic WIMP Coannihilation and the Current
  Dark Matter Search Strategy}},
  \href{https://doi.org/10.1007/JHEP10(2018)187}{\emph{JHEP} {\bfseries 10}
  (2018) 187} [\href{https://arxiv.org/abs/1806.07896}{{\ttfamily
  1806.07896}}].

\bibitem{Herrero-Garcia:2018koq}
J.~Herrero-Garcia, E.~Molinaro and M.~A. Schmidt, \emph{{Dark matter direct
  detection of a fermionic singlet at one loop}},
  \href{https://doi.org/10.1140/epjc/s10052-018-5935-5}{\emph{Eur. Phys. J. C}
  {\bfseries 78} (2018) 471}
  [\href{https://arxiv.org/abs/1803.05660}{{\ttfamily 1803.05660}}].

\bibitem{Ibarra:2022nzm}
A.~Ibarra, M.~Reichard and R.~Nagai, \emph{{Anapole moment of Majorana fermions
  and implications for direct detection of neutralino dark matter}},
  \href{https://doi.org/10.1007/JHEP01(2023)086}{\emph{JHEP} {\bfseries 01}
  (2023) 086} [\href{https://arxiv.org/abs/2207.01014}{{\ttfamily
  2207.01014}}].

\bibitem{Ibarra:2024mpq}
A.~Ibarra, M.~Reichard and G.~Tomar, \emph{{Probing dark matter electromagnetic
  properties in direct detection experiments}},
  \href{https://doi.org/10.1088/1475-7516/2025/02/072}{\emph{JCAP} {\bfseries
  02} (2025) 072} [\href{https://arxiv.org/abs/2408.15760}{{\ttfamily
  2408.15760}}].

\bibitem{Denner:1994nn}
A.~Denner, G.~Weiglein and S.~Dittmaier, \emph{{Gauge invariance of Green
  functions: Background field method versus pinch technique}},
  \href{https://doi.org/10.1016/0370-2693(94)90162-7}{\emph{Phys. Lett. B}
  {\bfseries 333} (1994) 420}
  [\href{https://arxiv.org/abs/hep-ph/9406204}{{\ttfamily hep-ph/9406204}}].

\bibitem{Papavassiliou:1994yi}
J.~Papavassiliou, \emph{{On the connection between the pinch technique and the
  background field method}},
  \href{https://doi.org/10.1103/PhysRevD.51.856}{\emph{Phys. Rev. D} {\bfseries
  51} (1995) 856} [\href{https://arxiv.org/abs/hep-ph/9410385}{{\ttfamily
  hep-ph/9410385}}].

\bibitem{Hashimoto:1994ct}
S.~Hashimoto, J.~Kodaira, Y.~Yasui and K.~Sasaki, \emph{{The Background field
  method: Alternative way of deriving the pinch technique's results}},
  \href{https://doi.org/10.1103/PhysRevD.50.7066}{\emph{Phys. Rev. D}
  {\bfseries 50} (1994) 7066}
  [\href{https://arxiv.org/abs/hep-ph/9406271}{{\ttfamily hep-ph/9406271}}].

\bibitem{BINOSI20091}
D.~Binosi and J.~Papavassiliou, \emph{Pinch technique: Theory and
  applications},
  \href{https://doi.org/https://doi.org/10.1016/j.physrep.2009.05.001}{\emph{Physics
  Reports} {\bfseries 479} (2009) 1}.

\bibitem{Salam:1957st}
A.~Salam, \emph{{On parity conservation and neutrino mass}},
  \href{https://doi.org/10.1007/BF02812841}{\emph{Nuovo Cim.} {\bfseries 5}
  (1957) 299}.

\bibitem{Fujikawa:1980yx}
K.~Fujikawa and R.~Shrock, \emph{{The Magnetic Moment of a Massive Neutrino and
  Neutrino Spin Rotation}},
  \href{https://doi.org/10.1103/PhysRevLett.45.963}{\emph{Phys. Rev. Lett.}
  {\bfseries 45} (1980) 963}.

\bibitem{Kayser:1982br}
B.~Kayser, \emph{{Majorana Neutrinos and their Electromagnetic Properties}},
  \href{https://doi.org/10.1103/PhysRevD.26.1662}{\emph{Phys. Rev. D}
  {\bfseries 26} (1982) 1662}.

\bibitem{Shrock:1982sc}
R.~E. Shrock, \emph{{Electromagnetic Properties and Decays of Dirac and
  Majorana Neutrinos in a General Class of Gauge Theories}},
  \href{https://doi.org/10.1016/0550-3213(82)90273-5}{\emph{Nucl. Phys. B}
  {\bfseries 206} (1982) 359}.

\bibitem{Bernabeu:2000hf}
J.~Bernabeu, L.~G. Cabral-Rosetti, J.~Papavassiliou and J.~Vidal, \emph{{On the
  charge radius of the neutrino}},
  \href{https://doi.org/10.1103/PhysRevD.62.113012}{\emph{Phys. Rev. D}
  {\bfseries 62} (2000) 113012}
  [\href{https://arxiv.org/abs/hep-ph/0008114}{{\ttfamily hep-ph/0008114}}].

\bibitem{Papavassiliou:2002te}
J.~Papavassiliou, J.~Bernabeu and J.~Vidal, \emph{{On the definition and
  observability of the neutrino charge radius}},
  \href{https://doi.org/10.1016/S0920-5632(02)01905-9}{\emph{Nucl. Phys. B
  Proc. Suppl.} {\bfseries 114} (2003) 197}
  [\href{https://arxiv.org/abs/hep-ph/0210312}{{\ttfamily hep-ph/0210312}}].

\bibitem{Papavassiliou:2003rx}
J.~Papavassiliou, J.~Bernabeu, D.~Binosi and J.~Vidal, \emph{{The Effective
  neutrino charge radius}},
  \href{https://doi.org/10.1140/epjcd/s2003-03-920-7}{\emph{Eur. Phys. J. C}
  {\bfseries 33} (2004) S865}
  [\href{https://arxiv.org/abs/hep-ph/0310028}{{\ttfamily hep-ph/0310028}}].

\bibitem{Nieves:1981zt}
J.~F. Nieves, \emph{{Electromagnetic Properties of Majorana Neutrinos}},
  \href{https://doi.org/10.1103/PhysRevD.26.3152}{\emph{Phys. Rev. D}
  {\bfseries 26} (1982) 3152}.

\bibitem{Kayser:1983wm}
B.~Kayser and A.~S. Goldhaber, \emph{{{CPT} and {CP} Properties of Majorana
  Particles, and the Consequences}},
  \href{https://doi.org/10.1103/PhysRevD.28.2341}{\emph{Phys. Rev. D}
  {\bfseries 28} (1983) 2341}.

\bibitem{Kayser:1984ge}
B.~Kayser, \emph{{CPT, CP, and c Phases and their Effects in Majorana Particle
  Processes}}, \href{https://doi.org/10.1103/PhysRevD.30.1023}{\emph{Phys. Rev.
  D} {\bfseries 30} (1984) 1023}.

\bibitem{Nowakowski:2004cv}
M.~Nowakowski, E.~A. Paschos and J.~M. Rodriguez, \emph{{All electromagnetic
  form-factors}}, \href{https://doi.org/10.1088/0143-0807/26/4/001}{\emph{Eur.
  J. Phys.} {\bfseries 26} (2005) 545}
  [\href{https://arxiv.org/abs/physics/0402058}{{\ttfamily physics/0402058}}].

\bibitem{Mohapatra:2004}
R.~N. Mohapatra and P.~B. Pal, \emph{Massive Neutrinos in Physics and
  Astrophysics}. WORLD SCIENTIFIC, 3rd~ed., 2004,
  \href{https://doi.org/10.1142/5024}{10.1142/5024},
  [\href{https://arxiv.org/abs/https://www.worldscientific.com/doi/pdf/10.1142/5024}{{\ttfamily
  https://www.worldscientific.com/doi/pdf/10.1142/5024}}].

\bibitem{Giunti:2014ixa}
C.~Giunti and A.~Studenikin, \emph{{Neutrino electromagnetic interactions: a
  window to new physics}},
  \href{https://doi.org/10.1103/RevModPhys.87.531}{\emph{Rev. Mod. Phys.}
  {\bfseries 87} (2015) 531} [\href{https://arxiv.org/abs/1403.6344}{{\ttfamily
  1403.6344}}].

\bibitem{Denner:1994xt}
A.~Denner, G.~Weiglein and S.~Dittmaier, \emph{{Application of the background
  field method to the electroweak standard model}},
  \href{https://doi.org/10.1016/0550-3213(95)00037-S}{\emph{Nucl. Phys. B}
  {\bfseries 440} (1995) 95}
  [\href{https://arxiv.org/abs/hep-ph/9410338}{{\ttfamily hep-ph/9410338}}].

\bibitem{Patel:2015tea}
H.~H. Patel, \emph{{Package-X: A Mathematica package for the analytic
  calculation of one-loop integrals}},
  \href{https://doi.org/10.1016/j.cpc.2015.08.017}{\emph{Comput. Phys. Commun.}
  {\bfseries 197} (2015) 276}
  [\href{https://arxiv.org/abs/1503.01469}{{\ttfamily 1503.01469}}].

\bibitem{Patel:2016fam}
H.~H. Patel, \emph{{Package-X 2.0: A Mathematica package for the analytic
  calculation of one-loop integrals}},
  \href{https://doi.org/10.1016/j.cpc.2017.04.015}{\emph{Comput. Phys. Commun.}
  {\bfseries 218} (2017) 66}
  [\href{https://arxiv.org/abs/1612.00009}{{\ttfamily 1612.00009}}].

\bibitem{MERTIG1991345}
R.~Mertig, M.~Böhm and A.~Denner, \emph{Feyn calc - computer-algebraic
  calculation of feynman amplitudes},
  \href{https://doi.org/https://doi.org/10.1016/0010-4655(91)90130-D}{\emph{Computer
  Physics Communications} {\bfseries 64} (1991) 345}.

\bibitem{Shtabovenko:2016sxi}
V.~Shtabovenko, R.~Mertig and F.~Orellana, \emph{{New Developments in FeynCalc
  9.0}}, \href{https://doi.org/10.1016/j.cpc.2016.06.008}{\emph{Comput. Phys.
  Commun.} {\bfseries 207} (2016) 432}
  [\href{https://arxiv.org/abs/1601.01167}{{\ttfamily 1601.01167}}].

\bibitem{Shtabovenko:2020gxv}
V.~Shtabovenko, R.~Mertig and F.~Orellana, \emph{{FeynCalc 9.3: New features
  and improvements}},
  \href{https://doi.org/10.1016/j.cpc.2020.107478}{\emph{Comput. Phys. Commun.}
  {\bfseries 256} (2020) 107478}
  [\href{https://arxiv.org/abs/2001.04407}{{\ttfamily 2001.04407}}].

\bibitem{ELLIS2017103}
J.~P. Ellis, \emph{Tikz-feynman: Feynman diagrams with tikz},
  \href{https://doi.org/https://doi.org/10.1016/j.cpc.2016.08.019}{\emph{Computer
  Physics Communications} {\bfseries 210} (2017) 103}.

\bibitem{Appelquist:1974tg}
T.~Appelquist and J.~Carazzone, \emph{{Infrared Singularities and Massive
  Fields}}, \href{https://doi.org/10.1103/PhysRevD.11.2856}{\emph{Phys. Rev. D}
  {\bfseries 11} (1975) 2856}.

\bibitem{Denner:1992me}
A.~Denner, H.~Eck, O.~Hahn and J.~Kublbeck, \emph{{Compact Feynman rules for
  Majorana fermions}},
  \href{https://doi.org/10.1016/0370-2693(92)91045-B}{\emph{Phys. Lett. B}
  {\bfseries 291} (1992) 278}.

\bibitem{Denner:1992vza}
A.~Denner, H.~Eck, O.~Hahn and J.~Kublbeck, \emph{{Feynman rules for fermion
  number violating interactions}},
  \href{https://doi.org/10.1016/0550-3213(92)90169-C}{\emph{Nucl. Phys. B}
  {\bfseries 387} (1992) 467}.

\bibitem{Schechter:1981hw}
J.~Schechter and J.~W.~F. Valle, \emph{{Majorana Neutrinos and Magnetic
  Fields}}, \href{https://doi.org/10.1103/PhysRevD.25.283}{\emph{Phys. Rev. D}
  {\bfseries 24} (1981) 1883}.

\bibitem{Pal:1981rm}
P.~B. Pal and L.~Wolfenstein, \emph{{Radiative Decays of Massive Neutrinos}},
  \href{https://doi.org/10.1103/PhysRevD.25.766}{\emph{Phys. Rev. D} {\bfseries
  25} (1982) 766}.

\bibitem{Enqvist:1989vt}
K.~Enqvist, J.~Maalampi and A.~Masiero, \emph{{MAGNETIC MOMENTS IN THE SEESAW
  MODEL}}, \href{https://doi.org/10.1016/0370-2693(89)90343-2}{\emph{Phys.
  Lett. B} {\bfseries 222} (1989) 453}.

\bibitem{Sher:2002ij}
M.~Sher and S.~Nie, \emph{{Large electric dipole moments of heavy neutrinos}},
  \href{https://doi.org/10.1103/PhysRevD.65.093018}{\emph{Phys. Rev. D}
  {\bfseries 65} (2002) 093018}
  [\href{https://arxiv.org/abs/hep-ph/0201220}{{\ttfamily hep-ph/0201220}}].

\bibitem{Beg:1977xz}
M.~A.~B. Beg, W.~J. Marciano and M.~Ruderman, \emph{{Properties of Neutrinos in
  a Class of Gauge Theories}},
  \href{https://doi.org/10.1103/PhysRevD.17.1395}{\emph{Phys. Rev. D}
  {\bfseries 17} (1978) 1395}.

\bibitem{Maalampi:1985fa}
J.~Maalampi, K.~Mursula and M.~Roos, \emph{{On the Possible Evidence for a
  Radiatively Decaying Neutrino}},
  \href{https://doi.org/10.1103/PhysRevLett.56.1031}{\emph{Phys. Rev. Lett.}
  {\bfseries 56} (1986) 1031}.

\bibitem{Chattopadhyay:1986cj}
U.~Chattopadhyay and P.~B. Pal, \emph{{Radiative Neutrino Decay in Left-right
  Models}}, \href{https://doi.org/10.1103/PhysRevD.34.3444}{\emph{Phys. Rev. D}
  {\bfseries 34} (1986) 3444}.

\bibitem{Duncan:1987ki}
M.~J. Duncan, J.~A. Grifols, A.~Mendez and S.~U. Sankar, \emph{{Restrictions on
  the Neutrino Magnetic Dipole Moment}},
  \href{https://doi.org/10.1016/0370-2693(87)90260-7}{\emph{Phys. Lett. B}
  {\bfseries 191} (1987) 304}.

\bibitem{Mohapatra:1987nx}
R.~N. Mohapatra, \emph{{Left-right Symmetry and Finite One Loop Dirac Neutrino
  Mass}}, \href{https://doi.org/10.1016/0370-2693(88)90610-7}{\emph{Phys. Lett.
  B} {\bfseries 201} (1988) 517}.

\bibitem{Zhang:2024ijy}
Z.-Y. Zhang, J.-L. Yang, H.-B. Zhang and T.-F. Feng, \emph{{Transition magnetic
  moment of Majorana neutrinos in the triplets next-to-minimal MSSM}},
  \href{https://doi.org/10.1103/PhysRevD.110.015035}{\emph{Phys. Rev. D}
  {\bfseries 110} (2024) 015035}
  [\href{https://arxiv.org/abs/2406.18323}{{\ttfamily 2406.18323}}].

\bibitem{Ruan:2024fgf}
L.~Ruan, S.-M. Zhao, M.-Y. Liu, X.-Y. Han, X.~Wang and X.-X. Dong,
  \emph{{Neutrino transition magnetic moment in U(1)$_{X}$SSM*}},
  \href{https://doi.org/10.1088/1674-1137/adc978}{\emph{Chin. Phys.} {\bfseries
  49} (2025) 083104} [\href{https://arxiv.org/abs/2407.05741}{{\ttfamily
  2407.05741}}].

\bibitem{Bolanos-Carrera:2023ppu}
A.~Bola\~nos Carrera, M.~Guiot-Lomelì and G.~Tavares-Velasco, \emph{{Neutrino
  charge radius and electromagnetic dipole moments via scalar and vector
  leptoquarks}},
  \href{https://doi.org/10.1140/epjc/s10052-024-12495-1}{\emph{Eur. Phys. J. C}
  {\bfseries 84} (2024) 217}
  [\href{https://arxiv.org/abs/2308.07493}{{\ttfamily 2308.07493}}].

\bibitem{Barbieri:1988fh}
R.~Barbieri and R.~N. Mohapatra, \emph{{A Neutrino With a Large Magnetic Moment
  and a Naturally Small Mass}},
  \href{https://doi.org/10.1016/0370-2693(89)91423-8}{\emph{Phys. Lett. B}
  {\bfseries 218} (1989) 225}.

\bibitem{Babu:1990wv}
K.~S. Babu and R.~N. Mohapatra, \emph{{Large transition magnetic moment of the
  neutrino from horizontal symmetry}},
  \href{https://doi.org/10.1103/PhysRevD.42.3778}{\emph{Phys. Rev. D}
  {\bfseries 42} (1990) 3778}.

\bibitem{Babu:1990hu}
K.~S. Babu and R.~N. Mohapatra, \emph{{Radiative fermion masses and large
  neutrino magnetic moment: A Unified picture}},
  \href{https://doi.org/10.1103/PhysRevD.43.2278}{\emph{Phys. Rev. D}
  {\bfseries 43} (1991) 2278}.

\bibitem{Barr:1990um}
S.~M. Barr, E.~M. Freire and A.~Zee, \emph{{A Mechanism for large neutrino
  magnetic moments}},
  \href{https://doi.org/10.1103/PhysRevLett.65.2626}{\emph{Phys. Rev. Lett.}
  {\bfseries 65} (1990) 2626}.

\bibitem{Pal:1991qr}
B.~K. Pal, \emph{{Zee model on Majorana neutrino mass and magnetic moment}},
  \href{https://doi.org/10.1103/PhysRevD.44.2261}{\emph{Phys. Rev. D}
  {\bfseries 44} (1991) 2261}.

\bibitem{Lindner:2017uvt}
M.~Lindner, B.~Radov\v{c}i\'c and J.~Welter, \emph{{Revisiting Large Neutrino
  Magnetic Moments}},
  \href{https://doi.org/10.1007/JHEP07(2017)139}{\emph{JHEP} {\bfseries 07}
  (2017) 139} [\href{https://arxiv.org/abs/1706.02555}{{\ttfamily
  1706.02555}}].

\bibitem{Glashow:1970gm}
S.~L. Glashow, J.~Iliopoulos and L.~Maiani, \emph{{Weak Interactions with
  Lepton-Hadron Symmetry}},
  \href{https://doi.org/10.1103/PhysRevD.2.1285}{\emph{Phys. Rev. D} {\bfseries
  2} (1970) 1285}.

\bibitem{Rosado:1999yn}
A.~Rosado, \emph{{Physical electroweak anapole moment for the neutrino}},
  \href{https://doi.org/10.1103/PhysRevD.61.013001}{\emph{Phys. Rev. D}
  {\bfseries 61} (2000) 013001}.

\bibitem{Cabral-Rosetti:2002zyl}
L.~G. Cabral-Rosetti, M.~Moreno and A.~Rosado, \emph{{Dirac neutrino anapole
  moment}}, \href{https://doi.org/10.1063/1.1489777}{\emph{AIP Conf. Proc.}
  {\bfseries 623} (2002) 347}
  [\href{https://arxiv.org/abs/hep-ph/0206083}{{\ttfamily hep-ph/0206083}}].

\bibitem{Zralek:1997sa}
M.~Zralek, \emph{{On the possibilities of distinguishing Dirac from Majorana
  neutrinos}}, {\emph{Acta Phys. Polon. B} {\bfseries 28} (1997) 2225}
  [\href{https://arxiv.org/abs/hep-ph/9711506}{{\ttfamily hep-ph/9711506}}].

\bibitem{Kayser:1981nw}
B.~Kayser and R.~E. Shrock, \emph{{Distinguishing Between Dirac and Majorana
  Neutrinos in Neutral Current Reactions}},
  \href{https://doi.org/10.1016/0370-2693(82)90314-8}{\emph{Phys. Lett. B}
  {\bfseries 112} (1982) 137}.

\bibitem{Passarino:1978jh}
G.~Passarino and M.~J.~G. Veltman, \emph{{One Loop Corrections for e+ e-
  Annihilation Into mu+ mu- in the Weinberg Model}},
  \href{https://doi.org/10.1016/0550-3213(79)90234-7}{\emph{Nucl. Phys. B}
  {\bfseries 160} (1979) 151}.

\bibitem{Hahn:1998yk}
T.~Hahn and M.~Perez-Victoria, \emph{{Automatized one loop calculations in
  four-dimensions and D-dimensions}},
  \href{https://doi.org/10.1016/S0010-4655(98)00173-8}{\emph{Comput. Phys.
  Commun.} {\bfseries 118} (1999) 153}
  [\href{https://arxiv.org/abs/hep-ph/9807565}{{\ttfamily hep-ph/9807565}}].

\end{thebibliography}\endgroup

\end{document}